\documentclass{aa}
\usepackage{amsmath}
\usepackage[varg]{txfonts}
\usepackage{hyperref}
\usepackage{cleveref}

\usepackage{natbib}
\bibpunct{(}{)}{;}{a}{}{,} % to follow the A&A style

\def\lsim{\mathrel{\rlap{\lower4pt\hbox{\hskip1pt$\sim$}}
    \raise1pt\hbox{$<$}}}                % less than or approx. symbol
\def\gsim{\mathrel{\rlap{\lower4pt\hbox{\hskip1pt$\sim$}}
    \raise1pt\hbox{$>$}}}                % greater than or approx. symbol

\def\plotonekindaspecial#1{\centering \leavevmode
    \includegraphics[angle=0,width=1.8\columnwidth]{#1}}

\def\plottwo#1#2{\centering \leavevmode
    \includegraphics[angle=0,width=0.98\columnwidth]{#1} \hfil
    \includegraphics[angle=0,width=0.98\columnwidth]{#2}}

\def\plotone#1{\centering \leavevmode
    \includegraphics[angle=0,width=1.0\columnwidth]{#1}}

\def\eps@scaling{1.0}% 
\newcommand\epsscale[1]{\def\eps@scaling{#1}}% 

\usepackage{marginnote}
\usepackage{color}
\begin{document}

\title{The VIMOS Ultra Deep Survey: Emerging from the dark, a massive proto-cluster at $z\sim4.57$\thanks{Based on data obtained with the European Southern Observatory Very Large
Telescope, Paranal, Chile, under Large Program 185.A-0791.}}
%\title{The VIMOS Ultra Deep Survey: Emerging from the Dark, a Massive Proto-Cluster Discovered in the Early Universe\thanks{Based on data obtained with the European Southern Observatory Very Large
%Telescope, Paranal, Chile, under Large Program 185.A-0791.}}
%\title{The VIMOS Ultra Deep Survey: Emerging from the Dark, a Massive Proto-Cluster at the Edge of the Reionization Era\thanks{Based on data obtained with the European Southern Observatory Very Large
%Telescope, Paranal, Chile, under Large Program 185.A-0791.}}
% This dong is considerably more competitive with other dongs. 
\author{B.~C.\ Lemaux\inst{1,2} \and O. Le F\`evre\inst{2} \and O. Cucciati\inst{3,6} \and B. Ribeiro\inst{2} \and L. A. M. Tasca\inst{2} \and G. Zamorani \inst{3} \and O. Ilbert\inst{2} 
\and R. Thomas\inst{2,4} \and S. Bardelli\inst{3} \and P. Cassata\inst{4} \and N.~P. Hathi\inst{2, 11} \and J. Pforr\inst{2,12} 
%\and V. Le Brun\inst{1} 
%\and L. Pentericci\inst{5} \and E. Vanzella\inst{2} 
\and V. Smol\v{c}i\'{c}\inst{7}
\and I. Delvecchio\inst{7}
\and M. Novak\inst{7}
\and S. Berta\inst{7}\thanks{Visiting Scientist}\inst{, 8}
%\and C. Laigle\inst{9,10}
\and H.~J. McCracken\inst{9,10}
\and A. Koekemoer\inst{11}
\and R. Amor\'in\inst{13,14}
\and B. Garilli\inst{5}
\and D. Maccagni\inst{5}
\and D. Schaerer\inst{16,15}
\and E. Zucca\inst{3}
%\and P. Capak\inst{15}
%\and L.~P. Cassar\`a\inst{4}
%\and A. Cimatti\inst{6}
%\and A. Durkalec\inst{1,17}
%\and A. Fontana\inst{5}
%\and M. Giavalisco\inst{16}
%\and A. Grazian\inst{5}
%\and A. Koekemoer\inst{11}
%\and M. Scodeggio\inst{4}
%\and V. Sommariva\inst{6,5}
%\and M. Talia\inst{6}
%\and L. Tresse\inst{1}
%\and D. Vergani\inst{11,2}
%\and S. Charlot\inst{10}
%\and N. Scoville\inst{15}
}
%}

% \affil{Aix Marseille Universit$\acute{\rm{e}}$, CNRS, LAM (Laboratoire d'Astrophysique de Marseille) UMR 7326, 13388, Marseille, France}
\institute{Department of Physics, University of California, Davis, One Shields Ave., Davis, CA 95616, USA \email{bclemaux@ucdavis.edu}
\and
%2
Aix Marseille Universit\'e, CNRS, LAM (Laboratoire d'Astrophysique de Marseille) UMR 7326, 13388, Marseille, France 
\and
%3
INAF--Osservatorio Astronomico di Bologna, via Gobetti, 93/3, I-40129, Bologna, Italy           %, via Ranzani,1, I-40127, Bologna, Italy
\and
%4
Instituto de Fisica y Astronom\'ia, Facultad de Ciencias, Universidad de Valpara\'iso, Gran Breta$\rm{\tilde{n}}$a 1111, Playa Ancha, Valpara\'iso Chile
\and
%5
INAF--IASF, via Bassini 15, I-20133,  Milano, Italy
\and
%6
%INAF--Osservatorio Astronomico di Roma, via di Frascati 33, I-00040,  Monte Porzio Catone, Italy
%\and
%7
University of Bologna, Department of Physics and Astronomy (DIFA), V.le Berti Pichat, 6/2 - 40127, Bologna, Italy
\and
%8
Department of Physics, Faculty of Science, University of Zagreb, Bijeni\v{c}ka cesta 32, 10000
\and
%9
Max-Planck-Institut für extraterrestrische Physik (MPE), Postfach 1312, 85741, Garching, Germany
\and
%10
Sorbonne Universit\'e, UPMC Universit\'e Paris 6 et CNRS, UMR 7095, Institut d'Astrophysique de Paris, 98 bis Boulevard Arago, 75014 Paris, France
\and
%11
Institut d'Astrophysique de Paris, UMR 7095 CNRS, Universit\'e Pierre et Marie Curie, 98 bis Boulevard Arago, F-75014 Paris, France
%\and
%12
%INAF--IASF Bologna, via Gobetti 101, I--40129,  Bologna, Italy
\and
%12
Space Telescope Science Institute, 3700 San Martin Drive, Baltimore, MD 21218, USA
\and
%13
European Space Agency, ESA/ESTEC, Keplerlaan 1, NL-22001 AZ Noordwijk, The Netherlands
\and
%14
Cavendish Laboratory, University of Cambridge, 19 J.J. Thomson Ave., Cambridge CB30HE, UK 
\and 
%15
Kavli Institute for Cosmology, University of Cambridge, Madingley Road, Cambridge CB30HA, UK 
%14
\and
Department of Astronomy, University of Geneva
ch. d'\'Ecogia 16, CH-1290 Versoix, Switzerland
%\and
%15
%Geneva Observatory, University of Geneva, ch. des Maillettes 51, CH-1290 Versoix, Switzerland
%\and
%16
%Department of Astronomy, California Institute of Technology, 1200 E. California Blvd., MC 249-17, Pasadena, CA 91125, USA
%\and
%17
%Astronomy Department, University of Massachusetts, Amherst, MA 01003, USA
%\and
%18
%Astronomical Observatory of the Jagiellonian University, Orla 171, 30-001 Cracow, Poland 
}

\date{Received March 25th, 2017/ Accepted January 15th, 2018} % 14 March 2014, resubmitted July 23rd, 2014, accepted July 29th, 2014
%\email{brian.lemaux@lam.fr}
\abstract{Using spectroscopic observations taken for the Visible Multi-Object Spectrograph (VIMOS) Ultra-Deep Survey (VUDS) we report here on the discovery of PCl J1001+0220, a 
massive proto-cluster of galaxies located at $z_{spec}\sim4.57$ in the COSMOS field. With nine spectroscopic members, the proto-cluster was initially 
detected as a $\sim12\sigma$ spectroscopic
overdensity of typical star-forming galaxies in the blind spectroscopic survey of the early universe ($2 < z\lsim6$) performed by VUDS. It was further 
mapped using a new technique developed which statistically combines spectroscopic and photometric redshifts, the latter derived 
from a recent compilation of incredibly deep multi-band imaging performed on the COSMOS field. Through various methods, the descendant mass of PCl J1001+0220 
is estimated to be $\log(\mathcal{M}_{h}/\mathcal{M}_{\odot})_{z=0}\sim14.5-15$ with a large amount of mass apparently already in place at $z\sim4.57$. An exhaustive 
comparison was made between the properties of various spectroscopic and photometric member samples and matched samples of galaxies inhabiting less dense environments 
at the same redshifts. Tentative evidence is found for a fractional excess of older galaxies more massive in their stellar content amongst 
the member samples relative to the coeval field, an observation which suggests the pervasive early onset of vigorous star formation for proto-cluster galaxies. 
No evidence is found for the differences in the star formation rates (SFRs) of member and coeval field galaxies either through 
estimating by means of the rest-frame ultraviolet or through separately stacking extremely deep Very Large Array 3 GHz imaging for both samples. Additionally,
no evidence for pervasive strong active galactic nuclei (AGN) activity is observed in either environment. Analysis of \emph{Hubble Space Telescope} images of 
both sets of galaxies as well as their immediate surroundings provides weak evidence for an elevated incidence of galaxy-galaxy 
interaction within the bounds of the proto-cluster. The stacked and individual spectral properties of the two samples are compared, with a definite
suppression of Ly$\alpha$ seen in the average member galaxy relative to the coeval field ($f_{esc, \, \rm{Ly}\alpha} = 1.8^{+0.3}_{-1.7}$\% and 
$4.0^{+1.0}_{-0.8}$\%, respectively). This observation along with other lines of evidence leads us to infer the possible presence of a large, cool,
diffuse medium within the proto-cluster environment evocative of a nascent intracluster medium forming in the early universe.}  
%This is still quite abstract, I would prefer something more definite.}
%\abstract{ At first I thought my first dong was competitive but this dong far exceeds most dongs in terms of competitiveness. There are only two
%dongs in the universe known to humankind which exceed this one in length and girth}
\titlerunning{VUDS discovery of a high-redshift proto-cluster}
\authorrunning{B.~C.\ Lemaux et al.}
\maketitle

\section{Introduction}

The last decade and a half has seen a revolution in the study of overdensities in the early universe. While the study and careful characterization of large 
associations of galaxies in the local universe has been possible for nearly a century, and in the intermediate redshift universe for a 
significant fraction of that time (e.g., \citealt{shapleynames26, shapley30, zwicky37,abell58,zwicky61}), the study of their progenitors presented 
several practical problems which have prevented their study until relatively recently. The primary problem, 
inherent to the study of nearly all galaxy populations in the early universe, is the extreme apparent faintness of galaxy populations at these distances. 
While some phenomena exist in the early universe, such as quasars or radio galaxies, which are so powerful and intrinsically bright that they have been able to serve as beacons to early searches near 
the epoch of \ion{H}{I} reionization ($z\sim5.5-10$, \citealt{thebob01, becker15,planck16}), the bulk of the galaxy 
population residing in the early universe does not contain such phenomena \citep{gmileznfriend08, ouchi08,lem14b, ueda14, marghe16}. As such, the first 
searches for these more typical primeval galaxies were largely doomed to failure \citep{davisnwilkinson74,partridge74, prichetnhartwick87, parkes94}. It was not until
the advent of the 10m-class ground-based telescopes largely used in conjunction with the \emph{Hubble} Space Telescope that the prospect of detecting and 
characterizing moderate samples of such galaxies became even remotely feasible (e.g., \citealt{steidel99, alice03, giavalisco04,stanway04,sangeeta04,sangeeta05, vanzella05, dong05}). With this, 
the prospect of finding and characterizing analogs of the progenitors of the massive clusters and superclusters of galaxies scattered throughout the local universe began 
to come within the realm of possibility.  

However, other issues remained. At least in the local and intermediate-redshift universe ($z\la1.5$), massive overdensities of galaxies are relatively rare 
\citep{piffaretti11,campanelli12} requiring searches over large swaths of the sky. While redshifts derived from imaging data alone can, with careful calibration, 
be used to select relatively pure and complete samples of overdensities at these redshifts, and while \emph{en masse} small-scale clustering of at least certain types of populations
may be reproducible with carefully crafted photometric redshift schemes (e.g., \citealt{menard13, skiddishsam13, aragon-calvo15}), the internal structure and dynamics 
of such overdensities are impossible to 
characterize on an individual basis without considerable investment of telescope time to obtain spectroscopic followup observations (see, e.g., 
\citealt{oke98, dressler99, halliday04,gal08, dressler13} and references therein). As both of these issues are potentially exacerbated at higher redshift ($z\ga1.5$), searches for overdensities 
at such redshifts were potentially 
required to both be larger and to compensate for larger levels of impurity and incompleteness than their lower redshift analogues. The large amounts of telescope time necessary to 
detect and confirm even the most modest number of individual high-redshift galaxies made such large-scale searches prohibitive. Further, the study of galaxy overdensities, 
clusters, superclusters, and, to a lesser extent, groups of galaxies, in the local and intermediate-redshift universe takes a somewhat regularized form. Such overdensities 
are known to contain certain markers that allow them to be detected readily, though with varying degrees of purity and completeness. To some 
degree, nearly all massive overdensities in the low- to intermediate-redshift universe are marked by a sequence of bright, redder galaxies (e.g., \citealt{gladdersnyee05, 
superB14, superB15}) and by a hot medium, detectable, at least in principle, in the X-ray through bremsstrahlung emission and at mm/cm wavelengths through the Sunyaev-Zel'dovich 
effect (e.g., \citealt{pierre04, pierre16, muchovej07,piffaretti11,hasselfield13, thecreature13,clerc14, breem15}). Additionally, the main component of the mass of these overdensities
typically takes a regular, triaxial shape, which, combined with their immense mass and favorable geometric conditions, allowed for at least some of these
overdensities to be detected through weak or strong gravitational lensing (e.g., \citealt{paczynski87, soucail87, fort94, clowe98,gladders03, jee06,hoekstra13,schrabback16}). 
While exceptions exist, it was these signposts that allowed for 
the success of early systematic searches for such overdensities, even with relatively moderate means. Analogs of the progenitors of these populations 
do not, however, necessarily share these signposts. The increasingly short time allowed for galaxies to form and evolve, to generate and heat an overarching medium, and for 
an overdensity to build up even a remote fraction of its eventual total mass, requires that the utility of all the methods mentioned above must
necessarily decrease and eventually fail as searches move to higher redshift.

The first searches for high-redshift overdensities, historically termed proto-clusters, attempted to overcome these deficiencies, with considerable success, 
by utilizing the previously mentioned beacons, quasars, other radio-emitting active galactic nuclei (AGN), and extremely prodigious star-forming galaxies 
also known as sub-mm galaxies \citep{ivison00,Lpentz00, venemens02,GMilez04, zheng06, overzier08}. The initial exploration of the surroundings of such phenomena took the 
form of large-scale deep narrowband imaging which allowed for the determination of the density of specific types of star-forming galaxies known as Lyman$\alpha$ 
$\lambda$1215.7\AA\ emitters (LAEs) as searches for such galaxies in this manner were relatively economical in terms of telescope time. Despite the success of these searches, 
it was not clear that such beacons were a requisite condition for a proto-cluster to form in the early universe and, if not, whether the proto-clusters found 
around such phenomena were typical. Further, only preliminary estimates were available of how the clustering of LAEs related to the clustering of galaxies as a whole 
(e.g., \citealt{ouchi03}), the former being a relatively small subset of the overall population \citep{schenker12, paolo15}, which, at 
least when selected through narrowband imaging, are potentially biased towards specific types of galaxies (e.g., \citealt{imfukingericgawiser06, trainor15}). While later surveys 
employed identical techniques to target other recombination lines, such as, for example, the H$\alpha$ $\lambda$6563\AA\ 
feature (e.g., \citealt{hatch11, matsuda11}), both these surveys and those of LAEs potentially suffered from lack of spectroscopic confirmation of the 
presumed features. Though some earlier methods 
detected and attempted to characterize proto-clusters through the photometric selection of a more representative population of high-redshift galaxies known 
as Lyman Break Galaxies (LBGs), the lack of redshift 
resolution of early LBG selection techniques ($\Delta z \sim1$) meant that only a few such selected overdensities held up to scrutiny (e.g., 
\citealt{steidel98}).

In recent years, however, searches have begun to shift to large samples of $L^{\ast}$-type galaxies using a variety of different techniques. While 
systematic spectroscopic characterization is still a luxury, large deep photometric surveys combined with dedicated spectroscopic followup of interesting 
patches of the sky, as well as archival searches of compilations of observations, are starting to produce results, with regular, well-defined selection 
criteria for proto-clusters \citep{capak11, coolchiang14, dey16, wang16, francknmcgaugh16b, francknmcgaugh16a, tersetoshikawa16}.
In addition, contemporary searches are persisting over large regions of the sky using radio or sub-mm galaxies \citep{wylezalek13, conivingcasey15, vernesa17}
allowing for a comparison sample to those proto-clusters selected blind to such beacons, crucial to understanding whether or not the presence of such phenomena biases
the types of proto-cluster environments found in these searches (see, e.g., \citealt{hatch14}).
Correspondingly, major advances in the understanding of proto-clusters 
are coming from the study of both N-body simulations and semi-analytic models (e.g., \citealt{coolchiang13, muldrew15, orsi16}). Both observations and models are converging 
on the realization that proto-clusters are extremely large in volume, contain diverse galaxy populations, and are more or less readily detectable depending on the galaxy 
population observed.  
Developing selection criteria applicable to all types of proto-clusters, a precise, well-characterized methodology of defining membership, and a consistent 
method to construct comparable lower-density galaxy samples, all of which were elusive in earlier searches for proto-clusters due to the inhomogeneity in 
both the conception of what a proto-cluster is and the observations employed for the search, is crucial to advancing the study of environmentally-driven 
evolution in the early universe. With new large-scale deep photometric and spectroscopic surveys imminent, such as those planned with the Large Synaptic Survey Telescope 
(LSST), Euclid, Hobby-Eberly Telescope Dark Energy Experiment (HETDEX), and Subaru Prime Focus Spectrograph (PFS), all of which capable, in principle, of detecting 
extremely large numbers of proto-clusters, the need to determine an operational definition of these structures becomes imperative.  

The wide-scale detection and careful characterization of high-redshift overdensities is not simply an academic exercise. Many open questions remain regarding the formation 
and early development in the life of present-day massive clusters and the constituent galaxies which seed them. While several studies of $z\sim1$ clusters 
have placed the formation epoch of the massive red galaxies prevalent in such structures at $z_{f}\sim3-4$ \citep{rettura10,raichoor12,lem12, greg12}, 
observational evidence of rapidly forming massive galaxies in proto-clusters at these redshifts is sparse. Such activity, if observed, would contain natural 
pathways not only to form the requisite number of massive quiescent galaxies observed in intermediate-redshift clusters through gas depletion, but also 
to begin the transfer of baryonic mass away from the constituent galaxies and into the infant intracluster medium (proto-ICM) through merging 
activity, tidal stripping, stellar feedback, and AGN activity. The inferred presence of a dense medium in the few $z\sim2-3$ overdensities where such an investigation 
has been possible (e.g., \citealt{gobat11, olga14, lee16, wang16, whathappensifIdonthavemyticket17}) strongly suggests that the processes by which the earliest (proto-)cluster galaxies
are quenched occur coevally with the formation of this medium. The observation of pervasive AGN activity in many high-redshift proto-clusters 
also suggests a possible mechanism to pre-heat the proto-ICM, a necessary requirement to reproduce observed cluster scaling relations (see, e.g.,  
\citealt{hilton12, kravtsovnborgani12} and references therein). The large amount of diffuse gas possibly contained within the proto-ICM and in areas 
surrounding proto-clusters may cause such regions to be the last in the universe to undergo \ion{H}{I} reionization (e.g., \citealt{ciardi03}), though
it is also possible that, due to the increased density of star-forming galaxies, they are the first (e.g., \citealt{castellano16}). 

%But why would we care? When did environment start to impress itself of the evolution of galaxies, it should be early, but how early and in what form? Also,
%galaxy overdensities at $z\ga5-6$ can begin to put constraints on reionization models through all of that stuff George Becker was talking about. So, that's enough
%right? Here, we develop a new method that's widely applicable to any field with deep spectroscopy and photometry and used it to find n individual proto-cluster 
%observed just after the epoch of reionization through a blind spectroscopic survey of galaxies at $2<z\le6$. That's it. 

In this paper we report on the systematic search for high-redshift overdensities in the Cosmic Evolution Survey (COSMOS; \citealt{scoville07a}) field using new 
observations from the VIsible Multi-Object Spectrograph (VIMOS; \citealt{dong03}) taken as part of the VIMOS Ultra-Deep Survey (VUDS; \citealt{dong15}). These
observations were combined with new and archival deep multi-band imaging and used in conjunction with a new density-mapping technique to discover PCl J1001+0220, a massive 
proto-cluster at $z\sim4.57$ emerging just $\sim250-500$ Myr after the canonical end of \ion{H}{I} reionization. 
The structure of the paper is as follows. Section \ref{obsnred} provides an overview of the spectroscopic and imaging data available in the COSMOS field, as well as the derivation
of physical parameters of galaxies in our sample, with particular attention paid to observations from the VUDS survey. Section \ref{protocluster} describes
the search methodology employed used to discover PCl J1001+0220 and to quantify its statistical overdensity. 
In Section \ref{memprop} we describe the estimation of the total mass of PCl J1001+0220 and an investigation of the 
properties of its spectroscopically-confirmed members compared with properties of galaxies inhabiting lower-density environments. Finally, Section \ref{conclusions}
presents a summary of our results. Throughout this paper all magnitudes, including those in the infrared (IR), are presented in the AB system \citep{okengunn83, fukugita96}.
All equivalent width (EW) measurements are presented in the rest frame with negative EWs defined as features observed in emission.
Unless otherwise noted, distances are given in proper rather than comoving units. We adopt a 
concordance $\Lambda$ Cold Dark Matter ($\Lambda$CDM) cosmology with $H_{0}$ = 70 km s$^{-1}$ Mpc$^{-1}$, $\Omega_{\Lambda}$ = 0.73, and $\Omega_{M}$ = 0.27. While abbreviated for convenience, 
throughout the paper stellar masses are presented in units of $h_{70}^{-2} \mathcal{M}_{\ast}$, star formation rates (SFRs) in units of $h_{70}^{-2} \mathcal{M}_{\ast}$ yr$^{-1}$, 
total masses in units of $h^{-1}_{70} \mathcal{M}_{\ast}$, ages in units of $h^{-1}_{70}$ Gyr or $h^{-1}_{70}$ Myr, absolute magnitudes in units of $M_{AB}+5\log(h_{70})$, 
proper distances and areas in units of $h_{70}^{-1}$ kpc/Mpc and $h_{70}^{-2}$ kpc$^2$/Mpc$^2$, respectively, where $h_{70}\equiv H_{0}/70$ km$^{-1}$ s Mpc.  
 
\section{Observations}
\label{obsnred}

To date, the Cosmic Evolution Survey (COSMOS) field ([$\alpha_{J2000}$, $\delta_{J2000}$] = [10:00:28.6,+02:12:21.0]) is arguably the most well-studied patch 
of the entire sky. It has been observed at wavelengths which span nearly the entirety of the electromagnetic spectrum, a large fraction of which we 
utilize in this paper. In Appendix A we summarize the various datasets available in the COSMOS field.

\subsection{Spectroscopic data}
\label{spectra}

The primary impetus for the current study comes from the vast spectroscopic data available in the COSMOS field, with a nearly exclusive reliance on recent 
VIMOS spectroscopic observations taken as part of VUDS. We therefore begin here with a brief discussion of the spectroscopic survey whose data are used for this study. 

\subsubsection{The VIMOS Ultra-Deep Survey}
\label{dongdong}

The observations from which nearly the entirety of our results are derived were drawn from VIMOS
observations taken for VUDS, a massive 640-hour ($\sim$80 night) spectroscopic campaign reaching extreme 
depths ($i^{\prime}\la25$) over 1 $\Box^{\circ}$ in COSMOS, the 02h field of the VIMOS Very Large Telescope Deep Survey (VVDS-02h)\footnote{Depending on 
the author, this field can also be referred to as the first field (D1) of the Canada-France-Hawai'i Telescope Legacy Survey (CFHTLS-D1)}, and the Extended 
Chandra Deep Field South (E-CDF-S). The design, goals, and survey strategy of VUDS are described in detail in \citet{dong15} and are thus described here only briefly.

The primary goal of the VUDS survey is to measure the spectroscopic redshifts of a large sample of galaxies at redshifts $2\la z \la6$. To this end, target
selection was performed primarily through photometric redshift cuts, 
occasionally supplemented with a variety of magnitude and color$-$color criteria. These selections were used primarily to maximize the number of galaxies with redshifts likely 
in excess of $z\ga2$ (see discussion in \citealt{dong15}) and, in the COSMOS field, to complement the largely color-color selected zCOSMOS-Deep sample. This selection 
has been used to great effect, as a large fraction (72.4\%) of the galaxies with secure spectroscopic redshifts (see below) in VUDS-COSMOS are at $z>2$.
The main novelties of the VUDS observations are the depth of the spectroscopy, the large wavelength coverage that is afforded by the 50400s integration time 
per pointing and per grating with the low-resolution blue and red gratings on VIMOS ($R\sim230$), and the large sky area covered by the survey. The combination of wavelength 
coverage and depth along with the high redshift of the sample allows not only for spectroscopic 
confirmation of the Lyman$\alpha$ (Ly$\alpha$) emitter (LAE) galaxies, galaxies which dominate other high-redshift spectroscopic samples, but also for redshift determination 
from Lyman series and interstellar medium (ISM) absorption in those galaxies that exhibit no or weak emission line features. Thus, the VUDS data allow for a selection of a 
\emph{spectroscopic} volume-limited sample of galaxies at redshifts $2\la z \la6$, a sample that probes as faint as $\sim$$M^{\ast}_{FUV}$ at the redshifts of interest 
for the study presented in this paper \citep{paolo15}. 

The flagging code for VUDS is discussed in \citet{dong15}. For some of the analysis presented in the paper, we adopt spectroscopic flags = X2, X3, \& X4, where X=0-3\footnote{X=0 is reserved for target galaxies, 
X=1 for broadline AGN, X=2 for non-targeted objects that fell serendipitously on a slit at a spatial location separable from the target, and X=3 for those non-targeted 
objects that fell serendipitously on a slit at a spatial location coincident with the target. For more details on the probability of a correct redshift for a given flag, 
see \citet{dong15}.}, for which the probability of the redshift being correct is $\ga$75\% (hereafter ``secure spectroscopic redshifts''). For 
other portions of this paper we will take a statistical approach incorporating the likelihood of each spectroscopic redshift rather than relying on binary logic 
(see \S\ref{voronoi}). In total, spectra of 4303 unique objects were obtained as part of VUDS in the COSMOS field, 
with 2687 of those resulting in secure spectroscopic redshifts (a $\sim$60\% redshift success rate, a value consistent with that of the full survey). The spectroscopic
sampling of VUDS in the COSMOS field is uniform in that no VUDS pointing appreciably overlaps another in the field. 

While we could, in principle, use galaxies
targeted by VUDS in the other two fields of the survey (VVDS-02h and E-CDF-S), effectively doubling the control sample used in this study, we choose not to for two reasons. The first 
is that the COSMOS photometry is more varied and, in general, deeper than that of the other two fields. Our lack of need for excess statistical power of the 
control sample outweighs the potential for adding in any secondary effects due to unseen bias due to inputting different photometry in the spectral energy distribution (SED) fitting procedure. The second and more important reason is that the lack of comparable photometric 
depth in the other two VUDS fields causes the reconstruction of the density field (see \S\ref{voronoi}), at least at the current level of implementation, to be more 
suspect for both fields at the redshift of this study ($z\ga4.5$), meaning we cannot robustly discriminate between low- and moderate-density environments at these redshifts.  
% took out flag=-10, #s are 4594 with flag=-10 included and 2874 with flag=9 included 
For further discussion of the survey design, observations, reduction, redshift determination, and the properties of the full VUDS sample, see \citet{dong15}.
See also \citep{tizzytasca16} for the first VUDS data release, which is available through the Centre de donn\'eeS Astrophysiques de Marseille (CeSAM) database\footnote{http://cesam.lam.fr/vuds/DR1/}. 

\subsubsection{Other spectroscopic data}

In order to maximize the effectiveness of our search for overdensities in the region of the COSMOS field covered by VUDS, we
additionally drew spectroscopic redshifts from the zCOSMOS-Bright\footnote{http://www.eso.org/sci/observing/phase3/data\_releases/zcosmos\_dr3 \_b2.pdf} 
\citep{lilly07,lilly09} and zCOSMOS-Deep (Lilly et al.\ \emph{in prep}, \citealt{dutifuldiener13, dutifuldiener15}) surveys. Accounting for duplicate objects where a 
more secure spectroscopic redshift was available from VUDS, a total of 19485 secure spectroscopic redshifts for unique objects are available from zCOSMOS of which 2034 
are at $z>2$ and only a small percentage ($\sim$1\%) reach $z>3$. A small number of additional redshifts were also taken from \citet{conivingcasey15}, 
\citet{coolchiang15}, and \citet{dutifuldiener15} at $z\sim2.5$, all of which were    % XXX I should add the Wang/Elbaz proto-cluster, though not going to affect this work XXX
considered secure. 
While we include mention of the additional redshifts of these surveys here to demonstrate the precision and accuracy of our photometric redshifts 
(see following Section) and to contextualize the main subject of this paper in terms of the full search for overdensities (see \S\ref{voronoi}), it is important to 
note that none of these galaxies enter into the main portion of the analysis presented in this paper (i.e., $4\le z \le 5$) and, thus, differential selection 
effects resulting from different targeting strategies are irrelevant.

The variety of other spectroscopic observations taken across the COSMOS field (see \citealt{ilbert13} for a 
review) were not incorporated into this analysis either because they were at redshifts that are too low to be pertinent to this study or the redshifts were not public at the time of 
publishing. The one exception is the DEep Imaging and Multi-Object Spectrometer (DEIMOS; \citealt{fab03}) campaign undertaken in the COSMOS field by \citet{pcap11}
targeting luminous galaxies at $4.5<z<6.5$. However, these observations are offset too far from the main target in this study, the closest slit being located 7.2$\arcmin$ 
($\sim$3 Mpc) away from the center of our target (see \S\ref{protoclusterspec}). While these observations, which contain a roughly equivalent number of 
$z>4$ galaxies as VUDS-COSMOS, could be, in principle, useful for bolstering the high-redshift (coeval) field sample defined in \S\ref{mainproperties}               % 147 for VUDS with no flag=1/9, 148 from Capak et al. 2011, but they say nothing about flags/Q, just that the redshifts are secure  
concerns over differential bias between this campaign and that of VUDS far outweigh the gain in sample size. As such, we do not include these observations as part of this study. 

\subsection{Synthetic model fitting}
\label{SED}

Despite the high density and immense depth of the spectroscopic coverage in the COSMOS field, the majority of the objects in the field that are detectable to the
depth of our imaging data were not targeted with spectroscopy. For these objects, information can only be obtained through fitting to their SEDs in the observed-frame optical/NIR broadband photometry. For this study we adopt four different forms of SED fitting. The four methods are used in complementary 
fashion throughout the paper and the results from each method are compared internally expiating any relative bias.

\subsubsection{Photometric redshifts}
\label{photozs}

To estimate photometric redshifts for objects with redshifts left unconstrained from spectroscopic observations, we draw from the fitting performed 
on the most recent version of the COSMOS2015 catalog (v1.3, \citealt{laigle16}) which employs the use of 
{\sc Le Phare}\footnote{http://cfht.hawaii.edu/$\sim$arnouts/LEPHARE/lephare.html} \citep{stephane99, dreadolivier06, dreadolivier09} on point-spread function 
(PSF)-matched photometry from FUV to [8.0]. Photometric redshifts, used as priors for a second round of fitting and magnitudes, originally estimated from 
3$\arcsec$ apertures, are corrected to total magnitudes following the method of \citet{thibaud16} and used to derive associated physical quantities, for example, stellar masses, 
mean luminosity-weighted stellar ages, extinctions, and SFRs. The parameters
used for deriving these quantities are identical to those given in \citet{lem14a} and to those used for the following two methods in this section. For further details on
this process as well as the assumptions and parameterization input for the SED-fitting process, see \citet{laigle16}.

\begin{figure}
\epsscale{1}
\plotone{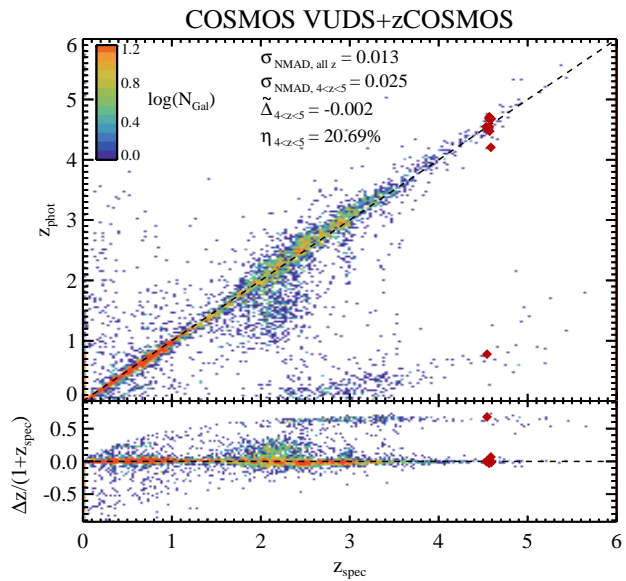}
\caption{Comparison of COSMOS2015 photometric redshifts ([3.6]$<$25.4) vs. secure spectroscopic redshifts (see \S\ref{spectra}) from the zCOSMOS-bright, 
zCOSMOS-deep, and VUDS surveys. Stars and known AGNs are included in the comparison as their identity is not generally known \emph{a priori}. A scale bar indicates the
density of objects in each region of the phase space. Filled red diamonds denote VUDS $z_{spec}$ members of PCl J1001+0220. The $\sigma_{NMAD}$ for the whole sample and for 
$4<z<5$ along with the median offset ($\Delta_{z} = (z_{phot}-z_{spec})/(1+z_{spec})$) and the percentage of catastrophic outliers
($|z_{phot}-z_{spec}|/(1+z_{spec}) > 0.15$) for the latter redshift range are shown in the main panel. Residuals are shown in the bottom panel. We note that the spectroscopic 
sample at $z>4$ has a median [3.6] magnitude nearly one magnitude brighter ($\Delta m=0.85$) than the $z_{phot}$ sample adopted in this study meaning that these values 
are likely lower limits. We note also that all known catastrophic outliers at these redshifts result from the SED 
fitting confusing the Lyman break for the Balmer break and not the reverse, thus resulting in what is likely a pure but incomplete sample at $4<z<5$. At $z<2$ the two zCOSMOS
samples dominate over VUDS providing 96.0\% of all secure spectroscopic redshifts. At $2\le z \le3$ the combined zCOSMOS samples and the VUDS sample are roughly matched in 
number with $\sim$2000 spectroscopic redshifts each. At higher redshifts, $z>3$ and $z>4$, the VUDS sample dominates providing 83.9\% and 97.3\% of the secure 
spectroscopic redshifts, respectively.} 
% numbers taken from the .sav file 
\label{fig:speczvsphotz}
\end{figure}

In Figure \ref{fig:speczvsphotz} we show a comparison of the photometric redshifts derived
from the COSMOS2015 catalog with the cut given above and the associated secure spectroscopic redshifts. The
normalized median of the absolute deviations, $\sigma_{NMAD}$ \citep{hoaglin83}, for the full sample (21781 objects) along with the $\sigma_{NMAD}$, the
median photo-$z$ offset ($\Delta_{z}/(1+z_s)$), and the catastrophic outlier rate ($|z_p-z_s|/(1+z_s) > 0.15$; $\eta$) for $4<z_{spec}<5$ (131 galaxies) are shown
in the main panel of
Figure \ref{fig:speczvsphotz}. If we instead adopt an alternative approach sometimes used to estimate photometric redshift precision and fit a Gaussian
to the $(z_{phot}-z_{spec})/(1+z_{spec})$ distributions we recover  $\sigma_{z/(1+z)}=0.008$ and 0.0145 for the full sample and the subsample at
$4<z<5$, respectively. These numbers are consistent with previous estimates of photometric redshift precision in the COSMOS field at similar redshifts
(e.g., \citealt{vernesa17}). 

Throughout the paper we conservatively adopt the $\sigma_{NMAD}$ estimate as the formal uncertainty on photometric redshifts. % XXX I think I should just average the two estimate, but whatever I do, make sure it's reflected in the density plot and with the estimates of likelihood of true membership, etc. XXX DONE, I used NMAD, all consistent 
This estimate, however, is still likely to be a lower limit to the true spread in the photo-$z$ population selected in this study for the following reasons. As the accuracy
and precision of photometric redshifts is a function of magnitude (see, e.g., \citealt{dreadolivier06, dreadolivier09}), the difference of more than a magnitude between
the median [3.6] magnitude of the spectral sample at $4<z_{spec}<5$ and the [3.6]-limited $z_{phot}$ sample selected in this study at similar redshifts
(see below), 23.3 versus 24.4, respectively, is disconcerting. While we attempt to mitigate any effects of underestimating the uncertainties
by using 1.5$\sigma_{NMAD}$ in instances where we use the global uncertainty, impose an additional $K_s<24.0$ criterion on all $z_{phot}$ objects
used to generate maps for this study (which forces the median [3.6] magnitude of $z_{phot}$ objects at these redshift to a value similar to that of the spectral
sample (23.2 vs. 23.3, respectively)), and adopt methods that rely on individual $z_{phot}$ errors (see \S\ref{voronoi}), this caveat should be kept in mind throughout the study.
An additional important subtlety of this analysis is understood through examining the spectroscopic redshift distribution of galaxies with $4<z_{phot}<5$ and
secure spectroscopic redshifts, which are almost always (77.1\% of the time) at $4<z_{spec}<5$.         % 84/109, COSMOS.speczvsphotozonly.goodflags.IRAC1le25.4.cat 
In all (24) cases where a galaxy is at $4<z_{spec}<5$ and the photometric redshift estimation failed catastrophically, the Ly$\alpha$/Lyman $\lambda$912\AA\ break was mistaken
for the Balmer/4000\AA\ break placing the galaxy at lower ($z\sim0.7$) redshifts (see also discussion in \citealt{pcap11} and in \S3.5 of \citealt{dong15}). The directionality
of these failures implies that the photometric redshift sample used at these redshifts, while somewhat incomplete, has a high level of purity at least for those galaxies
that we are able to test with our spectroscopy.
 
For analyses presented in this paper related to galaxy evolution we use only those objects with a $\ge3\sigma$ 
detection in [3.6] ([3.6] $<25.4$), cuts which apply both to spectroscopic and photometric objects. The median number of filters used in the fitting for the 
samples presented in this paper for which this cut is applied is 28 and the minimum number is 10. For the study of the properties of galaxies in different 
environments, such a cut is preferable to a, for example, $K_{s}$-selected sample for a variety of reasons. The primary redshifts of interest in this study are 
in the range $4 < z < 5$, a redshift range for which the $K_{s}$ filter begins to be sensitive primarily to rest-frame wavelengths blueward of the 
Balmer/4000\AA\ break. While the entire COSMOS2015 catalog is selected by a $z^{++}YJHK_s$ ``chi-squared" image \citep{szalay99}, the added requirement 
of a significant IRAC detection imposes that the sample selected at these redshifts be roughly stellar mass limited and minimally affected by a star-formation 
driven Malmquist bias. 
%Additionally, a [3.6] selection minimizes internal bias potentially resulting 
%from the UltraVISTA ``ultra-deep" strips, one of which cuts directly through the main area of interest for this study. These four strips, which cover 0.62 $\Box^{\circ}$ and run 
%vertically (N-S) at equally spaced intervals across the 1.5 $\Box^{\circ}$ full mosaic, probe significantly deeper than the full (``deep") survey ($K_{s, UD}\sim24.75$, 5$\sigma$). 
%While some differential effects will inevitably remain from lower formal errors calculated for photometric redshifts in the ultra-deep strips at a given $K_{s}$ magnitude 
%(everything else being equal), results involving photometric redshifts are generally treated statistically in this study, which should largely mitigate these effects. 
Additionally, it was shown by \citet{caputi15} that IRAC bright, $K_{s}$-faint sources ([4.5]<23, $K_{s}$>24) comprise $\ge$50\% of galaxies with large stellar masses 
($\log(\mathcal{M}_{\ast}/M_{\odot})\ge 10.75$) at these redshifts, a phenomenon we have verified to hold within the photometric redshift range adopted in this paper. 
Such galaxies would be missed if we had instead opted for a $K_{s}$-selected sample which would have resulted in a $\sim35$\% incompleteness at these stellar masses. 

For analyses that involve mapping of the field either in photometric redshifts or through a combination of photometric and spectroscopic redshifts (see \S\ref{voronoi})
we instead opt for a $K_{s}<24$ and [3.6] $<25.4$ sample. Such a cut allows us to keep a majority of high-stellar-mass galaxies in this redshift in the sample while  
forcing the average brightness of objects with $4 < z_{phot} < 5$ to those similar to the spectroscopic sample such that we can reasonably apply the statistics 
derived in this Section to this sample. Additionally, this cut yields a spectroscopic sampling rate (i.e., the number of objects targeted by all surveys vs. 
the total number of objects) within the area covered by VUDS to a value roughly twice that of a [3.6]-limited sample (16.3 vs. 9.8\%) and at $\sim$10\% for objects
in the magnitude range which place them as potential members (21 $<$ [3.6] $<$ 25.4). This point will be especially important when we statistically combine the
$z_{spec}$ and $z_{phot}$ samples in \S\ref{voronoi}. It should be noted, however, that all analyses presented in this study are relatively unaffected by making 
cuts which are less well motivated for both portions of the analysis as long as some sort of reasonable $K_{s}$ is applied to the galaxy sample used
to make various maps.

\subsubsection{Estimation of physical parameters}

For galaxies with secure spectroscopic redshifts, we used three different methods to derive associated parameters. The first was to use the package 
{\sc Le Phare} in a method identical to the one described in \citet{lem14a}. For this fitting we drew upon v2.0 of the PSF-matched photometry given 
in \citet{capak07} using all bands blueward and including \emph{Spitzer}/IRAC [4.5]. The \emph{Spitzer}/IRAC cryogenic bands ([5.8]/[8.0]) were excluded from the 
fitting due to relatively large ($\sim1-2.5$ mag) offsets seen in these bands in $\sim$30\% of all VUDS-COSMOS galaxies with respect to the best-fit model estimated 
without the use of these bands. We note that this issue appears limited to the \citet{capak07} catalog and does not apply to fitting performed with the COSMOS2015 photometry.
The process of PSF-homogenizing all UV/optical/ground--based NIR images, source detection, the inclusion of the \emph{Spitzer}/IRAC data,
and the conversion of all magnitudes to ``pseudo--total" magnitudes is described in detail in \citet{ilbert13}.  
Details of this fitting including the input parameters and the effect of various assumptions made for this fitting are given in
\citet{lem14a}. %Identical fitting was also performed on those galaxies without a secure spectroscopic redshift by fixing their redshift to the 
%photometric value. 

For some analyses, a similar fitting was performed on VUDS rest-frame 
near-ultraviolet (NUV) spectra in conjunction with identical photometry as in the {\sc Le Phare} analysis using {\sc GOSSIP+}, a modified version of the Galaxy Observed-Simulated SED 
Interactive Program (GOSSIP; \citealt{franzetti08}). The details of the modifications made for {\sc GOSSIP+} as well as the fitting process as it pertains to VUDS spectra 
and the advantages of the fitting over traditional photometric SED fitting, particularly in regards to estimating galaxy ages, are given in \citet{romain17a,romain17b}. 
Both methodologies, {\sc Le Phare} and {\sc GOSSIP+},  use some form of a $\chi_{\nu}^2$ minimization in order to recover the best-fit model from which the physical parameters 
are estimated. In addition, both methodologies employ a nearly identical set of input parameters including \citet{bc03} (hereafter BC03) stellar templates, models 
generated from exponentially declining and delayed star formation histories (SFHs), dust extinctions, and a \citet{chab03} initial mass function (IMF). For 
a full listing of the parameters adopted for the fitting see \citet{lem14a} and \citet{romain17a}.

The final method employs the three-component SED-fitting code {\sc SED3FIT}\footnote{Publicly available at \newline \url{http://cosmos.astro.caltech.edu/page/other-tools}} \citep{berta13}, 
which combines the emission from stars, dust heated by star formation, and a possible AGN emission component. The fiducial package of galaxy 
templates is taken from the Multi-wavelength Analysis of Galaxy PHYSical properties ({\sc MAGPHYS}; \citealt{dacunha08}) code, which relies on an energy balance between the
dust-absorbed stellar continuum and the reprocessed emission in the mid- to far-IR. Stellar templates are taken from the BC03 library, whose output
SEDs are modulated by the effects of dust attenuation and the stellar heating of dust \citep{charlotnfall00, dacunha08}. To this fiducial set of models, the {\sc SED3FIT} 
code implements a set of libraries which include both an accretion disc component and a warm dust component surrounding the AGN in a smooth toroidal structure 
(\citealt{feltre12}; see also \citealt{fritz06}) following the method of \citep{berta13}. For each galaxy in our sample, we run both the {\sc SED3FIT} and the {\sc MAGPHYS}
codes to the full COSMOS2015 photometry \citep{laigle16}, from FUV to 500$~\mu$m, using a common set of galaxy templates. The results of these two runs are then compared 
statistically (see \S\ref{multilam}) to estimate the relative contribution of various components to the global UV-IR SED (see \citealt{ivan14} for details). 
Though the \emph{Herschel} SPIRE/PACS data for the COSMOS field are relatively shallow for individual galaxies, stacked photometry of various galaxy 
sub-samples in this paper allows for the placing of meaningful constraints on both AGN activity and the amount of obscured star formation (see \S\ref{multilam}).

%The final method employs the Multi-wavelength Analysis of Galaxy PHYSical properties ({\sc MAGPHYS}; \citealt{dacunha08}) code in an attempt to estimate the contribution
%of active galactic nuclei (AGN) in our sample. The {\sc MAGPHYS} code is unique with respect to the other types of fitting presented in this study in that it is able to 
%incorporate input from the rest-frame NUV to sub-mm and simultaneously balance the energy of higher energy photons absorbed by dust and their re-emission in the mid- 
%to far-IR. The fiducial package contains BC03 stellar templates whose output SEDs are modulated by the effects of dust attenuation and the stellar heating of dust 
%\citep{charlotnfall00, dacunha08}. To this fiducial set of models we have incorporated within the framework of the code a set of libraries which include 
%both an AGN component and a warm dust component surrounding the AGN in a toridal geometry (\citealt{fritz06}, see also \citealt{feltre12}) following the method of 
%\citet{berta13}. For each galaxy in our sample, we run {\sc MAGPHYS} twice to COSMOS2015 photometry, once                          
%with the AGN component and once without it, fitting to all bands from FUV to 500$\mu m$. The results of these two runs are then compared statistically (see \S\ref{multilam})
%to estimate possible contributions of an AGN to any part of the UV-IR SED. For more details on {\sc MAGPHYS} and the implementation used here see \citet{dacunha08} and 
%\citet{berta13} and references therein. 

\section{Discovery of a $z\sim4.57$ proto-structure in the COSMOS field}
\label{protocluster}

In this Section we describe the discovery and characterization of the highest redshift overdensity seen by VUDS. We begin by  
focusing exclusively on the spectroscopic overdensity, as it was in these data that the proto-structure was first detected. However, the observed characteristics 
and magnitude of the overdensity of VUDS proto-structures are sensitive to a variety of different aspects of the distribution of the VUDS targets spatially, in galaxy type, 
and in brightness. As such, we additionally classify this overdensity through the use of photometric redshifts and by combining photometric and spectroscopic 
redshifts in a probabilistic manner. These methods paint a coherent picture of a highly significant overdensity assembling in the early universe.

\subsection{Overdensity as seen by spectroscopy}
\label{protoclusterspec}

\begin{figure*}
\epsscale{1}
\plottwo{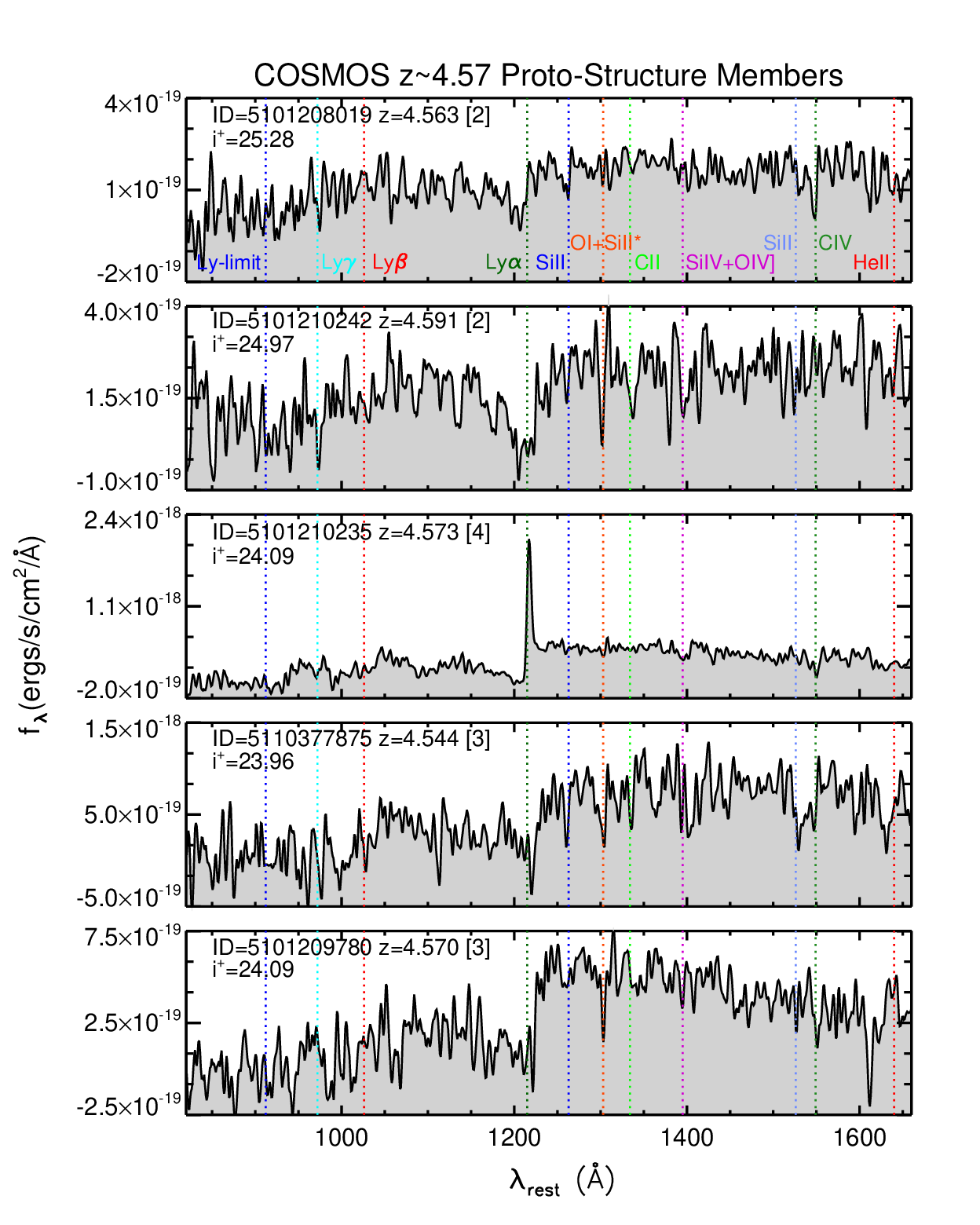}{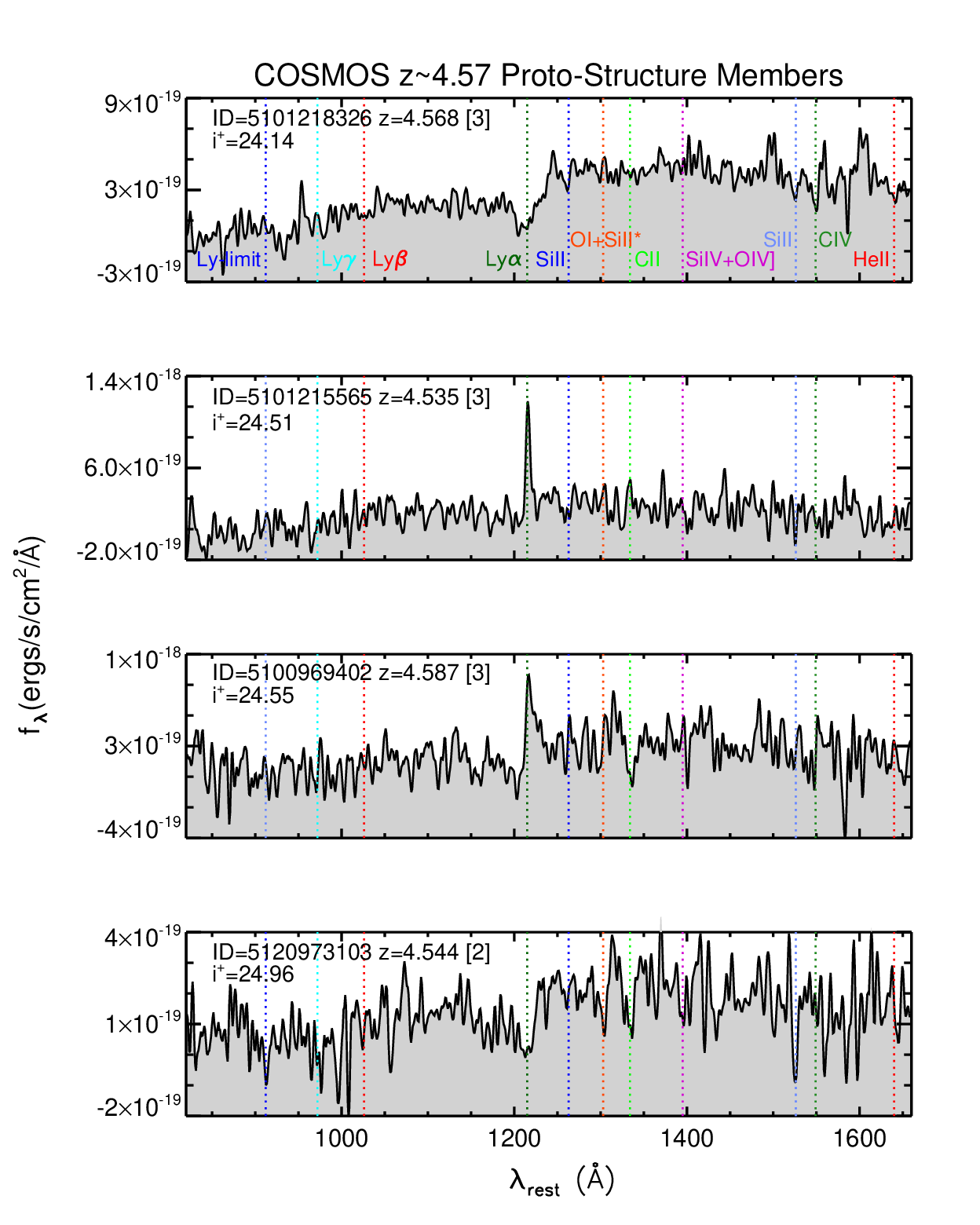}
\caption{Individual rest-frame VUDS spectra of the nine $z_{spec}$ members of the $z\sim4.57$ proto-structure detected in COSMOS with secure spectroscopic redshifts (see \S\ref{dongdong} 
for the meaning of this phrase). 
%The signal and noise spectra are plotted as the black solid and magenta dashed lines, respectively. The noise spectra are taken directly 
%from the reduction process. 
The spectra were smoothed with a Gaussian filter of $\sigma$=1.5 pixels ($\sim1.5$\AA\ in the rest-frame). The ID of each member galaxy along with its spectroscopic   % sigma=1.5pix in the observed frame, 1.4A/pix in the rest-frame 
redshift, redshift quality flag, and observed $i^{+}$ magnitude are shown on the upper left hand corner of each panel. The typical flux density uncertainty 
as estimated by the NMAD scatter of each spectrum over the wavelengths plotted here is 1.5$\times$10$^{-19}$ ergs s$^{-1}$ cm$^{-2}$ \AA$^{-1}$. 
Colored dotted lines indicate important spectral features. We highlight that in many cases the Ly$\alpha$ emission feature is absent or an identical redshift would have been recovered 
even in its absence.} 
\label{fig:specmosaic}
\end{figure*}

The initial search for overdensities, generically and non-threateningly termed ``proto-structures" here and throughout this study, in the COSMOS field took an almost 
identical  form to that described in \citet{lem14a}. We briefly describe the process here. The 4895 galaxies in the COSMOS field with secure     % 1966 VUDS, 2037 zCOSMOS, 14 Casey+Diener,9 HETDEX, Chiang et al. 2015, I had only z>2 galaxies before, ok, 1.5 <= z < 2, 35 from VUDS, 834 from zCOSMOS, total is  
spectroscopic redshifts at $z>1.5$ available to us were combined into a single catalog. In the case of duplicates, we gave preference to the redshifts with the most 
secure flag, or, in the case of equal flags, to VUDS redshifts. This catalog was used to generate spectral 
density maps of the entire COSMOS field from $1.5\le z \le 5$ in 25 Mpc slices applying the nearest-neighbor method of \citet{guter05}. 
These maps were searched by eye for overdensities and all those overdensities which contained seven galaxies with concordant redshifts (i.e., in the same slice) 
within $R_{proj} \le 2$ Mpc were considered as proto-structures. For each proto-structure, new maps were generated iteratively decreasing the 
redshift range until the overdensity was maximized. Following the generation of the final maps, luminosity-weighted and unit-weighted 
$z_{spec}$ member centers were determined iteratively following the method of \citet{superB14} using $R_{proj} \le 2$ Mpc and the $K_{s}$ or [3.6] bands for 
luminosity weighting. In total, 26 such proto-structures were found in the COSMOS field rank ordered in increasing redshift, of which 
one of the most significant was reported in \citet{olga14}. The highest redshift of these and also one of the most significant in the entire COSMOS field, 
a proto-structure at $z\sim4.57$ spanning 7.5 Mpc\footnote{Equivalent to $\Delta v$$\sim$3800 km s$^{-1}$ at these redshifts.} along the LOS and containing 
nine $z_{spec}$  member galaxies (see Fig. \ref{fig:specmosaic}), serves as the subject of this paper.

The method to determine the significance of the spectral overdensity of each of these proto-structures was identical to that of \citet{lem14a} and \citet{vernesa17}. Briefly, for 
each proto-structure, a volume equivalent to that used to define that proto-structure, that is, ``filter", was randomly placed at 1000 spatial locations
securely ($\ge2$ Mpc) inside the borders of the VUDS+zCOSMOS spatial coverage\footnote{At $z>4$ the spatial locations were limited to the spatial 
coverage of VUDS as essentially no galaxies with secure spectroscopic redshifts from zCOSMOS exist at these redshifts}, avoiding gaps in coverage, and at a random 
central redshift. Because of the rapidly varying selection and effectiveness of the zCOSMOS and VUDS surveys as a function of redshift, random central redshifts 
for the filters were limited to certain redshift ranges depending on the redshift of the proto-structure: $1.5 \le z \le 2$, $2 \le z \le 3$, $3 \le z \le 4$, or 
$4 \le z \le 5$. The angular spatial constraints listed above ensure that the density of spectroscopic targeting varies by less than a factor of two over the entirety of the        % see spectral_sig_test_zgt4.pro on laptop, 305 -> 585 
area sampled, a negligible variation for this exercise. For each realization, galaxies in the spectroscopic catalog falling within the filter were counted, and the
 resulting distribution was fit to a Poissonian or Gaussian function, the former being used when the average number of galaxies falling within the filter was small. 

\begin{figure}
\epsscale{1}
\plotone{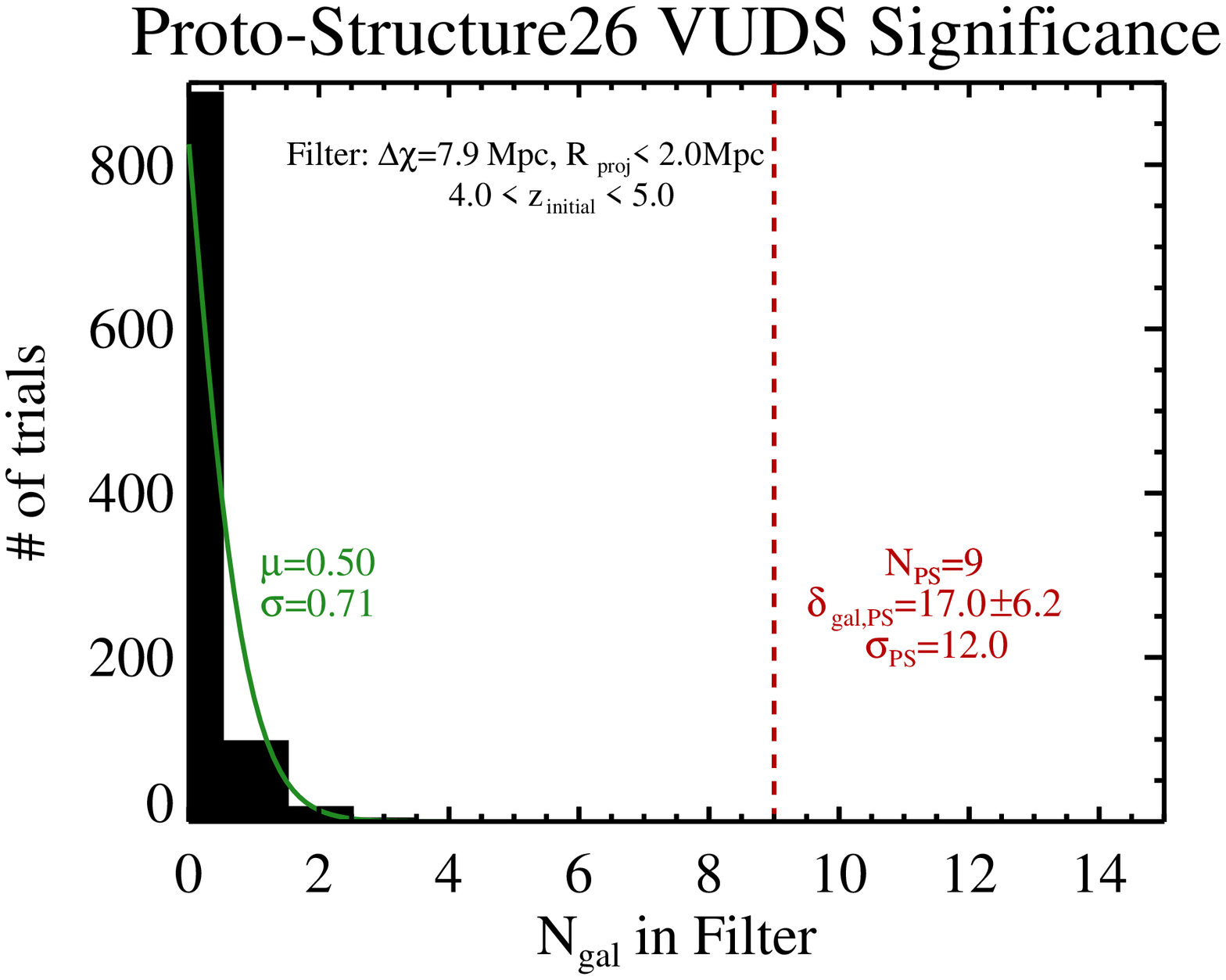}
\caption{Spectroscopic overdensity of the $z\sim4.57$ proto-structure as seen by VUDS. The black histogram shows the incidence of galaxies with secure spectroscopic
redshifts falling within a ``filter" of size given at the top of the plot for 1000 trial observations of the COSMOS field. For each observation the filter is
placed at a random (uniformly generated) spatial location over the VUDS footprint and given a random (uniform) redshift center in the range $4<z<5$.
The filter size is chosen to be identical to that adopted for this proto-structure. The green solid line is a Poissonian fit to the histogram with best-fit parameters
given to the right of the line. The number of $z_{spec}$ members of the proto-structure, $N_{PS}$, is denoted by the vertical dashed line. The spectroscopic overdensity
($\delta_{gal}=(N_{PS}-\mu)/\mu$) and detection significance ($\sigma_{PC}=(N_{PS}-\mu)/\sigma$) of the proto-structure is listed to the right of the line.}
%Coeval field is $4<z<5$. Uhhh... pretty significant. Put in photo-z one if so inclined, but mebbe not valid for these redshifts.}
\label{fig:specsig}
\end{figure}

This distribution and the resulting fit for the $z\sim4.57$ proto-structure is shown in Figure \ref{fig:specsig} and 
the resulting overdensity values given in Table \ref{tab:PSprop}. While there exist large formal uncertainties in the magnitude of the calculated 
spectral overdensity, $\delta_{gal}=17.0\pm6.2$, where $\delta_{gal}\equiv (N_{PS} - \mu)/\mu$, 
the overdensity is highly significant, representing a 12$\sigma$ fluctuation of the spectroscopic density field, that is, $\sigma_{PS}\equiv(\rm{N}_{PS} - \mu)/\sigma$, 
where $\mu$ and $\sigma$ are the mean and standard deviation, respectively, associated to the Poissonian fit performed above and N$_{PS}$ is the number of 
spectroscopically-confirmed proto-structure members. Such an overdensity 
is even more impressive given that the VUDS spectroscopic coverage does not continue eastward of RA$\sim$150.4, meaning only $\sim$70\% of the area bounded
by $R_{proj}\le2$ Mpc from the number-weighted $z_{spec}$ member center of the             
proto-structure was covered by VUDS. Given the relatively small number of confirmed $z_{spec}$ members, the non-uniform VUDS coverage, the sparsity of the 
coeval field, and the complicated nature of the VUDS selection at these redshifts, we refrain from pushing these values further and wait for 
further context presented later in this Section.

\subsection{Overdensity as seen by photometric redshifts}
\label{zphotoverdens}

\begin{figure*}
\epsscale{1.5}
\plotonekindaspecial{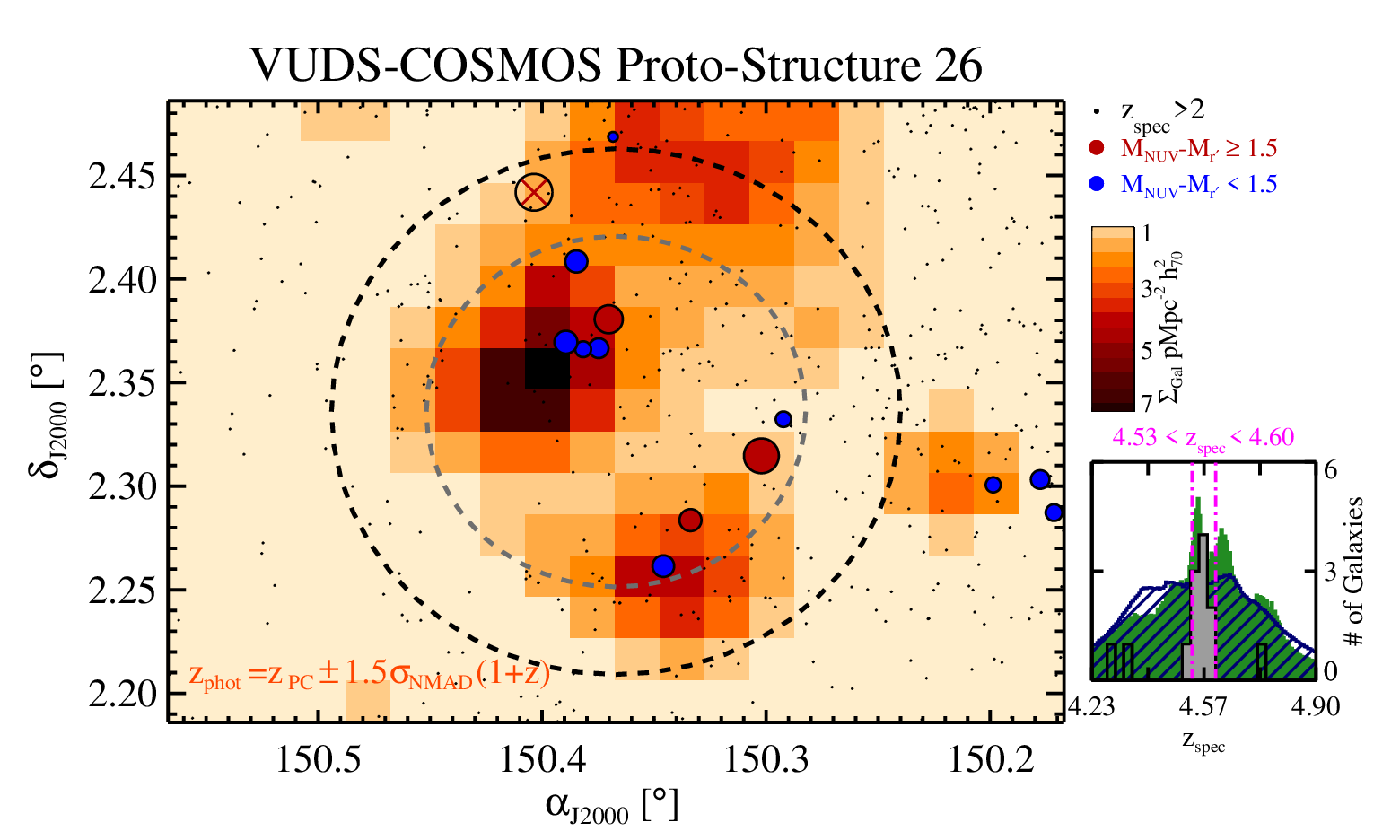}
\caption{Sky plot of the highest redshift proto-structure detected from VUDS spectroscopy (PCl J1001+0220). Plotted in the background is the 
smoothed (2.5 pixels FWHM) density map for all objects with $z_{phot}=4.57\pm 1.5\sigma_{NMAD}\times(1+4.57)$ with the scale in gal/Mpc 
indicated by the color bar to the right. Additionally, all galaxies from VUDS+zCOSMOS with $z>2$ with secure spectroscopic redshifts are plotted
in the background as small gray dots. The density of these points drops precipitously to the east of the proto-cluster center owing to a lack
of VUDS coverage of this area. Galaxies with secure and less secure (see \S\ref{spectra}) spectroscopic redshifts consistent with the redshift
range of the proto-structure ($4.53 < z< 4.60$) are represented by filled circumscribed circles and Xs, respectively. Blue and red symbols
differentiate galaxies at the redshift of the proto-structure by their rest-frame $M_{NUV}-M_{r^{\prime}}$ colors (younger and older than
200 Myr, respectively, see \S\ref{mainproperties}) and are logarithmically scaled ($\log_{4}$) by their stellar mass. The dashed gray and black circles
($R_{proj}= 2$ and 3 Mpc from the proto-structure center, respectively) set the boundaries for $z_{spec}$ and $z_{phot}$ members,
respectively. Plotted on the bottom right is a histogram of all secure spectroscopic redshifts within $R_{proj}<2$
Mpc of the proto-structure center. Plotted in the background is the combined PDF of $z_{phot}$ members (green filled) and non-members
(dark blue hatched) with redshifts consistent with the redshift of the proto-structure.}
%\caption{Dong only 1.2 Gyr after the big bang. Damn this dong is dangling different densities drilling deep distances. Green histogram
% is the combined PDF of the photo-z members define by the redshift and projected spatial cut, blue is that of the coeval field.}
\label{fig:radecphotozdens}
\end{figure*}

In Figure \ref{fig:radecphotozdens} we show the sky location of the $z_{spec}$ members of the $z\sim4.57$ proto-structure plotted against the backdrop of a 
density map of objects with photometric 
redshifts (hereafter $z_{phot}$) from the COSMOS2015 catalog, consistent with that of the proto-structure, that is, $z_{phot}\sim4.57\pm 1.5\sigma_{NMAD}\times(1+4.57)$, 
subject to the criteria given in \S\ref{SED} and following the nearest-neighbor method of \citet{guter05}. In order to determine the level and significance of the 
overdensity, if any, \texttt{SExtractor} \citep{BertinArn96} was run on the full COSMOS density map at the redshift of the proto-structure 
along with those at a variety of other redshifts. All peaks in the $z_{phot}$ density map that formally exceeded $5\sigma$ detections were cataloged. Spurious 
peaks were identified as those that had the requisite spectroscopic coverage but which lacked a corresponding spectroscopic overdensity at $\ge3\sigma_{PS}$, where
$\sigma_{PS}$ is defined earlier in this Section. As in \citet{lem14a}, 
a Gaussian was fit to the distribution of \texttt{SExtractor} significances of spurious peaks\footnote{Photometric redshift peaks were only considered
spurious if the peak fell within the spatial coverage of full spectroscopic catalog and failed to show a significant spectral overdensity.}, with the resulting parameters used to estimate the 
true (spurious-corrected) significance of measured overdensities. Here, and throughout this Section, we expand the projected area considered to be part of 
the proto-structure to 3 Mpc. While the results presented in this Section do not change appreciably if we instead consider the region which 
bounds the spectroscopic overdensity 
($R_{proj}\le 2 $ Mpc), we choose a larger radius here because it is better matched to the spatial size of proto-structures both in simulations 
(e.g., \citealt{coolchiang13, muldrew15, contini16,orsi16}) and in observations (e.g., \citealt{lem14a, dey16, tersetoshikawa14, tersetoshikawa16}) and due to our
ignorance of the true center of the proto-structure\footnote{Though all of these arguments equally apply to the spectral sample, in practice, partially due
to lack of VUDS coverage in the eastern portion of the proto-cluster, the spectral member sample remains unchanged if we instead enforce $R_{proj}\le 3 $ Mpc for that sample.}. 
Furthermore, the absence of VUDS observations eastward of $\alpha_{J2000}\sim150.40^{\circ}$ means that our spectroscopy certainly does not probe the full extent of the 
proto-structure. Thus, we take here an inclusive approach. There appears both a significant raw (20.7) and spurious-corrected (4.7) % I really ought to check this again 
overdensity in the region centered on the $z\sim4.57$ VUDS-COSMOS proto-structure. It is important to note this overdensity is not necessarily redundant evidence 
of an overdensity of galaxies as VUDS certainly does not target nor detect most $L^{\ast}$ galaxies at this redshift and the number of galaxies required 
to create such an overdensity, assuming the catastrophic outlier rate given in \S\ref{SED}, considerably exceeds the number of $z_{spec}$ members.

However, despite the use of arguably the best $z_{phot}$ measurements made to date, such measurements are moderately prone to fail catastrophically 
(see Fig. \ref{fig:speczvsphotz}) and, even if correct, to adopt the approach of a binary $z_{phot}$ membership criteria 
given the large extent in redshift space required by the $z_{phot}$ precision ($\Delta_z=0.42$) is perhaps overly             
coarse. While we will employ a more complex statistical combination of photometric and spectroscopic redshifts in the following Section to quantify the overdensity, 
we attempt one other method here to determine the genuineness of the $z_{phot}$ overdensity. The full probability distribution functions (PDFs) generated by 
\texttt{Le Phare} for each $z_{phot}$ member were re-constructed using the effective uncertainties in the $z_{phot}$ values and, when necessary, those of secondary
peaks in the $z_{phot}$ PDF which exceeded 5\% of the overall probability density. These re-constructed PDFs were combined into a single composite PDF shown in 
the bottom right panel of Figure \ref{fig:radecphotozdens} as a green filled histogram. Simultaneously, the same process was performed for all objects in the 
COSMOS2015 catalog, subject to the criteria set in \S\ref{SED}, with $z_{phot}=4.57\pm 1.5\sigma_{NMAD}\times(1+4.57)$ but outside of the projected spatial 
cut (coeval $z_{phot}$ field). The combined distribution of these objects is shown in the bottom right panel of Figure \ref{fig:radecphotozdens} as the blue hashed histogram. The combined 
PDF of the coeval $z_{phot}$ field sample appears roughly flat from $4.35\le z_{phot} \le 4.67$, a redshift range which contains $\sim$85\% of
relatively peaked $z_{phot}$ measurements ($\sigma_{z_{phot}} < 0.3$) in this sample and $\sim$80\% of all objects in the coeval $z_{phot}$ field. In stark contrast, 
the combined PDF of $z_{phot}$ members has its largest peak within the spectroscopic redshift bounds of the proto-structure. Indeed, by this analysis $z_{phot}$ members 
are nearly twice as likely as coeval $z_{phot}$ field objects to be within the true redshift bounds of the proto-structure, 16\% versus 9\%, respectively. Such a result
strongly suggests that the overdensity observed amongst the $z_{phot}$ objects at the location of the proto-structure is genuine. 

\subsection{Overdensity as measured by monte-carlo voronoi tessellation}
\label{voronoi}

We have now determined unequivocally that an overdensity is observed in both spectroscopic and photometric redshifts in the area encompassing 
PCl J1001+0220\footnote{Note that the prefix ``PCl "is meant as shorthand notation for ``proto-cluster". The validity of using this nomenclature will be justified 
at the end of this Section.} However, both of these methods have their limitations. In the case of the former, though accurate high-precision measurements are made, 
only $\sim$10\% of potential ($z_{phot}$) members were targeted for spectroscopy, with the targeting having a complex dependence on galaxy type, the level
of current star formation, and intervening structure. For example, because VUDS generally targets objects with $z_{phot}>2.4-1\sigma_{z_{phot}}$, a true projected 
dearth of galaxies along the LOS at lower ($z\sim2-4$) redshifts means a higher chance of targeting higher-redshift galaxies and, thus, finding a proto-structure 
like PCl J1001+0220. Conversely, while the $z_{phot}$ methods presented in the previous Section allow for an essentially stellar-mass-limited sample 
largely independent of galaxy type, though subject to the caveats discussed in \S\ref{photozs}, estimates are model dependent and have moderate accuracy and 
coarse precision at these redshifts. A method which combines both samples statistically can, in principle, help to mitigate the shortcomings of each sample.  

To this end, we introduce here a modified version of the Voronoi tessellation measure of the density field employed by a variety of 
other studies in the COSMOS field which rely almost exclusively on photometric redshifts (e.g., \citealt{scoville13, darvish15, vernesa17}). The method that we 
employ here most resembles the ``weighted Voronoi tessellation estimator" introduced in \citet{darvish15}. In that study it was found that using this method 
to recover the underlying density field matched or exceeded the accuracy and precision of all other methods of density estimation. The one metric with comparable 
performance to the Voronoi approach, weighted adaptive kernel estimation, is sensitive to both the form and size of the kernel and, generally, employs 
a spatially symmetric kernel (along the transverse dimensions) which is not ideal for the complex transverse shape of proto-structures. 

Our version of this method is as follows. Beginning at $z=2$ and reaching up to $z=7$ in 7.5 Mpc steps (i.e., half the size of a slice), a 
suite of ten Monte Carlo realizations of the magnitude-cut nearest-neighbor-matched master spectroscopic and COSMOS2015 $z_{phot}$ catalogs were generated for each step. 
This number of realizations was a compromise between computational intensity and stability of the resultant density estimates. For each Monte Carlo 
realization, first we selected the spectroscopic sub-sample by drawing from a uniform distribution ranging from 0 to 100 and retaining those $z_{spec}$ measurements
where the drawn number exceeded 75, 95, and 99.5 for those measurements with flags X2/X9, X3, and X4, respectively. These thresholds follow the fiducial reliability estimates of the 
VUDS/zCOSMOS flagging system (see, e.g., \citealt{dong15}). Such a method allowed for the incorporation of the statistical reliability of these measurements. 
If the pre-determined threshold was not exceeded, $z_{spec}$ would be retained for that realization, otherwise it was replaced with the $z_{phot}$ information. 
For each object where the  
$z_{phot}$ information was used or was the only information available, the original $z_{phot}$ for that object was perturbed by sampling from an 
asymmetric Gaussian distribution with $\sigma$ values that correspond to the lower and upper effective $1\sigma$ $z_{phot}$ uncertainties.

%a random value chosen from sampling a Gaussian distribution was multiplied by 
%either the effective $1\sigma$ lower or upper uncertainty on $z_{phot}$ for that object depending on whether the sampled point was negative or positive. 
%For each object, this value (positive or negative) was then added to its original $z_{phot}$ to create a new $z_{phot,\ MC_{i}}$ for that realization. Thus, 
%for each realization there existed a unique set of redshifts for all objects in the COSMOS field.  

Voronoi tessellation was then performed on each realization at each redshift step on objects with redshifts falling within $\pm$ 7.5 of the central
redshift of each bin, that is, a bin width of $\Delta \chi=15$ Mpc or $\Delta z\sim0.03-0.3$ from $z\sim2-7$. The 7.5 
Mpc steps between slices along with the slice thickness ensure overlap between successive slices such that we do not miss overdensities by randomly choosing unlucky redshift bounds. 
For each realization of each slice, a grid of 75$\times$75 kpc was created to sample the underlying local density distribution. The local density
at each grid value for each realization and slice was set equal to the inverse of the Voronoi cell area (multiplied by $D_{A}^2$) of the cell that encloses the
central point of that grid. Final local densities, $\Sigma_{VMC}$, for each grid point in each redshift slice are then computed by median combining the values
of ten realizations of the Voronoi maps. The local overdensity value for each grid point is then computed as
$\log(1+\delta_{gal}) \equiv \log(1+ (\Sigma_{VMC}-\tilde{\Sigma}_{VMC})/\tilde{\Sigma}_{VMC})$, where $\tilde{\Sigma}_{VMC}$ is the median $\Sigma_{VMC}$
for all grid points over which the map was defined, that is, excluding a border region of 
$\sim1\arcmin$ in width to mitigate edge effects.  In preliminary tests 
of the method employed for our study we observed realistic mock catalogs of proto-clusters and proto-groups with a combination of deep spectroscopy and
photometric redshifts with COSMOS2015-level precision and accuracy. It was found that the \emph{overdensity} field estimated by our method after an identical 
magnitude cut resembling that applied to our true sample used for the mapping (i.e., $Ks<24$) falls within $<30$\% of the true (i.e., real space) overdensity field
essentially independent of the value of overdensity, the redshift, or the spectroscopic completeness, as long as the latter is in excess of $\sim$10\%, a value
similar to that of this study (see \S\ref{photozs}). While reconstruction is still possible when the spectroscopic completeness is lower, the relative 
accuracy drops immensely, with the reconstructed values deviating by up to $\sim200$\% away from the true value when the sampling drops to $\sim3$\%. 
The results of these tests will be presented in depth in a future work.

\begin{figure*}
\epsscale{1.2}
%\plottwo{LSS.CFHTLSD1.zeq3.3.VUDS+VVDS.beststructure.dec2013.wAGN.eps}{CFHTLS-D1.beststructure.ds9screengrab.Icandobetter.eps}
%\plotonekindaspecial{COSMOS.protostructure26.RAdecplot.wVoronoiMCoverdens.ps}
\plotonekindaspecial{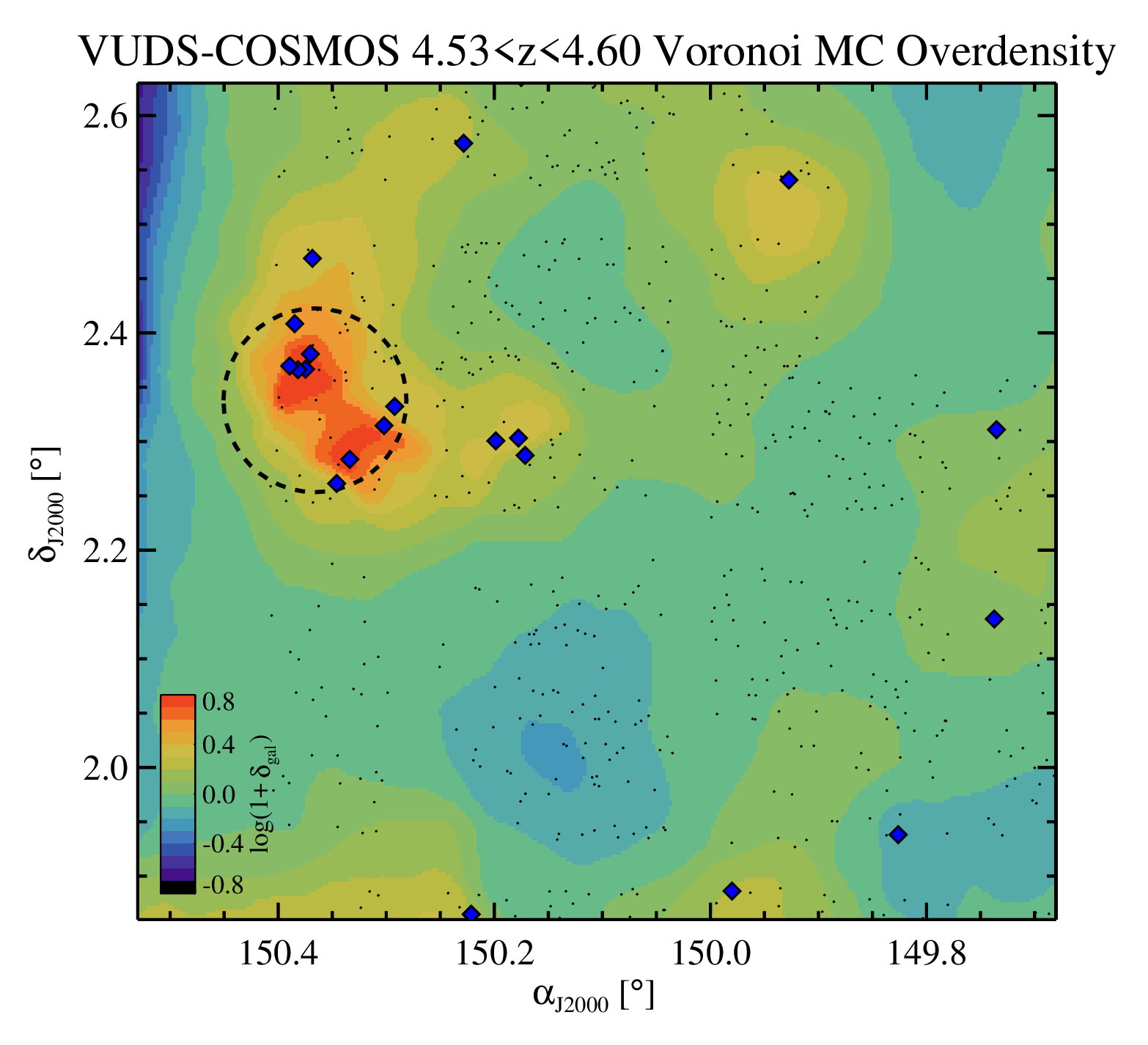}
\caption{Sky plot of the Voronoi Monte-Carlo overdensity map encompassing a large fraction of the VUDS coverage in the COSMOS field. The redshift bound over which the 
Voronoi tessellation is calculated is set to $4.53 < z < 4.60$ . Both spectroscopic and photometric redshifts are treated statistically following the prescription in
\S\ref{voronoi} and used generate 500 realizations of the overdensity map in the field. These 500 maps are then median combined and smoothed with a Gaussian of 
FWHM=4.5 pixel to create the plotted map. A color bar indicates the magnitude of the overdensity in each 75x75 kpc$^2$ pixel. The dashed line indicates the 
extent of our $z_{spec}$ member cut ($R_{proj} = 2$ Mpc). Small black dots denote galaxies with secure spectroscopic redshifts at $z>3$ from VUDS
and filled circumscribed blue diamonds denote those within the range of PCl J1001+0220. The PCl J1001+0220 proto-cluster exhibits coherent overdensity ($\log(1+\delta_{gal})>0.5$) 
over an area of 7.58 Mpc$^{2}$.}
% Dong only 1.2 Gyr after the big bang. Damn this dong is dangling different densities drilling deep distances. Dong Voronoi MC 500 iterations
%spec-z probabilistic SPLASH with some cuts. Look in the CMD directory to motivate those cuts, pretty good logic in one of the programs. Dong. Also, add 
%seg map boundaries. Also, make sure this is the final figure.}
\label{fig:radecvoronoi}
\end{figure*}
 
Once a proto-structure was identified, as in the case of PCl J1001+0220, zoom Monte Carlo Voronoi mappings were made by iteratively shifting the redshift 
bounds to find the values most appropriate to that particular proto-structure. For PCl J1001+0220, these values were found to be identical to the redshift 
range of the $z_{spec}$ members, $4.53 \le z \le 4.60$. Once the redshift bin was chosen, 500 Monte Carlo 
realizations were performed for that redshift slice, with the resultant final density and overdensity maps generated using an identical method to the mapping 
described earlier. The latter of these two maps for PCl J1001+0220 is shown in Figure \ref{fig:radecvoronoi}. Following the generation of these maps for 
PCl J1001+0220, \texttt{SExtractor} was run on the overdensity maps for the purpose of quantifying both the significance and the extent of the proto-structure using
$\log(1+\delta_{gal})>0.5$ as a detection threshold. The resulting detection of PCl J1001+0220 extends over an area of 7.58 Mpc$^{2}$ and has an average
$\log(\langle 1+\delta_{gal} \rangle)=0.63\pm0.03$. The uncertainty is determined from the dispersion around both the average density contained within the region
defining PCl J1001+0220 and the median density of the full slice as measured in all 500 realizations. We adopt this overdensity value and its associated uncertainty 
as being the most reliable estimate of the galaxy overdensity of PCl J1001+0220. Because this value 
exceeds the value for equivalent galaxy populations of nearly all simulated proto-clusters at these redshifts when viewed over an equivalent volume, 
($\sim$21 comoving Mpc$^{3}$, see Table 4 of \citealt{coolchiang13}), here and for the rest of the paper we abandon the term proto-structure when referring to
the PCl J1001+0220 proto-cluster. It is also perhaps revelant to note that at this galaxy overdensity, assuming the galaxy bias value given in \S\ref{weight} 
and pressureless spherical collapse, the entire proto-cluster as defined over the volume above collapses well before $z\sim0$.
Each overdensity value and central coordinate computed in this Section, as well as a variety of other information 
on the PCl J1001+0220 proto-cluster, is given in Table \ref{tab:PSprop}. 
 
\begin{table*}
\caption{General properties of PCl J1001+0220.\label{tab:PSprop}}
\centering
\begin{tabular}{cc}
\hline \hline
Spectral-number-weighted center & $[\alpha_{J2000},\delta_{J2000}]=[$10:01:27.8, 02:20:16.8$]$ \\[4pt]
Weighted Photo-$z$ density map barycenter & $[\alpha_{J2000},\delta_{J2000}]=[$10:01:24.8, 02:20:16.6$]$  \\[4pt]
Voronoi barycenter & $[\alpha_{J2000},\delta_{J2000}]=[$10:01:22.8, 02:20:16.1$]$ \\[4pt]
Number of spectral members & 9\tablefootmark{a} \\ [4pt]
Median redshift & $\tilde{z}=4.568$ \\[4pt]
Spectral overdensity & $\delta_{gal}=17.0\pm6.2$, $\sigma_{Proto-Struct.}=12.0$\\[4pt]
Photo-$z$ overdensity & $\sigma_{z_{phot},LSS}=$4.7\tablefootmark{b} (20.7)\\[4pt]
Voronoi average overdensity & $\langle \delta_{gal}\rangle = 3.30\pm0.32$ \\[4pt]
Galaxy velocity dispersion & $\sigma_{v}=1037.6\pm177.9$ km s$^{-1}$\\[4pt]
\hline
\end{tabular}
\tablefoot{
\tablefoottext{a}{This number refers to all spectroscopically-confirmed members with $R_{proj}<2$ Mpc and $4.53<z_{spec}<4.60$.} 
\tablefoottext{b}{The number outside the parentheses refers to the formal significance of the detection after accounting for spurious density peaks (see \S\ref{protocluster}).}
}
\end{table*}

\section{Properties of the PCl J1001+0220 proto-cluster}
\label{memprop}

With the identity of the PCl J1001+0220 proto-cluster now firmly established, in this section we attempt to contextualize this proto-cluster 
and to make a cursory exploration of the properties of its member galaxies using all the information at our disposal. 

\subsection{Weighing the PCl J10001+0220 proto-cluster}
\label{weight}

At lower redshifts ($z\le1.5$), a clear correlation exists between the total mass of a structure and the evolutionary state of its constituent galaxy 
population (e.g., \citealt{moran07,poggianti08,pog09, hansen09,lub09, lem12, vanderburg14, balogh16}). While dispersion in this correlation exists, and 
while the evolutionary state of a galaxy population clearly has a complex relationship with other factors either causally or circumstantially connected 
with global environment, 
such as local (over)density, stellar mass, and global dynamics, estimating the total mass for a structure is a necessary step in predicting the fate of 
that structure and its constituent galaxy population. At such redshifts reasonably precise estimates are at least achievable  
with current technology through strong or weak lensing, X-ray observations, and dynamics measurements from exhaustive spectroscopic campaigns. 
While widely employed, these estimates necessarily have (relatively) large systematic uncertainties originating from the large number of 
assumptions required to even attempt a measurement with these data. Furthermore, at the highest redshifts in this redshift range $z\sim1.5$, only the 
most exquisite data sets can be reasonably employed to estimate total masses, as sparser sampling, differential bias in sampling, and/or lower signal-to-noise ratio (S/N) 
data makes it impossible to begin to probe the validity of these assumptions. 

\begin{figure}
\epsscale{1}
%\plottwo{LSS.CFHTLSD1.zeq3.3.VUDS+VVDS.beststructure.dec2013.wAGN.eps}{CFHTLS-D1.beststructure.ds9screengrab.Icandobetter.eps}
\plotone{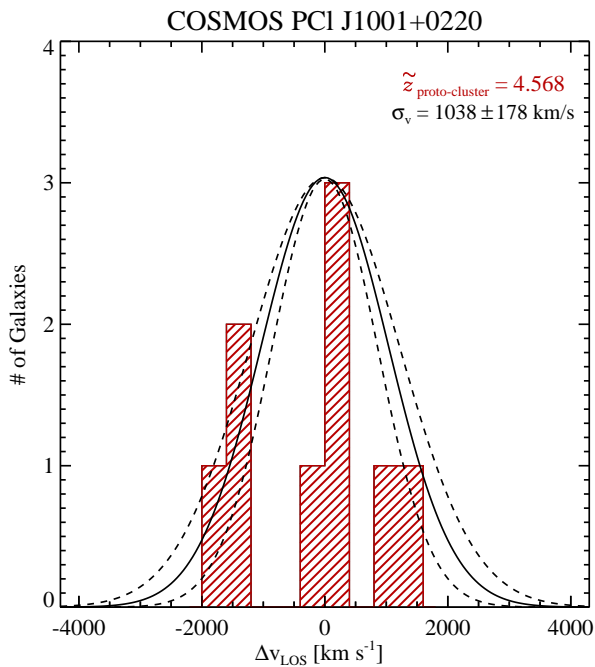}
\caption{Differential velocity distribution of the $z_{spec}$ members of PCl J1001+0220. The median
redshift of $z_{spec}$ member galaxies is shown in the top right corner of the plot. Also shown in the top right corner is the value of the best-fit
line of sight (LOS) velocity dispersion ($\sigma_{v}$, see \S\ref{protocluster} for details). The resulting Gaussian function generated by the
best-fit $\sigma_{v}$ is overplotted on the differential velocity histogram (solid black line) along with those functions generated from
$\sigma_{v}\pm\sigma_{\sigma_v}$.}
\label{fig:diffvel}
\end{figure}
 
At higher redshift, the situation becomes much more dire. Precision estimates of total masses through lensing and X-ray measurements become nearly impossible. 
Quiescent galaxies, whose global dynamics should, generally, better trace the underlying potential than other galaxy populations, become sparser 
and more difficult to detect spectroscopically, a requisite condition of any dynamics analysis. Furthermore, assumptions which generally come close 
to being valid for massive overdensities at lower redshift, for example, virialization, and hydrostatic equilibrium, almost certainly do not 
hold at these redshifts. Further, proto-clusters appear in simulations to have extremely complex morphologies, and overdensity estimates,
estimates which are subsequently translated into total masses, appear to be strongly dependent on viewing angle \citep{shattow13}. 
Given these difficulties, here we attempt a variety of different methods (as was done in \citet{lem14a}) to estimate the total 
mass of PCl J1001+0220. Because each of these methods requires a different set of assumptions, the above uncertainties are, to the best of our ability, minimized, 
allowing for at least the potential to triangulate the requisite accuracy to properly contextualize the PCl J1001+0220 proto-cluster 
when the estimates are averaged. Note that, for all proto-cluster estimates, the term ``total mass" is meant to refer to the composite mass 
of all halos which comprise the proto-cluster (or the eventual cluster) rather than the mass of the most massive halo in the (proto-)cluster. 
As pointed out in, for example, \citet{muldrew15}, the most massive halo for all high-redshift proto-clusters likely contains a small fraction 
of proto-cluster galaxies and a small fraction of the overall mass ($\la$10\% for both cases).

%To be briefly described...
Under the assumption of virialization and isotropic motion, the LOS dynamics of satellite galaxies provide a readily available proxy for the
total mass. Though the redshift of Cl1001+0220 makes it extremely unlikely that the observed members have reached a 
near-virialized state, and though the time at which a system is observed in its dynamical history dictates whether the dispersion of the 
observed LOS differential velocities is an overestimate or underestimate relative to it virial equivalent (see, e.g., \citealt{olga14}), neither 
of these possibilities precludes the possibility that such a measurement could provide useful information. Indeed, the member galaxies
in both simulations of cluster progenitors \citep{olga14} and in large compilations of observed proto-clusters \citep{faithlessFranck16b} exhibit velocity
dispersions, and, by consequence, dynamical masses, which correlate, weakly but signficantly, with masses derived through independent estimates. In 
Figure \ref{fig:diffvel} we plot the differential
LOS velocity dispersion of all nine members of Cl1001+0220 along with the biweight fit to distribution. The estimated LOS velocity dispersion 
of 1038$\pm$178 km s$^{-1}$ corresponds to a virial mass of: 

\begin{equation}  
\log(\mathcal{M}_{vir}/\mathcal{M}_{\odot})_{z\sim4.57, dyn} = 14.39\pm0.22
\label{eqn:Mvir}
.\end{equation}

\noindent While this is an immense total mass at this redshift, it is possible that galaxy dynamics provide a gross overestimate due the effects 
described above. The level by which this dynamical mass might exceed the true total mass can be crudely estimated by applying the average offset between 
dynamical masses and masses estimated by matter overdensities, in those cases where both were definitely measured, for the large ensemble of proto-clusters 
compiled by \citet{faithlessFranck16b}. Though it is not clear that overdensity masses necessarily more accurately reflect true total masses, the methods used in that
study differ from those used here, and there exists a large scatter in this distribution ($\widetilde{(\mathcal{M}_{dyn}/\mathcal{M}_{\delta_{m}}}) = 6.7^{+15.5}_{-4.6}$, 
median and effective $1\sigma$ scatter); such a correction can perhaps provide a reasonable range of values in which the true total mass resides. Applying this median 
correction and propagating its uncertainties into the resultant corrected value yields $\log(\mathcal{M}_{vir}/\mathcal{M}_{\odot})_{z\sim4.57, dyn, corr}=13.6^{+0.4}_{-1.0}$. 

A different approach is to count the amount of baryonic material within the proto-cluster bounds associated with the member galaxies
and attempt to relate that back to the overall mass of the structure, an approach which has been employed successfully at low redshift when the galactic 
baryonic content of cluster member galaxies is dominated by stars \citep{andreon12}. The approach we take here is similar to that of \citet{lem14a}.
Briefly, the total amount of stellar matter of the $z_{spec}$ members is counted and a completeness correction is made to this value for galaxies 
at stellar masses $\log(\mathcal{M}_{\ast}/M_{\odot})\ge 9.5$ (see \S\ref{mainproperties} for the reasoning behind this cut) based on the number of 
$z_{phot}$ members and non-members without secure spectral redshifts within the bounds of the proto-cluster and the likelihood of their being true 
members. An additional correction is made to correct for galaxies in the stellar mass range $8.0 < \log(\mathcal{M}_{\ast}/M_{\odot})< 9.5$ 
by integrating the stellar mass function of \citet{idolatrousiary17} appropriate for this redshift. Here we additionally make the assumption that stellar 
mass comprises 50\% of the baryonic content of galaxies by mass at these redshifts, a value broadly consistent with the few measurements made at 
or near these redshifts (e.g., \citealt{tacconi10, capak11, schinnerer16, scoville16}). We assume for the purposes of this calculation that
the proto-cluster is a closed system, with all gas being converted to stars by $z=0$ and that the completeness-corrected galaxy population which lies within 
$R_{proj}\le 3$ Mpc at $z\sim4.57$ comprises the entirety of the galaxy population which will eventually be contained within the cluster virial radius at $z=0$. 
The latter assumption is broadly consistent with simulations \citep{muldrew15} to $\sim$10\% accuracy, a factor which we account for in the calculation below. 
Any total to stellar mass conversion then provides a $z=0$ total mass, which we de-evolve to $z\sim4.57$ using the correction factors of \citep{muldrew15} appropriate 
for z$\sim$4.57 (0.20$\pm$0.03), a correction which is appropriate for descendents of all masses. Into this formalism we input the resulting completeness-corrected 
baryonic content of $\log(\Sigma\mathcal{M}_{\ast}/M_{\odot})=12.39^{+0.05}_{-0.07}$ to the $r_{200}$ stellar mass to $M_{500}$ total mass relation of 
\citet{andreon12} and scale the resulting $M_{500}$ to the virial radius ($R_{vir}=0.33$ Mpc) using the methods presented in \citet{lem14a} giving:

\begin{equation} 
\log(\mathcal{M}/\mathcal{M}_{\odot})_{z=4.57, \, \Sigma\mathcal{M}_{\ast, corr}} = 13.31^{+0.23}_{-0.27}  % updated 11/10/17 using method in referee report, multiplied original estimate by 0.2 +/- 0.03 and divided by 0.9, errors added in quadrature      
\label{eqn:MSM}
.\end{equation}

%$\log(\mathcal{M}_{vir}/\mathcal{M}_{\odot})_{z=4.57, X-ray} = $ TBD

\noindent We note that this estimate, within the context of this formalism, is almost certainly a lower limit to the total mass of the proto-cluster given that
there exists diffuse gas at $z\sim4.57$, which we do not account for here, of which some fraction will undoubtedly be accreted and used to create stars 
by $z=0$. This lack of precision may be mitigated somewhat by the fact that we also do not account for the loss of stellar and gaseous material in member galaxies due 
to, for example, tidal or ram pressure stripping or by merging activity as the proto-cluster evolves. Despite all the uncertainties related to this calculation, if the  
completeness-corrected baryonic content of the proto-cluster is instead used in conjunction with the universal dark-to-baryonic      % https://arxiv.org/pdf/1502.01589.pdf, table 3, column 5 (labeled 4) 
fraction of 6.42$\pm$0.19 \citep{planck16} to estimate the total mass, the result, $\log(\mathcal{M}/\mathcal{M}_{\odot})_{z=4.57, \, \Sigma\mathcal{M}_{\ast, corr}}$=13.20$^{+0.05}_{-0.07}$,
is nearly identical to the above calculations. Nevertheless, all three of these estimates place the descendant total mass of PCl J1001+0220 at or above $M_{h, z=0}=10^{14}$ once 
evolved to $z=0$ following the formalism of \citet{Mcbride09} and \citet{FakhourinMa10} (see \citealt{lem14a} for its implementation in this context) or \citet{muldrew15}. 
Again, however, these methods relied on a number of assumptions which, while perhaps applicable generally, may or may not be valid for this particular proto-cluster. 

The final method we employ here relies on the galaxy overdensity as measured from the Voronoi Monte Carlo technique in an attempt to relate this overdensity to the present day
total mass through large-scale simulations \citep{coolchiang13}. As such, though this method is model dependent, neither the galaxy velocities, the lack of complete spectroscopic sampling, nor the bulk properties of the galaxies themselves should affect this estimation. This estimation follows 
that of \citet{lem14a} nearly identically and, as such, we do not repeat the formalism here. The one major exception for this study is the replacement of 
the spectroscopic overdensity, used previously, with the Voronoi Monte Carlo overdensity making the estimate much less uncertain and much less dependent 
on sampling effects. As we have also changed the volume used to calculate this overdensity, a volume which is now set by the \texttt{SExtractor} detection 
(segmentation) map (see \S\ref{voronoi}) and the adopted redshift range of the proto-cluster, a correction is made following \citet{coolchiang13} to 
account for additional mass outside of the volume used (e.g., similar to an aperture correction, see also \citealt{muldrew15}). The correction factor is
taken to be 1/0.8, appropriate for the segmentation-estimated proto-cluster volume which has an effective radius ($R_{e}$) of 10.9 comoving Mpc (see \citealt{coolchiang13}
for more details). We have also adopted here a bias parameter of $b=3.60$, a reasonable estimate for the redshift and galaxy sample used here \citep{coolchiang13, ania17}. 
The final descendant mass of PCl J1001+0220 is then estimated to be: 

\begin{equation} 
\log(\mathcal{M}_{h}/\mathcal{M}_{\odot})_{z=0, (1+\delta_{gal})} = 14.48\pm0.03  
\label{eqn:Moverdens}
,\end{equation}

\noindent a descendant mass intermediate to those estimated by the other two methods. It is important to note, though, 
that this descendant mass is not cause for ambiguity in the overall nature of PCl J1001+0220; the galaxy overdensity value used here along with the 
area over which that overdensity was measured still places PCl J1001+0220 as a proto-cluster at $>$80\% confidence \citep{coolchiang13}. The 
$\mathcal{M}_{h, z=0, (1+\delta_{gal})}$ estimated above is broadly similar to the value derived from the formalism of \citet{orsi16}, a study which employs vastly different models and assumptions. %$\log(\mathcal{M}_{vir}/\mathcal{M}_{\odot})_{Orsi et al.\ 2015, radio} = 14.08\pm0.04$, if needed scale to the appropriate r* and mention that I scaled it
While it is necessary to assume for the latter that the proto-cluster contains a quasar or radio galaxy which has yet to be observed and that, further, the 
density of faint Ly$\alpha$ emitters scales directly with the density of $\sim$L$^{\ast}$ galaxies at these redshifts, the concordance in 
$\mathcal{M}_{h, z=0, (1+\delta_{gal})}$ estimates between these two methods is encouraging. Regardless of the method chosen, the large descendant mass 
estimated for PCl J1001+0220 requires, under the formalism of models predicting typical mass growth \citep{FakhourinMa10, muldrew15}, that considerable amount of mass
assembles ($\log(\mathcal{M}_{h}/M_{\odot})\ga12$) before the end of the epoch of reionization (i.e., $z\sim6$). As such, the progenitors of systems like
PCl J1001+0220 can be, under certain circumstances, useful in differentiating between different models of reionization (see \citealt{d'aloisio15, castellano16, davies16}
and references therein). Because none of our results are dependent on the 
exact total mass of PCl J1001+0220 either at $z\sim4.57$ or at $z=0$, we satisfy ourselves with the understanding that PCl J1001+0220 is a massive 
proto-cluster that will evolve into a cluster at low redshift with an overarching mass somewhere in between that of the Fornax and 
Coma clusters \citep{kentngunn82, collessndunn96,nasonova11}.   

%For the last one, the descendant halo masses were calibrated for a slightly smaller radius 
%6.97 vs. 8.94 co-moving, so this value is a lower limit to the true halo mass since a smaller area will give a higher average log(1+delta$_{gal}$). I can 
%either express this as an upper limit or I can tune the SEx threshold such that the measurement is made over an area with an reff = 7 co-moving Mpc. I did
%the latter and got an only slight difference, I think the dependence of the average log(1+delta$_{gal}$) on reff, at least at this level for this range of 
%radii, is pretty minimal. It's probably easier to frame in terms of this and say it's likely to be a slight underestimate. BUT, this overdensity is quoted 
%for overdensities of faint LAEs, so not sure how valid the comparison is here, i.e., how well does the log(1+delta$_{gal}$) from $L^{\ast}$ not necessarily 
%LAE galaxies couple to the log(1+delta$_{gal}$) of faint LAEs? Leave it as an open question. 

%A sentence or two saying that regardless the halo mass is large, probably somewhere between Fornax and Coma. The end.

\subsection{An investigation of environmentally-driven evolution at $z\sim4.57$}
\label{properties}

From the previous sections we have determined that the member galaxies of PCl J1001+0220 reside in a globally dense (i.e., proto-cluster) environment 
that exhibits a large-scale coherent elevated local density. Though strong correlations between galaxy properties and both local and global environment 
are observed at lower redshifts ($z\la2$, e.g., \citealt{goto03a,postman05, mcoopz07, muz12, lem12, cooke16, iary16, lem17}), such correlations, 
if they exist, are far from established at higher redshifts. In this section we use every method at our disposal to 
determine if the onset of environmental effects have imprinted themselves onto the galaxy population of PCl J1001+0220 just 1.3 Gyr after the Big Bang.  
 
\subsubsection{Luminosity, color, stellar mass, and age properties} 
\label{mainproperties}

We begin by comparing quantities derived from our various SED fitting techniques for PCl J1001+0220 member and coeval field samples defined in a 
variety of ways. We choose to use a variety of techniques as well as a variety of sample definitions to mitigate our ignorance of the true 
center of the proto-cluster, our ignorance of both the redshift and transverse extent of the member population, the lack of VUDS spectroscopic 
targeting in much of the eastern quadrants of the proto-cluster, and the various systematic and random uncertainties in SED fitting methods, 
uncertainties which are almost certainly exacerbated at these redshifts. Throughout these comparisons, the member and coeval field samples will always, 
by design, have the same median redshift and will be consistently cut at the various limits imposed. In Appendix B we describe in detail 
the definitions for the various samples used in the analysis in the remainder of this paper as well as the statistical approaches used to 
compare these samples. 

\begin{figure*}
\epsscale{1}
\plottwo{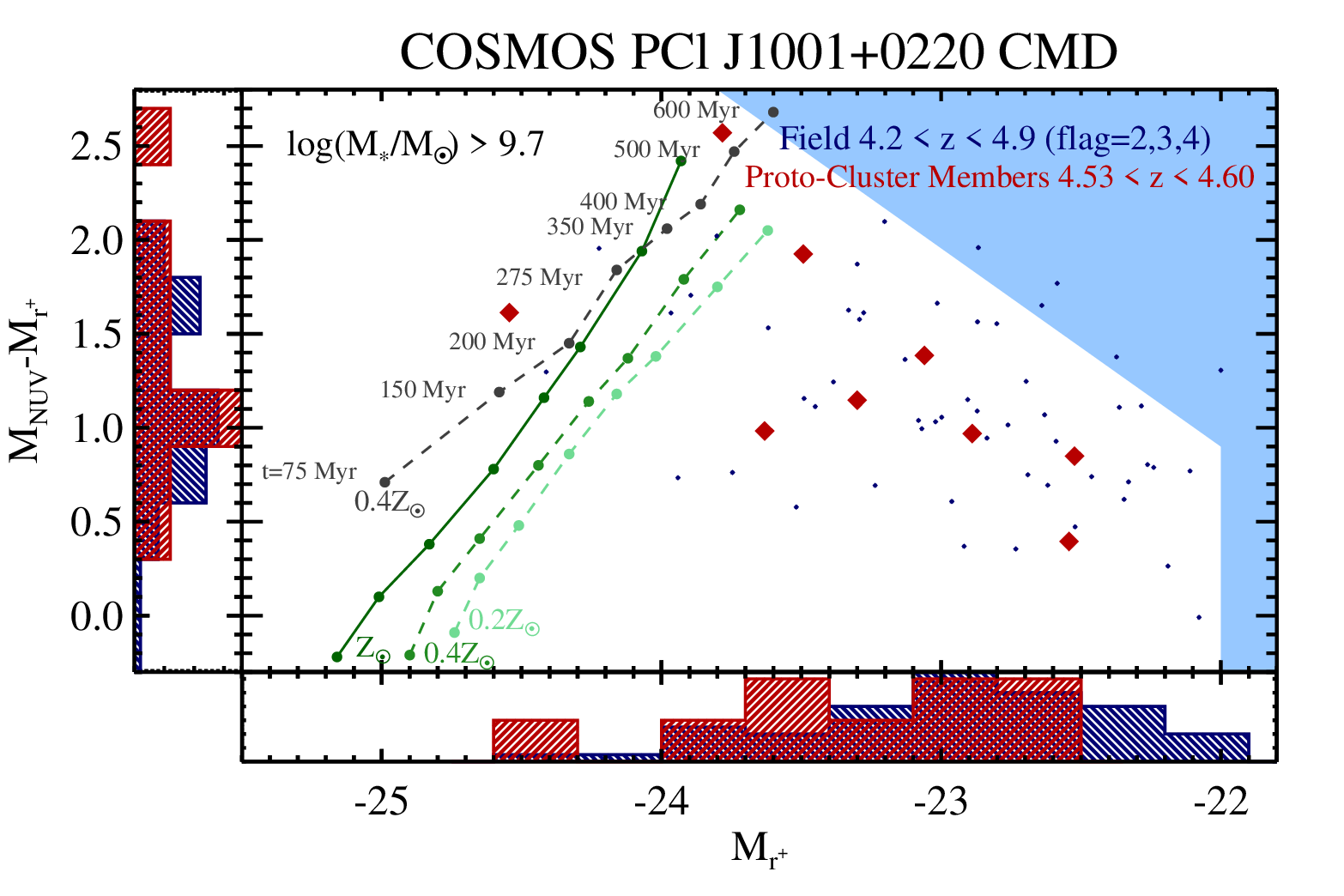}{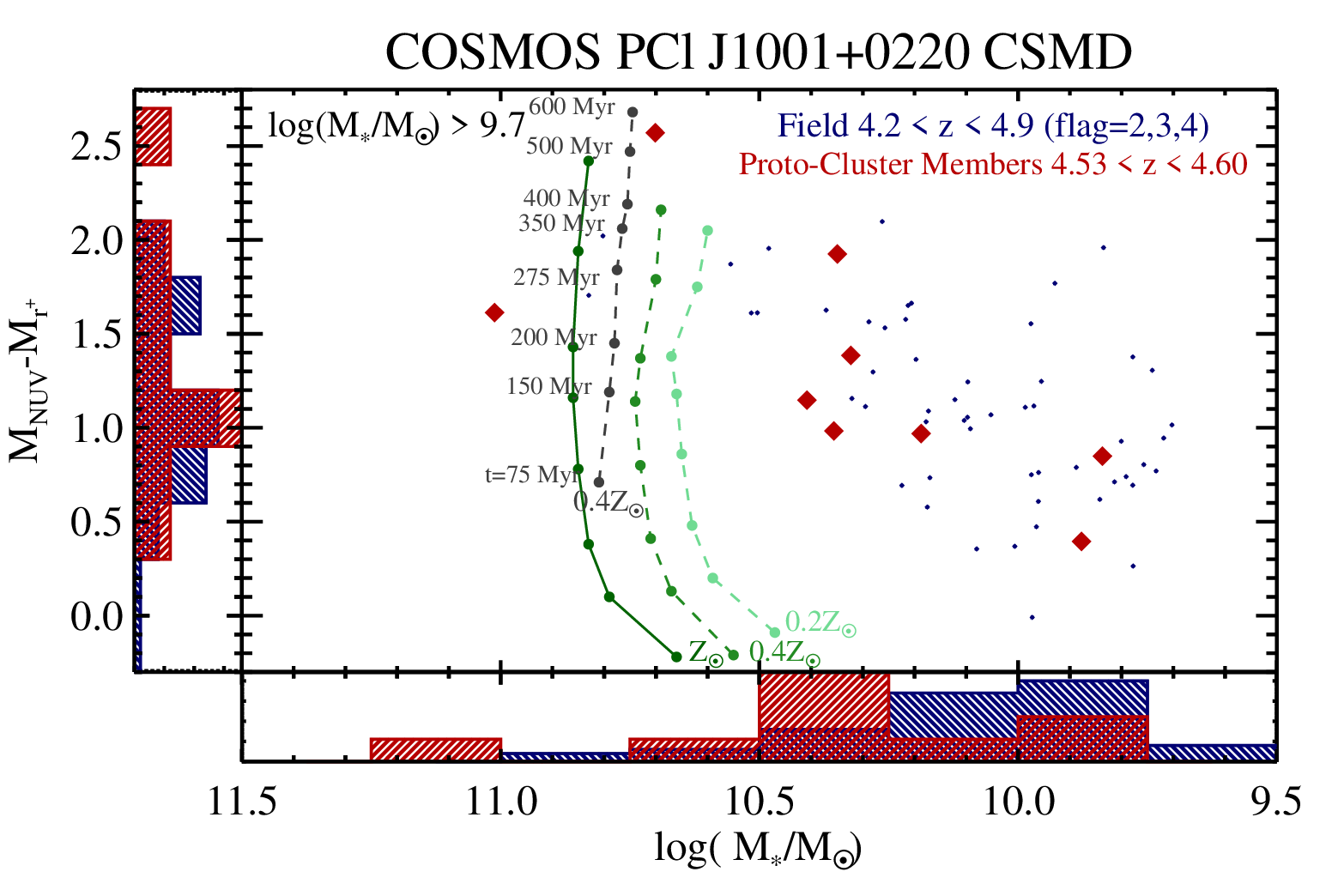}
\caption{\emph{Left:} Rest-frame NUV-$r^{+}$ vs. $r^{+}$ color-magnitude diagram for the $z_{spec}$ members of the PCl J1001+0220 proto-cluster (red diamonds)
and a matched coeval field sample at $4.23 < z < 4.88$ (small navy dots) at stellar masses $\log(\mathcal{M}_{\ast}/M_{\odot})\ge 9.7$. The stellar mass
cut is imposed here in an attempt to mitigate any induced differential bias between the field and proto-cluster members. Overplotted are the expected locations
in this phase space of a progentor of a low-redshift cluster $L^{\ast}$ galaxy (assuming passive evolution) estimated from synthetic spectra (see \S\ref{mainproperties}) at different
stellar-phase metallicities and with different SFHs. The gray dashed track is generated from an instantaneous burst while the three green tracks are
generated from an exponentially declining SFH with $\tau=100$ Myr at three different stellar-phase metallicities. Age ticks along each track indicate the time since
the inception of the star formation event. Color and magnitude histograms for each sample are normalized to each other such that they contain the same area.
The solid light blue area indicates the area of this phase space for which VUDS is no longer representative of the underlying photometric sample at these redshifts and
stellar masses (see \S\ref{mainproperties}). \emph{Right:}
Rest-frame $M_{NUV}$-$M_{r^{+}}$ vs. $\mathcal{M}_{\ast}$ color-stellar-mass diagram for the identical samples as in the left panel. The meanings of all symbols and lines
are unchanged. While some of the brightest, reddest, and most (stellar) massive galaxies in the entire spectroscopic sample at these redshifts fall within
PCl J1001+0220, only a moderately significant difference in the average brightness and a significant difference in the stellar mass distribution is detected
relative to the coeval field population.}
%appear to be skewed towards redder, brighter, and more (stellar) massive galaxies than the overall field population and
%the proto-cluster contains both the reddest and most massive galaxies in the entire VUDS sample at these redshifts.} % kstwo on the photo-z objects at this redshift (LSS +/- 1.5sigma_z*(1+z)) and the VUDS galaxies shows that at Mr < -22 nd MNUV-Mr < 2.7 the VUDS galaxies are representative of the photo-z sample
\label{fig:restframeCMDnCSMD}
\end{figure*}

In the left panel of Figure \ref{fig:restframeCMDnCSMD} we plot the rest-frame $M_{NUV}-M_{r^+}$ versus $M_{r^+}$ color-magnitude diagram (CMD) of the $z_{spec}$ PCl J1001+0220 
proto-cluster members against the backdrop of the coeval field for all galaxies with $\log(\mathcal{M}_{\ast}/M_{\odot})\ge 9.7$. The area of this phase space 
where VUDS begins to become unrepresentative of the underlying photometric sample is shaded out. 
These bands are chosen as they discriminate galaxies of different ages in the absence of a copious dust content \citep{stephane13, thibaud16}, which, as
we show below, is likely minimal, on average, in both samples. Additionally, by selecting these colors, we guarantee that each galaxy in the selected sample was
significantly detected in at least one observed-frame band that has reasonably similar wavelength coverage at $z\sim4.57$ to each of these rest-frame bands\footnote{All 
VUDS galaxies used in this paper are detected significantly in the $i^{+}$ band and are required to be detected significantly in the [3.6] band.}. Such 
a guarantee minimizes \emph{k}-correction necessary to calculate the rest-frame magnitudes and thus the model dependence of these magnitudes. The overplotted 
tracks show BC03 model-derived colors, absolute magnitudes, and stellar masses as generated by {\sc EZGal}\footnote{http://www.baryons.org/ezgal/}. These models, which employ 
a variety of different SFHs, stellar-phase metallicities, and epochs of the most recent major star formation event, were designed to match the 
brightness of a $z\sim0.6$ $L^{\ast}$ cluster 
galaxy \citep{depropris13} after passively evolving for $\sim$7 Gyr (i.e., the difference in lookback time between $z\sim4.57$ and $z\sim0.6$).  

%While differences appear to exist by eye between the various distributions of the member population relative to those of the coeval field, the relatively small number of 
%member galaxies frustrates a rigorous investigation of these differences. 

Despite having both what is, by far, the reddest galaxy in VUDS at these redshifts and several other redder galaxies fall within the $\sim$1\% of the 
volume of the VUDS-COSMOS 
field at these redshifts spanned by the proto-cluster, a statistical difference relative to the field is not observed. Neither a KS test nor a comparison of the 
average $M_{NUV}-M_{r^{+}}$ colors of the member and coeval field samples ($\widetilde{M_{NUV}-M_{r^{+}}}$=1.15$\pm$0.15 vs. 1.11$\pm$0.08, respectively) reveals any 
significant difference between the two populations. The proto-cluster also contains the brightest as well as three of the top ten brightest rest-frame optical galaxies 
in VUDS at these redshifts, and though, again, a KS test fails to find a significant difference ($\sim1\sigma$), a comparison of the average $M_{r^{+}}$ magnitudes yields 
marginally significant hints of a difference between the two populations ($\widetilde{M_{r^{+}}}$=-23.30$\pm$0.20 vs. -22.87$\pm$0.08, respectively). In the 
right panel of Figure \ref{fig:restframeCMDnCSMD} we plot a rest-frame color-stellar mass diagram (CSMD) of the same two samples. The trend in magnitude hinted in the 
CMD appears to be manifest in the CSMD more significantly. Two of the four most (stellar) massive galaxies in the entire spectroscopic sample at these redshifts, 
both of which appear to contain a suggestively old luminosity-weighted stellar population ($>200$ Myr), lie within the bounds on the proto-cluster. Galaxies in
this region of the phase space, $\log(\mathcal{M}_{\ast}/M_{\odot})>10.6$, $M_{NUV}-M_{r^{+}}>1.5$, have, broadly, already formed the requisite stellar material to 
passively evolve to lower redshift $L^{\ast}$ red-sequence galaxies. The presence of these two galaxies alone hint that this population is relatively more abundant in the 
proto-cluster than in the field (2/9 vs. 2/54), an inference which does not depend especially on the specific limits chosen here and one which we attempt to expand 
on below when comparing $z_{phot}$ members. A KS test ($2.5\sigma$) and a comparison of the average stellar masses of the member and coeval field galaxies 
($\log(\widetilde{\mathcal{M}_{\ast}}/M_{\odot})=10.34\pm0.08$ and $10.09\pm0.03$, respectively) additionally both suggest that the average member galaxy has built 
up a more massive stellar content relative to other galaxies in the field. We note that these conclusions and those below comparing stellar mass distributions of 
$z_{phot}$ members and non-members are unchanged if we instead compare inverse error-weighted means. 
 
%If we instead compare the mean values\footnote{Mean values are calculated as $2.5\log(\langle f_{\nu} \rangle)$, not 
%simply the mean of the magnitudes or colors.} of the two sub-samples, we find a clear excess in both quantities for the member population 
%($\langle M_{r^{+}} \rangle = -23.2\pm0.2$, $\langle M_{NUV}-M_{r^{+}} \rangle = 1.5\pm0.2$) relative to those of the coeval field ($\langle M_{r^{+}} \rangle = -22.8\pm0.1$, 
%$\langle M_{NUV}-M_{r^{+}} \rangle = 1.2\pm0.1$). Within the context of the models overplotted on Figure \ref{fig:restframeCMDnCSMD}, such a color difference corresponds
%to an age difference of 50-100 Myr at constant stellar-phase metallicity and star formation history (SFH). The right panel of Figure \ref{fig:restframeCMDnCSMD} 
%shows a rest-frame color-stellar mass diagram (CSMD) of the same two samples. The trend in magnitude hinted in the CMD appears to be manifest in the CSMD 
%as well. Two of the four most stellar massive galaxies in the entire spectroscopic sample at the redshifts of the coeval field lie in the $\sim$1\% of 
%the volume of this field at these redshifts spanned by the proto-cluster bounds. Additionally, both a KS test ($2.5\sigma$) and a comparison of the 
%median stellar masses of the member and coeval field galaxies ($\log(\langle \mathcal{M}_{\ast} \rangle/M_{\odot})=10.34\pm0.08$ and $10.09\pm0.03$, respectively) 
%strongly suggest that the average member galaxy has built up a more massive stellar content relative to other galaxies in the field.  
 
\begin{figure}
\epsscale{1}
\plotone{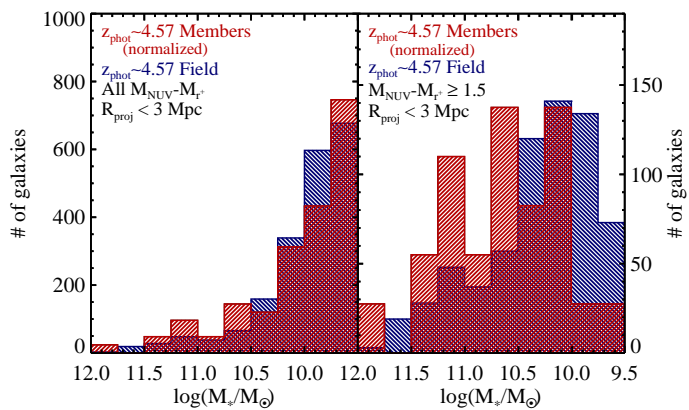}
\caption{\emph{Left:} Histograms of stellar mass as estimated by {\sc Le Phare} for all $z_{phot}$ members of the PCl J1001+0220 proto-cluster (\emph{left})
and only those $z_{phot}$ members with redder rest-frame colors ($M_{NUV}-M_{r^{+}}>1.5$, \emph{right}) plotted against the corresponding coeval field sample. 
Such a color cut selects galaxies with ages in excess of 200 Myr (see Fig. \ref{fig:restframeCMDnCSMD}). All spectroscopic knowledge is ignored when defining 
these samples. Photo-$z$ membership is defined by the projected radial cut given in the top left of 
the panel and $4.53-1.5\sigma_{NMAD}(1+4.53)<z_{phot}<4.60+1.5\sigma_{NMAD}(1+4.60)$ with the remaining galaxies in this $z_{phot}$ range comprising the coeval
field sample. Histograms in both panels are normalized such that they contain the same area. While the two distributions shown in the left panel are formally 
consistent, once a color cut is imposed, the two distributions are found to be disparate at a $>3\sigma$ level, with redder $z_{phot}$ proto-cluster members 
exhibiting a preference for significantly higher masses relative to the coeval field.} 
\label{fig:redSMhist}
\end{figure}

The above comparisons were made with only the spectral members and exclusively using one of the SED fitting techniques we have at our disposal. Though we have 
gone to extreme lengths to ensure that the VUDS sample is largely representative of the underlying population of galaxies at these redshifts within
the selected color and stellar mass range and that the SED method chosen is reliable at these redshifts, we now attempt two different reformulations of the analysis 
in order to understand if the trends hinted at in the above analysis are robust to changing the analysis framework. We begin by considering the $z_{phot}$ member sample 
and contrasting it with the $z_{phot}$ coeval field sample (see \S\ref{zphotoverdens} for the definition of each sample). While it is not necessary that differences seen in the 
previous analysis persist in significance or gain traction, as the photometric redshifts blur the lines between the proto-cluster and the coeval field considerably, 
any agreement would lend veracity to the claims made with the relatively small spectroscopic sample. In the left panel of Figure \ref{fig:redSMhist} we 
compare the stellar mass distribution of \emph{all} $z_{phot}$ members and the coeval field sample to the stellar mass limit of $\log(\mathcal{M}_{\ast}/M_{\odot})>9.5$.
A comparison of the median stellar masses ($\log(\widetilde{\mathcal{M}_{\ast}}/M_{\odot})\sim9.9$ in both cases) and  a KS test return no significant 
difference between the two samples. If, however, we instead restrict the sample to those galaxies with 
$M_{NUV}-M_{r^{+}}\ge1.5$, which, according to the models overplotted in Figure \ref{fig:restframeCMDnCSMD} have ages that are $\ge$200 Myr, the two
populations begin to differentiate themselves. Both a comparison of the average stellar mass of the two populations and a KS test return differences 
at $>3\sigma$, with the redder galaxies in the proto-cluster appearing at much higher stellar masses than their coeval field counterparts;
$\log(\widetilde{\mathcal{M}_{\ast}}/M_{\odot})=10.61\pm0.06$ versus $10.22\pm0.02$, respectively. The relative overdensity within the proto-cluster bounds 
of possible progenitors to low-redshift cluster $L^{\ast}$ galaxies hinted at in the spectral member sample is also hinted at here ($f_{red, L^{\ast}}$=14.6$\pm$4.2               % 12/82 vs. 164/1975, NUV-r > 1.5 and SM > 10^10.6, used IRAC1 < 25.4 and IRAC1 > 0
vs. 8.2$\pm$0.7\%). 
%Additionally, the third most massive $z_{phot}$ object in the entire field at these redshifts falls within the $\la$1\% of the 
%volume which encloses the proto-cluster and sits only $\sim$1 Mpc away from the luminosity-weighted center. 

\begin{figure}
\epsscale{1}
\plotone{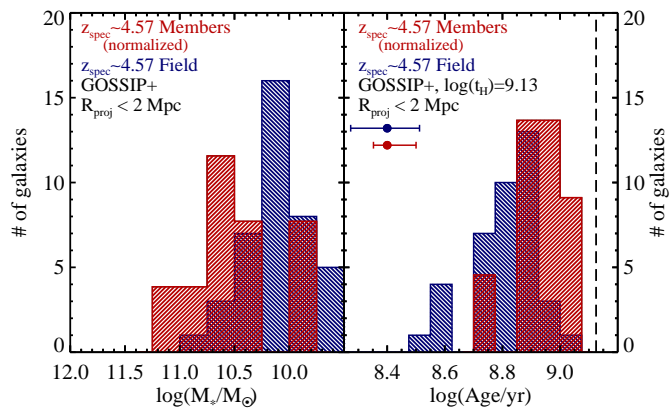}
\caption{\emph{Left:} Distribution of stellar masses of spec-$z$ members of PCl J1001+0220 (red histogram) and coeval field galaxies (blue histogram) as measured by {\sc GOSSIP+}. 
The projected radial cut used to define membership is indicated at the top left of the panel. \emph{Right:} Distributions of the formation ages of proto-cluster 
spec-$z$ members and coeval field galaxies, where age$_{f}$ is defined as the time elapsed since the galaxy began forming stars (see \citealt{romain17b}). 
Colors are identical to the left panel. The age of universe at the redshift of the proto-cluster ($z\sim4.57$) using our adopted cosmology is denoted 
by the dashed line. Two coeval field galaxies with log(age$_{f}$ yr$^{-1}$)$\sim$7.7 are not shown for clarity. Average effective 1$\sigma$ lower and upper uncertainties 
for age estimates of the galaxies in each sample are shown in the top left. There is an obvious excess of more (stellar) massive 
and older galaxies within the bounds of the proto-cluster; both sets of distributions are inconsistent with being drawn from the same underlying population at a $>2.5\sigma$ 
level and have average quantities which are inconsistent at an even larger significance. Additionally, the majority of proto-cluster members (5/9) appear to have formed 
within 500 Myr of the Big Bang as compared to only $\sim$10\% of coeval field galaxies.}      % 4/41 of the coeval field galaxies with measured mass-weighted ages 
\label{fig:massnages}
\end{figure}

As a final check on these results, we shift back to considering only the spectral sample and plot in Figure \ref{fig:massnages} the stellar masses 
and formation age (age$_{f}$\footnote{This age is
not to be confused with the previously mentioned age coming from {\sc EZGal}, which is essentially a luminosity-weighted age that is limited in resolution to the most 
recent star formation event.} hereafter) of the member and coeval field galaxies derived following the combination spectral and photometric fitting 
methodology of \citet{romain17b}. For all fits we required a ``fitting flag" of $\ge$2 corresponding to good or excellent fits to both the spectroscopy and 
photometry for each galaxy. 
The fitting method used to derive these parameters is considerably different in scope, in the range of models used, and in input than those used in any of the previous 
analysis presented in this section, and has been found to accurately and precisely recover the ages of simulated galaxies 
using VUDS-like observations under the assumption that the SFHs of galaxies at these redshifts can be approximated well by the range of SFHs assumed in our 
fitting process (see \citealt{romain17b} for more details). Again, a difference 
between the two populations is observed in the stellar mass distribution in that galaxies in the proto-cluster are, 
on average, more massive in their stellar content ($\widetilde{\log(\mathcal{M}_{\ast}/M_{\odot})}=10.51\pm0.07$ vs. $10.05\pm0.04$). 
Here, however, we additionally see a clear difference between the two populations in their age in that galaxies in the proto-cluster appear older 
($\widetilde{\rm{age}_{f}}=904\pm63$ vs. $640\pm41$ Myr) than the coeval field population. These differences are also significant at the $\ge2.5\sigma$ level 
for both parameters when using a KS test. The observed difference in the ages of the two sub-samples is in apparent contrast to the lack of significant difference 
between the two spectroscopic populations seen in the rest-frame colors. However, the 
methods used by \citet{romain17b} are considerably more advanced than that used to generate the models overplotted in Figure \ref{fig:restframeCMDnCSMD}, with 
the former method employing a vastly larger array of SFHs, treatment for the effects of extinction, and the virtue of numerous tests performed on its precision and 
accuracy. A similar comparison of $z_{phot}$ members and coeval field galaxies could not be performed as the method of \citet{romain17b} requires spectroscopy. Regardless, 
the common element of all three analyses presented here appears to be, at varying degrees of significance, a fractional excess within the proto-cluster of older, and, in some 
analyses, redder, galaxies which appear more massive in their stellar content. The large differences in stellar masses compared to what are relatively small differences 
in ages (or colors) necessarily implies that, on average, a proto-cluster galaxy must have a more violent SFH than their field counterparts, with 
stellar mass being built up more rapidly after their formation. As we will show in the following section, the two populations have indistinguishable instantaneous SFRs 
(integrated over $\sim100$ Myr), which, in conjunction with the results presented in this section, further implies that there is something necessarily different 
about the SFHs of the two populations. 

%\begin{figure}
%\epsscale{1}
%\plotone{COSMOS.protostruct26.pseudomassfunction.allcolors.ps}
%\caption{Pseudo-stellar-mass function. Spec n photo-z members vs. non. XXX Remember to do the stellar mass limit in the way Laigle did it with the IRAC1
%magnitudes in place of the K-band magnitudes and remember that her numbers dun mean shit for these redshifts since the K-band is shifted into the 
%Balmer/D4000 break. Check notes on her paper. There's prolly some other stuff to do from those notes. XXX}
%\label{fig:pseudoSMfunc}
%\end{figure}

\begin{figure}
\epsscale{1}
\plotone{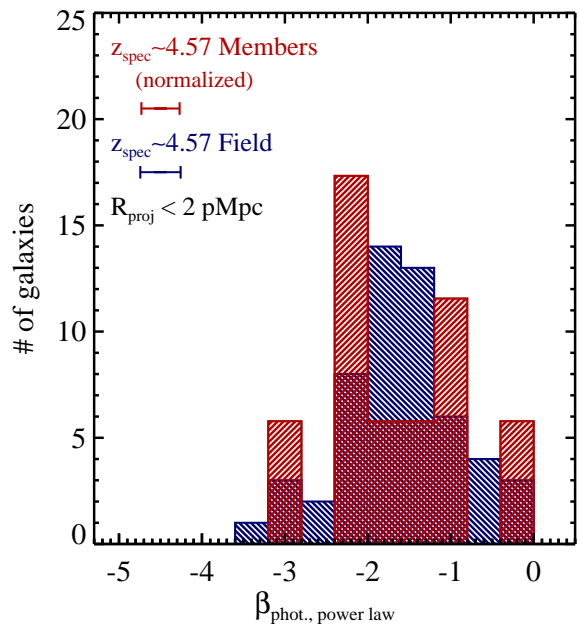}
\caption{Distributions of the estimate for the index of the power-law dependence between flux density and wavelength in the rest-frame UV ($\beta$-slope) measured from the
photometry of individual galaxies for PCl J1001+0220 spec-$z$ members and the coeval field (red and blue histograms, respectively). Measurements are made assuming a
single (unbroken) power-law dependence of $f_{\lambda}$ on $\lambda$ for each galaxy following the method of \citet{nimish16}. For galaxies at $z<4.5$, $\beta$ is
measured with the $i^{+}z^{+}YJ$ photometry spanning roughly 1300\AA\ $<\lambda_{rest}<$ 2300\AA. At $z>4.5$, the $i^{+}$ band is exchanged for the $H$ band
(covering roughly 1400\AA\ $<\lambda_{rest}<$ 3000\AA) to prevent Ly$\alpha$ emission from potentially contaminating the measurement. The average errors on individual
measurements for the two samples are shown at the top left ($\sigma_{\beta}\sim0.2$ in both cases). Galaxies where extremely blue $\beta$-slopes are measured 
($\beta<-2.5$) have average errors roughly twice the global average. The median $\beta$-slope for both samples is $\tilde{\beta} \sim-1.7$ and no significant difference 
is measured between the two distributions.}
\label{fig:betaslope}
\end{figure}

The slope, $\beta$, of the continuum flux density in the rest-frame ultraviolet ($f_{\lambda} \propto \lambda^{\beta}$) has been shown, for a given IMF and a 
given dust geometry and law, to correlate well with the level of dust in a galaxy (e.g., \citealt{meurer99, reddy10, javier16}). As the analysis throughout this section relied heavily on the difference 
in the colors and various ages of member and coeval field samples, and as the precision and meaning of such parameters can be heavily influenced by the presence of dust, an independent 
measure of the dust content is important to gain context on the reliability of these results and their interpretation. In Figure \ref{fig:betaslope} 
we plot the distribution of $\beta$ slopes measured for the member and coeval field samples. While, in principle such measurements could be made directly on the galaxy spectra, 
the high redshift of the sample along with the low resolution of the VUDS spectra and the high density of OH lines preclude a brute force measurement in practice. Future effort 
will be made to comprise a scheme to  measure the spectral $\beta$ slope on VUDS spectra at these redshifts. Here, instead, we measure $\beta$ on the COSMOS2015 photometry using 
the $I^{+}/Z^{++}/Y/J$ bands for galaxies in the range $4.23\le z \le 4.5$  
and the $Z^{++}/Y/J/H$ bands for galaxies in the range $4.5< z \le 4.88$ adopting the photometric methods of \citet{nimish16}. This choice of photometry avoids 
the Ly$\alpha$ feature for all cases and provides a nearly contiguous coverage of all ten bandpasses defined by \citet{calzetti94}. In neither the average value of the members versus
the coeval field galaxies ($\tilde{\beta} \sim-1.7$ in both cases) nor a KS test do the two populations appear different in their UV continuum slopes. 
Furthermore, the average value measured in the two populations implies that the member and coeval field galaxies have little dust content 
(\citealt{finkel12, castellano14, javier16}), though a non-negligible fraction appear to have at least moderate levels of dust 
($\beta>$-1.4 or roughly $E_{s}(B-V)>0.15$, see \citealt{javier16}). While there are many caveats for the interpretation of this measurement 
(see, e.g., \citealt{castellano14}) and there are perhaps more reliable ways to calculate this quantity 
than the method employed here (see, e.g., \citealt{finkel12}), we require only the precision here necessary to claim that there is no apparent difference in the dust content of the two 
populations. Having achieved that aim, we now move to the properties of galaxies as measured by photometry in wavelengths not exclusively in the UV/optical/NIR regime. 

\subsubsection{Multiwavelength properties and AGN content}
\label{multilam}

One of the most intriguing lines of inquiry of high-redshift environmental studies relates to the question of whether the proto-cluster environment preferentially enhances, 
diminishes, or has no effect on the frequency and/or level of star formation and galactic nuclear activity in galaxies relative to matched samples in field environments. An
overwhelming variety of results have been obtained for studies of high-redshift (proto-)clusters (e.g., \citealt{dig10, pcap11, martini13, olga14, lem14a, conivingcasey15, dutifuldiener15, wang16}) 
underscoring that trends are highly dependent on the sample selected, the nature of the overdensity or overdensities studied, the level of care taken in appropriately matching the high- and low-density samples, and 
the wavelength of light and scheme used to proxy star formation or AGN activity. Though the amount of information available to us at $z\sim4.57$ is somewhat limited, we draw 
here on nearly the entirety of the wealth of data in the COSMOS field to attempt to investigate the relative level of activity in the PCl J1001+0220 proto-cluster. 

\begin{figure}
\epsscale{1}
\plotone{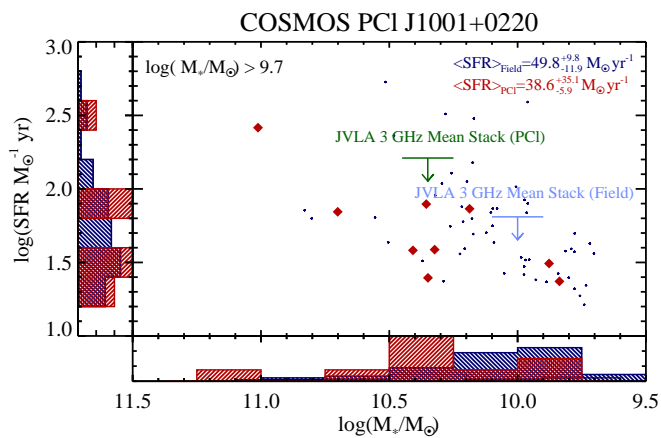}
\caption{Distribution of the PCl J1001+0220 spec-$z$ members (red diamonds and histogram) and coeval field galaxies (navy points and histogram) in the SFR-$\mathcal{M}_{\ast}$ plane
as estimated by {\sc Le Phare}. Histograms are normalized as in other figures such that the area is identical for both samples. The upper limit to the radio-derived SFR of
spec-$z$ member and coeval field galaxies coming from stacked 3 GHz JVLA data is shown as the green and light blue lines/arrows, respectively. The lack of detection 
for both proto-cluster members and coeval field galaxies suggests that prodigious dusty star formation ($IRX \lsim 10$) is no more common in the proto-cluster 
than in the overall field at these redshifts and is not particularly common in either environment.}  
%XXX Radio line tentatively final, confirm with Vernesa. Confirmed XXX
\label{fig:massSFR}
\end{figure}

Shown in Figure \ref{fig:massSFR} are the {\sc Le Phare} SED-fit stellar masses and SFRs of the ($z_{spec}$) proto-cluster members plotted against those of the coeval field population 
subject to the constraints applied in the previous section. By eye, the two SED-fit SFR distributions appear similar, with a KS 
test confirming this observation, finding the two distributions to be statistically indistinguishable. As in the previous analyses, we do not rely solely on a KS test to 
attempt to distinguish the two populations. Shown in the top right-hand corner of Figure \ref{fig:massSFR} are the median SED-fit SFRs of each population along with their 
bootstrapped uncertainties generated following the method of \citet{lem14a}. While there appears a slight suppression of the level of star formation activity within
the proto-cluster bounds, the uncertainties are far too large to claim any level of significance. As all galaxies in VUDS are, by selection, star forming, these galaxy 
populations cannot be used to comment on the frequency (fraction) of star-forming galaxies in each environment. Similarly ambiguous results are found when we compare 
the star formation activity of the $z_{phot}$ proto-cluster member sample relative to that of the $z_{phot}$ coeval field sample. Using the definitions from
\S\ref{zphotoverdens} results in median SED-fit SFRs
%\footnote{Because of the relatively large number of galaxies in each sample and the extremely non-Gaussian 
%distributions of their SFRs, we opt here to use the median and the uncertainty on the median given by 1.253$\times\sigma$/$\sqrt(N)$. Our conclusions are unchanged
%if we instead adopt a mean and bootstrap errors.} 
of 46$\pm$12 and 32$\pm$3 $\mathcal{M}_{\odot}$ yr$^{-1}$ for the $z_{phot}$ proto-cluster member and coeval field populations, respectively. 
A lack of significant differences in this quantity remains if we instead consider only those $z_{phot}$ objects in the magnitude-color-$\mathcal{M}_{\ast}$ phase space             % Values are 254 +/- 103 & 89 +/- 11 M_solar/yr and f_q = 15.4 +/- 10.9 & 13.2+/-2.3 for PC and field using pmassmed gt 9.7 and pMNUV-pMr ge 1.5, but that has nothing to do with anything. Using Mr < -22, pmassmed gt 9.7, and -0.3 < pmassmed gt 9.7 < 2.8 gives numbers which are consistent.
where VUDS is representative (see \S\ref{mainproperties}).  

The previous analysis relies on SED-fit SFRs, a method that has not been thoroughly tested for accuracy at the redshifts explored in this paper, a deficiency 
which is amplified due to the relatively small size of the galaxy samples considered here. Further, the SED-fit SFRs are extinction corrected using photometry that has only a modest dynamic range in wavelength in the rest-frame, the bulk of which lies 
in the rest-frame UV at these redshifts. As such, we attempt here to measure the average SFR of the proto-cluster member and coeval field samples using a tracer not subject to 
such uncertainties. Such a complementary approach is important as it has been suggested that at least some (proto-)clusters are dominated by dusty star-forming galaxies
(e.g., \citealt{conivingcasey15,wang16}). For this purpose we are required to shift to wavelengths further to the red. There are only four viable possibilities for 
the COSMOS field. 
The first are deep \emph{Spitzer}/MIPS 24$\mu$m observations used to routinely probe galaxy populations at $0 < z < 2$ in COSMOS \citep{emeric09}. However, at $z\sim4.57$,
these observations are too far blue to reliably probe young stellar emission processed and re-radiated by dust. The second are deep \emph{Herschel}/PACS and SPIRE observations
taken on COSMOS, but even at $z\sim3$, the stack of 100s to 1000s of galaxies are needed for a significant detection \citep{javier16}. The third comes from imaging 
taken with the James Clerk Maxwell Telescope (JCMT) at 850$\mu$m to a 5$\sigma$ depth of 6.0 mJy/beam observed as part of the SCUBA-2 Cosmology Legacy Survey 
(S2LCS; \citealt{geach17}). The final possibility comes from 
recently obtained extremely deep (currently 11.5 $\mu$Jy/beam, 5$\sigma$) 3 GHz imaging from the Karl G. Jansky Very Large Array (JVLA) of the entirety of the COSMOS 
field (\citealt{mladen15, vernesa17b}). The rest-frame frequency which these observations correspond to at $z\sim4.57$ is a probe of synchrotron 
emission generated from supernovae and is found to correlate well with star formation activity (see \citealt{condon92} and references therein) in the absence of a radio AGN. 
The latter possibility is addressed later in this section. No significant individual detections in either the S2LCS nor the JVLA imaging were found amongst the 
proto-cluster member or coeval field samples. 

In order to attempt to place more stringent constraints on the prevalence of dusty star formation activity we performed stacking on the JVLA imaging, which was
preferred over the S2LCS imaging due to its comparable depth in SFR and considerably smaller PSF. 
Proto-cluster member and coeval field galaxies were stacked separately at 3 GHz by inverse-variance-weighted mean combining postage stamps centered on the optical centroid
of each galaxy. While poor relative astrometry, if randomly rather than systematically poor, can render this exercise pointless, the astrometry of the JVLA mosaic was checked 
against the COSMOS2015 catalog and was found to be accurate to an RMS of 0.1$\arcsec$. As this is half the size of the JVLA pixel scale and considerably smaller 
than the restored circular PSF of 0.75$\arcsec$, it is unlikely that astrometric smearing would lead to degradation of any stacked signal.
Despite the extreme depth of the data, the relatively large (combined) sample size, and the relatively high mean optical/UV SFRs estimated for each sample, 
in neither case was a significant ($>3\sigma$) detection measured. The 3$\sigma$ $f_{\nu}$ limit for the proto-cluster member 
sample was found to be 2.24 $\mu$Jy. This limit translates to a $3\sigma$ $SFR_{1.4 GHz}$ limit of $<$157 $\mathcal{M}_{\odot}$ yr$^{-1}$ following a $k$-correction of 
$\alpha=0.68$ \citep{lem14b} and the SFR conversion of \citet{bell03} converted to a \citet{chab03} IMF. This limit, overplotted in      
Figure \ref{fig:massSFR}, is $\sim$4$\times$ the average UV/optical SFR
of the proto-cluster member sample. However, since this ratio is dependent on the dust-correction scheme employed by the SED fitting process, we instead calculate the 
IR excess ($IRX\equiv L_{IR}/L_{FUV}$) from the average $M_{FUV}$ and the radio limits assuming the $q_{TIR}$ value from \citet{lem14b} for star-forming          
% median M_FUV = -21.5970 for the coeval field sample, -21.774 for the proto-cluster member sample, means are -21.43 and -21.72, respectively. This corresponds
% to L_FUV = 9.2952012e+10 and 1.03968e+11 for the two samples respectively (median), see radio_instructions.notes. Ivan calculated a 8-1000 um limit of 10^11.2 and 10^11.8
% from MAGPHYS (Herschel stacks were way to unconstraining), which gives an IRX of <1.7 and <6.1, respectively. 
galaxies. This exercise yields an $IRX\lsim10$ for the proto-cluster member sample, a value which precludes pervasive dusty starburst activity amongst the 
proto-cluster member galaxies. A similarly stringent limit of $IRX\lsim6$ is found for the same sample if we replace the stacked radio limit with the 
$L_{IR}$ limit as derived by {\sc MAGPHYS} from the average proto-cluster member SED. An additional attempt was made to stack with \emph{Herschel}/SPIRE
but no significant detection was found and the corresponding limits were too shallow. 

The 54 galaxies of the coeval field sample were stacked at 3 GHz in a similar manner finding no detection and 
resulting in a 3$\sigma$ $f_{\nu}$ limit of 0.95 $\mu$Jy or a $SFR_{1.4 GHz,\, 3\sigma}$ limit of $<$64 $\mathcal{M}_{\odot}$ yr$^{-1}$ at $z\sim4.6$. Such limits begin
to rival the average UV/optical SFRs of the coeval field sample (see Figure \ref{fig:massSFR}) and the resulting $IRX$ limit of $\le4.8$ approaches the value 
expected for galaxies with $\beta$ slopes similar to those observed in the coeval field sample (e.g., \citealt{javier16}). As above, a similarly stringent value of 
$IRX\le2$ is found if we instead impose the {\sc MAGPHYS} $L_{IR}$ limit in conjunction with the average $M_{FUV}$. While not extremely constraining 
for the populations studied in this paper, this exercise has allowed us to definitively show that the SED-fit SFRs do not vastly underestimate the true 
SFRs for these populations and that, despite the high redshift and relatively high local densities, the proto-cluster is not dominated by dusty, prodigiously star-forming 
galaxies, nor, for that matter, by radio AGN activity. However, neither this analysis nor the $\beta$-slope analysis presented in the previous section 
precludes the possibility that some subdominant portion of the member population is undergoing obscured star formation activity (as in, e.g., \citealt{hatch17}). 
Such analysis will be followed further as the member and coeval field sample in and around PCl J1001+0220 is increased 
and for other, lower redshift VUDS proto-structures discovered in the COSMOS field.

To investigate the relative presence of AGN activity amongst the proto-cluster member and coeval field populations we matched both samples to the \emph{Chandra}-COSMOS point
source catalogue \citep{elvis09,civano16} finding no significant detection ($>3\sigma$). No significant detections were found amongst the $z_{phot}$ proto-cluster members either. 
Such matching was also performed
on the JVLA 3 GHz imaging finding no significant detection to a 3$\sigma$ power density limit of $log(P_{\nu} \, W^{-1}$ Hz$) \ga 24.4$.  % this is the 3sigma flux density limit translated to a luminosity at z~4.57 using a k-correction of x1.4^(-0.68)/(3*5.6)^(-0.68), i.e., an alpha of 0.68, i.e., 13.5e-29/5.57*(lumdist(4.57)*3.086d24)^2*4d*!pi/10d^7*(1.4)^(-0.68)/(16.7)^(-0.68), where 13.5e-29/5.57 is fnu,obs/(1+z) = fnu, rest 
Additionally, Chandra images\footnote{The Chandra web tool for X-ray stacking, {\sc CSTACK}
(http://cstack.ucsd.edu/cstack/), developed by T. Miyaji was employed for stacking} were stacked using the optical position of the sources as a centroid yielding no significant 
stacked signal to a 90\% limit of $log(L_{X, \, 2-10 keV}$ ergs$^{-1}$ s$)<43.54$ and $43.21$ for the proto-cluster member and coeval field populations, respectively. An 
identical stack was performed on $z_{phot}$ proto-cluster members yielding no significant detection to a depth of $log(L_{X, \, 2-10 keV}$ ergs$^{-1}$ s$)<43.10$. 
Note that the proto-cluster lies at the boundary between the deep and shallower portions of the C-COSMOS survey with the effective exposure times varying by a factor 
of $\sim$5 over area enclosed by both the spectral and $z_{phot}$ members. Further, we applied the mid-infrared criteria developed by \citet{donley12} to select powerful 
(both obscured and unobscured) AGN, finding that 
none of the sources analyzed here exhibit an AGN-like trend which could possibly be ascribed to a circumnuclear torus. Note that, in this context,
the terms ``obscured" and ``unobscured" AGN refer to sources with rest-frame optical emission dominated by the host galaxy or by the AGN, respectively. 
The lack of individual and stacked X-ray and
mid-infrared emission along with the complete absence of any broadline AGN activity amongst the VUDS sample at these redshifts suggest that no significant radiatively
efficient AGN activity is taking place in these sources. 

To test this hypothesis in a statistical sense, we constructed the average SED separately for the proto-cluster member, 
coeval field, and $z_{phot}$ members. These average SEDs were then fit with {\sc SED3FIT} \citep{berta13} 
via $\chi_{\nu}^2$ minimization. Through this SED-fitting decomposition technique we found that the fraction of the rest-frame IR (8-1000 $\mu$m) and UV light contributed by the 
torus/AGN is, in all three cases, negligible ($\le$1\%). In the mid-IR, defined as an integration over 5-40 $\mu$m in the rest-frame, where the relative contribution 
from an obscured AGN is maximized, the fraction of light contributed by the torus/AGN remains at only a few percent ($\le$5\%). This line of analysis, while 
almost certainly imprecise due to only upper limits being available on many of the most discriminating bands, allows us to make definitive claims on the prevalance 
of AGN activity amongst the proto-cluster galaxies. It appears that, within the limits of our data, there is no evidence for any powerful 
(i.e., quasar level) unobscured or obscured radiatively efficient or inefficient AGN activity in any of the known or potential member galaxies, a conclusion
which is robust to changes of factors of a few in the above estimates. This fact precludes 
the possibility that such activity is a requisite condition for the presence of a proto-cluster and suggests that the proto-cluster does not necessarily foster such 
activity (e.g., \citealt{hatch14}), though the validity of the latter suggestion is subject to a variety of timescale concerns which we do not attempt to address in this paper.
 
%These SEDs imply the need for an AGN contribution at these
%wavelengths. (XXX Which stacks include photo-z members? Need to make explicit. Also, could include an example fit 
%from Stephano but bug in ode plots upper limits in flux density as points with huge error bars, would have to re-do 
%from raw files XXX) 

As a final thought related to multiwavelength activity in and around PCl J1001+0220, it is interesting to note that a prodigiously star forming submillimeter 
galaxy (SMG, $SFR_{TIR}=500-2000$ $\mathcal{M}_{\odot}$ yr$^{-1}$) or galaxy-galaxy merger detected in AzTEC 1.1 mm observations was spectroscopically 
confirmed by \citet{capak08} to be at $z=4.547\pm0.002$, in proximity to PCl J1001+0220. Such a discovery is interesting since potential links between 
proto-structures and SMGs have been proposed by many authors (e.g., \citealt{daddi09,aravena10,vernesa17,wilkinson17}). However, this SMG is well outside the bounds of either the 
$z_{spec}$ or $z_{phot}$ proto-cluster member samples ([$\alpha_{J2000},\delta_{J2000}$] = [150.22715, 2.5764]) regardless of the choice of centers. As a consequence, 
PCl J1001+0220 could not have been found solely by looking for an overdensity surrounding this SMG. While the SMG may eventually merge with the 
proto-cluster core, a process that would take $\sim6$ Gyr assuming the LOS velocity to be equivalent to the infalling velocity and a purely radial orbit, 
it is sufficient to say that the proto-cluster and the SMG do not appear causally related at the epoch at which they are observed. 

\begin{figure*}
\epsscale{1}
%\plotone{ClusterMembers.HSTF814W.new.eps}
\plotone{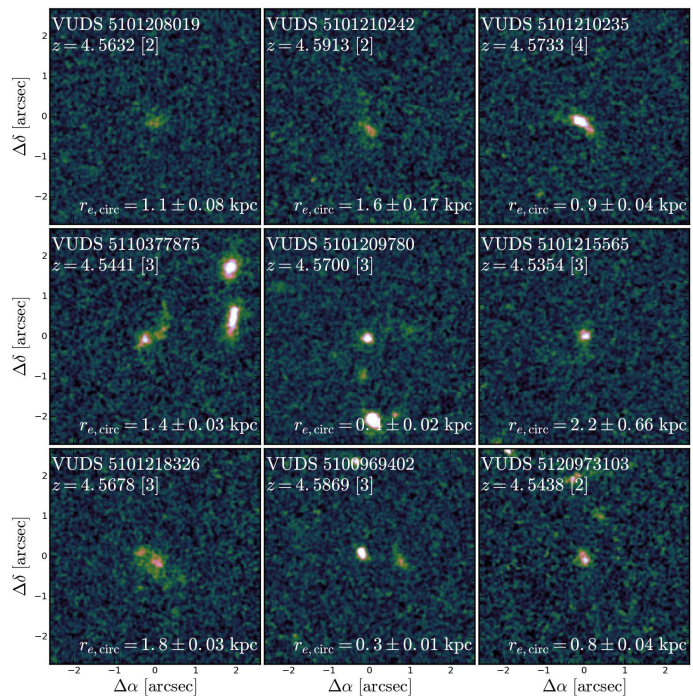}
\caption{\emph{Hubble}/ACS $F814W$ 5$\times$5$\arcsec$ ($\sim$33$\times$33 kpc) postage stamps of the nine secure spec-$z$ members of PCl J1001+0220. 
At the redshift of the proto-cluster, $F814W$ almost exactly corresponds to the rest-frame FUV band. The redshift, SFR, circularized effective radius ($r_{e, circ}$), 
and stellar mass for each galaxy is shown in each panel. The nine galaxies exhibit a large diversity of morphologies, from a single, compact ($r_{e, circ}<0.5\arcsec$) 
bright star-forming clump to larger ($r_{e, circ}\sim1-2\arcsec$), multi-component diffuse structures devoid of concentrated light. Many of the proto-cluster members 
are observed with single or multiple projected companions, a rate which possibly exceeds that of the coeval field (see text).} 
\label{fig:HSTimages}
\end{figure*}

\subsubsection{Morphological properties}
\label{morph}

In the penultimate section of the comparisons between proto-cluster members and coeval field galaxies we return to rest-frame UV light to inspect the morphology 
of these galaxies as they appear in the COSMOS \emph{HST}/$F814W$ imaging. Because the PCl J1001+0220 is not covered by the CANDELS/COSMOS survey, these images 
are the only images for which we can get reliable morphological information. %(though sizes may be obtainable in ground-based images, see \citealt{bruno16}).
This band is nearly identical to the rest-frame FUV at these wavelengths\footnote{The $k-$correction for a 100 Myr old exponentially declining BC03 model with 
stellar-phase metallicity 40\% of solar is $\sim$0.05.} and, thus, any morphological information derived from these images will be highly subject to dust effects. 
While we showed earlier in \S\ref{mainproperties} that there does not appear to be any bulk difference between the two populations in this regard in terms of 
integrated dust content, differing dust geometry may still differentially affect the measurements of the two samples. As such, we limit ourselves here to 
the most basic of morphological quantities and observations.  
 
%By eye and non-parameteric morphology and then merging information, coeval field vs. members.
Plotted in Figure \ref{fig:HSTimages} are the \emph{HST}/$F814W$ postage stamps of the nine PCl J1001+0220 member galaxies. Given in the bottom
left-hand corner of each postage stamp is the effective circularized radius ($r_{e, circ}$) obtained from \textsc{GALFIT} for each galaxy following the methods of \citet{bruno16}.
Proto-cluster members are observed to span a wide variety of FUV morphologies from irregular diffuse galaxies with multiple clumps to compact, high surface-brightness galaxies. 
Correspondingly, the observed $r_{e, circ}$ values span nearly an order of magnitude from 0.3 to 2.2 kpc with an average of 1.2 kpc. The (luminosity-)dominant stellar populations 
probed in these images are those associated with the current star formation activity in member galaxies. Thus, the vast diversity of morphologies and $r_{e, circ}$ values 
unambiguously demonstrate that while these galaxies appear to share a common overarching environment, their stellar mass assembly is not proceeding in a common manner. 
This large variety of visual morphologies and $r_{e, circ}$ values is, however, also shared by the general VUDS field population at these redshifts 
\citep{bruno16}. While it should be noted that a stellar mass limit was imposed on the sample in \citet{bruno16}, this imposition was only made on galaxies in 
their lower redshift bins meaning that the galaxies in redshift bin referenced here, $3.5\le z \le 4.5$, the highest redshift studied in \citet{bruno16}, can be 
validly compared to the coeval field sample. Such a large diversity in UV morphologies in all cases 
is perhaps not unexpected for galaxies at these redshifts as either continuous moderate-level ($\sim20$ $\mathcal{M}_{\odot}$ yr$^{-1}$) or shorter-lived, 
vigorous star formation over the course of the last $\sim$1 Gyr is required to build up the average stellar mass of member and coeval field galaxies. 

\begin{figure}
\epsscale{1}
\plotone{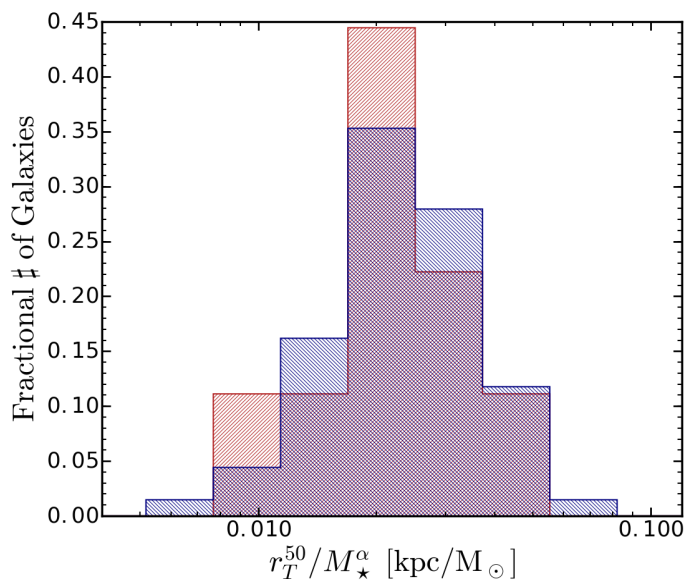}
\caption{Distribution of non-parametric total sizes, $r_{T}^{50}$, as measured in the \emph{Hubble}/ACS $F814W$ imaging of spec-$z$ member galaxies of PCl J1001+0220
compared to that of the coeval (red and blue histograms, respectively). As in previous figures, histograms are normalized to have equal areas. Due to the relationship between
size and stellar mass, size measurements are modulated by the estimated stellar mass of each galaxy ($M_{\star}^{\alpha}$) where $\alpha$ is derived from
a fit to the observed size-stellar mass relation in the combined proto-cluster and coeval field sample ($\alpha=0.18\pm0.03$). The average value for both
samples is identical within the uncertainties, though a slight increase ($\sim25$\%) in the spread of the proto-cluster distribution is observed 
which we argue in \S\ref{morph} that, if real, may be due to increased galaxy-galaxy interactions.} 
\label{fig:sizemass}
\end{figure}

While the same range in visual and (one form of) parametric morphologies is observed in both the coeval field and the proto-cluster member population, 
the latter has a $\sigma_{NMAD}\sim25$\% higher and in a sample comprised of only nine galaxies. %In \citet{bruno16} it was found that 
%the $r_{e, circ}$ distribution of galaxies at these redshifts was well represented by a log-normal function of $\mu=1.06$, $\sigma=0.24$. 
To test the likelihood of the dispersion in the observed size distributions occurring by random sampling of the VUDS field galaxies at this redshift, we randomly drew 
1000 samples of nine galaxies from the $r_{e, circ}$ distribution observed in \citet{bruno16} for the redshift bin most comparable to our 
own. In only $\sim$10\% of the trials did the variance of the $r_{e, circ}$ values in the randomly drawn sample exceed that of the proto-cluster member 
sample. This result is identical if we instead perform this exercise drawing from a Gaussian fit to the distribution of $r_{e, circ}$ values of the coeval field
sample. However, because galaxy size is correlated with the mass of their stellar content (e.g., \citealt{allen15}), it is perhaps the potentially disparate stellar mass 
distribution of the member population with respect to the field that drives the observed scatter (see \S\ref{mainproperties}). Further, the lack of an azimuthally 
symmetric profile for a large fraction of both the member and coeval field samples, a phenomenon common to galaxies at high redshift (e.g., \citealt{mortlock13, guo15}), 
likely decreases the utility of both the $r_{e, circ}$ measurement and the scatter on this quantity. To mitigate both issues, an effective non-parametric half-light radius, 
$r^{50}_{T}$, was measured for each galaxy in both the member and coeval field samples following the methodology of \citet{bruno16} and normalized in such a way as to remove the 
correlation between size and stellar mass (see Fig. \ref{fig:sizemass}). None of the results in this section change appreciably if we instead adopt $r^{100}_{T}$, the 
effective non-parametric radius containing 100\% of the flux of each galaxy. 
%While the distributions and average value of the coeval field and member populations
%for our sample. The latter was achieved by least squares fitting of a linear model in log space the size and stellar 
%masses of combined member and coeval field sample, 
%, i.e., using the relation $r^{50}_{T}\propto\mathcal{M}_{\ast}^{\alpha}$ with $\alpha=0.18\pm0.03$ derived from a fit to the combined member and coeval field sample. The 
%results of these measurements are shown in Figure \ref{fig:sizemass}. 
%Again the proto-cluster members appear to exhibit a large spread in size, with an average value of $\widetilde{r^{50}_{T}}=1.8$ kpc. While the distributions of $r^{50}_{T}/\mathcal{M}_{\ast}^{\alpha}$ for the two samples appear similar by 
%eye and from a KS test, 
Performing a similar exercise as was done for the $r_{e, circ}$ distributions again on those of $r^{50}_{T}/M_{\star}^{\alpha}$ results in $\sim$10\% of 
trials exceeding the observed variance of proto-cluster member population, a result which mildly suggests more stochasticity in the assembly 
stage of member galaxies relative to the field.  
 
%Fitting a Gaussian
%to the observed distribution of $r^{50}_{T}/\mathcal{M}_{\ast}^{\alpha}$ for coeval field galaxies and performing an identical exercise as was done for $r_{e, circ}$ 
%results in a similarly small number of random trials ($\sim10$\%) for which the variance in the $r^{50}_{T}/\mathcal{M}_{\ast}^{\alpha}$ values of nine randomly        
%drawn galaxies field-like galaxies exceeds that of the member population mildly suggesting more stochasticity in the assembly stage of member galaxies relative to the 
%field.  

Assuming this increased scatter to be real, we explore here the possibility of increased galaxy-galaxy interactions and potential merging activity 
within the proto-cluster as possible driving mechanisms for this increased stochasticity.  
In order to estimate the level and type of galaxy-galaxy interactions experienced by the average galaxy in the member and coeval field samples, for each galaxy in 
each sample we counted the number of unique objects in the magnitude range $24\le F814W \le 26$ other than the galaxy itself whose centroids lay within 
$r_{proj}<25$ kpc from the galaxy center. This radius was adopted as it is commonly used for galaxy pair studies and has been 
found through simulations and observations to contain galaxies with a high likelihood of eventual coalescence (see, e.g., \citealt{dong2000,dirtydeR09,jlotz11, carlos13, carlos15} 
and references therein). It is worth noting that galaxy-galaxy interactions can have visible effects over much larger separations ($r_{proj}\sim50-100$ kpc, e.g., \citealt{patton11}) 
and can still merge given a long enough timescale (3-8 Gyr, \citealt{kdubz08}). Here, however, we are concerned only with the most severe of interactions and thus 
limit ourselves to the more restrictive criterion. %We make no attempt here to eliminate companion objects which have known spectroscopic redshifts or 
%(clean) photometric redshifts inconsistent with those of the central galaxy as to do so would incite unnecessary spatial bias to this comparison. However, 
The bright end of the magnitude limit imposed corresponds to $M_{FUV}\la-22.3$ at $z\sim4.5$ ($\sim3L^{\ast}$), a population extremely unlikely to be observed 
with any abundance in the small areas searched here \citep{anykey15}. The fainter end of the magnitude range given above was set as a rudimentary 
attempt to control for surface brightness effects and detection limits as $F814W\sim26$ corresponds to the 50\% completeness limit for $r_{h}=0.25\arcsec$ sources 
\citep{scoville07b}. Finally, because the ACS observations probe the rest-frame FUV, there is some concern that confusion could occur between multiple 
distinct galaxies and bright star-forming regions contained within a single galaxy. The method for disambiguating these cases in VUDS is discussed 
in detail in \citet{bruno16b}. 
%Since we are calculating a relative rather than an absolute quantity, this level of control is sufficiently precise for our purposes.  

Subsequent to these cuts, the simplest possible comparison is made: the average number of companions per galaxy in the two samples corrected by the average density of sources in this magnitude
range in the COSMOS2015 catalog. In total, 0.4$\pm$0.3 and 0.1$\pm$0.1 background-corrected projected companions per galaxy are found for the proto-cluster member and coeval 
field sample, respectively. A discrepancy of similar directionality and significance is seen when comparing the $z_{phot}$ member and field sample, 0.17$\pm$0.08 versus 0.08$\pm$0.02, 
respectively. 
%While all projected pairs necessarily fall within the traditional $<4:1$ major galaxy-galaxy 
%merging criterion based on the ratio of their $F814W$ magnitudes, it is important to note that the relationship between $M_{FUV}$ and stellar mass is an extremely complex 
%one and it is not clear that this equivalency can be applied. Further, we have no ability to contain the prevalence of more minor merging events or interactions as such 
%galaxies fall below the 50\% completeness limit of the COSMOS ACS images. 
While these results depend on the choice of the magnitude and $r_{proj}$ criteria adopted, the cuts chosen here are the most conservative that could be made with
respect to projection effects for which this analysis could reasonably be performed. For these choices, the analysis is suggestive of possible elevated levels of strong 
galaxy-galaxy interactions within the proto-cluster bounds. This line of thought will be followed further at the conclusion of future followup spectroscopic 
observations of this and other VUDS proto-clusters. 

% field correction made by selecting all galaxies at 24 < F814W < 26 in the SPLASH catalog (added F814W to the .sav file) between 149.6 < RA < 150.4 and 1.7 < dec < 2.8. 
% a = where(pf814w lt 26 and pf814w gt 24 and pRA gt 149.6 and pRA lt 150.4 ), then
% print, float(n_elements(a))/(sphdist(149.6, 2.25, 150.4, 2.25, /degrees)*3600.*sphdist(150, 1.7, 150, 2.8, /degrees)*3600.) = 0.014357528 objects/arcsec^2
% Searched over r_proj < 25 kpc, using the average redshift of 4.57 -> print, !pi*(25./Da(4.57,0.7))^2 = 42.5933 arsec^2 => 0.611534 objects/search region
 
\subsubsection{Spectral properties}

The final comparison between proto-cluster members and their field counterparts comes in measuring rudimentary individual and stacked (hereafter ``coadded") 
spectral properties. We begin with the former and concentrate on the features most commonly studied for high-redshift galaxies. The Ly$\alpha$ feature has been
found to have a complex relationship with environment in the few studies with a reasonable chance to test this possibility. In some cases an enhancement of the Ly$\alpha$ 
feature has been found. Larger Ly$\alpha$-emitting regions were found to be associated, on average, with Ly$\alpha$ emitters (LAEs) situated in overdense environments 
\citep{matsuda12, momose16, coolchiang15}. It has been suggested that the presence of such overdensities are possibly a requisite condition in the 
higher redshift universe in order to observe multiple strong, physically associated LAEs (e.g., \citealt{castellano16}). Indeed, some proto-cluster or proto-cluster 
candidates have been found solely through the presence of large numbers of relatively strong LAEs (e.g., \citealt{lpent00, venemans02, kuiper11, coolchiang15}) with some 
studies finding an increase of strong narrowband-selected LAEs with increasing local galaxy density (e.g., \citealt{zheng16}). Such an increase in the size of Ly$\alpha$ 
halos could be related to both the column density and kinematics of the neutral hydrogen in and around the galaxies (see \citealt{loopylucia17} and references therein) and, 
thus, a possible signature of a nascent medium. 

However, other results have found a lack of enhancement or suppression of Ly$\alpha$ amongst galaxies in the higher-density environments. Narrowband-selected
LAEs are found in general to be less clustered than those galaxies without strong Ly$\alpha$ emission at high redshifts (e.g, \citealt{ouchi10, bielby16}), 
though this trend is predicted to be a strong function of redshift and Ly$\alpha$ line luminosity \citep{orsi08}. In a study of 
the member galaxies of a $z\sim3.3$ proto-cluster, \citet{lem14a} found no definitive evidence for differences in individual or coadded Ly$\alpha$ measurements 
compared to a corresponding coeval field population. In a study of a narrow-band-selected $z=3.78$ proto-cluster, \citet{dey16} found no environmental dependence
of EW(Ly$\alpha$) between proto-cluster LAE candidates and coeval field candidate LAEs, though a marginal increase in the average Ly$\alpha$ luminosity was seen 
closer to the core of the proto-cluster. In a spectroscopic study of a proto-cluster LAEs at $z\sim3.67$, \citet{tersetoshikawa16} found  
proto-cluster member galaxies to exhibit less negative EW(Ly$\alpha$) relative to galaxies in a similarly selected field sample indicating a suppression of the Ly$\alpha$       % XXX you know there's a IA679 Subaru image of COSMOS (lambda_c= 6781.1 width=336.0A) to a 3sigma depth of 26th mag. At z=4.57, Lya = 6773A... XXX 
escape fraction in the proto-cluster. 

\begin{figure*}
\epsscale{1}
%\plottwo{COSMOS-VUDS.PS26vsfield.unitweight.logSMge9.ps}{COSMOS.protostruct26.EWnLumLyahist.ps}
\plottwo{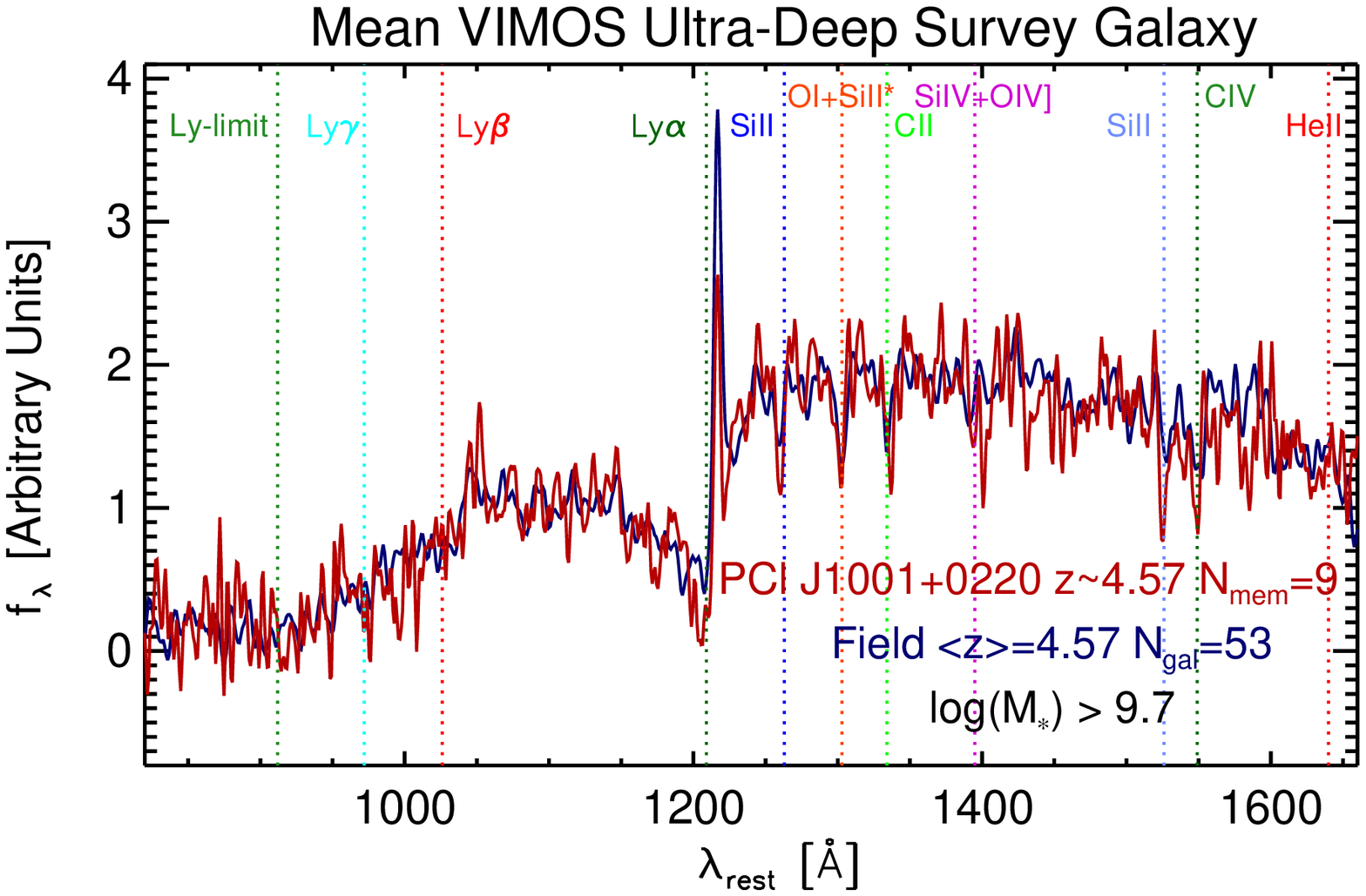}{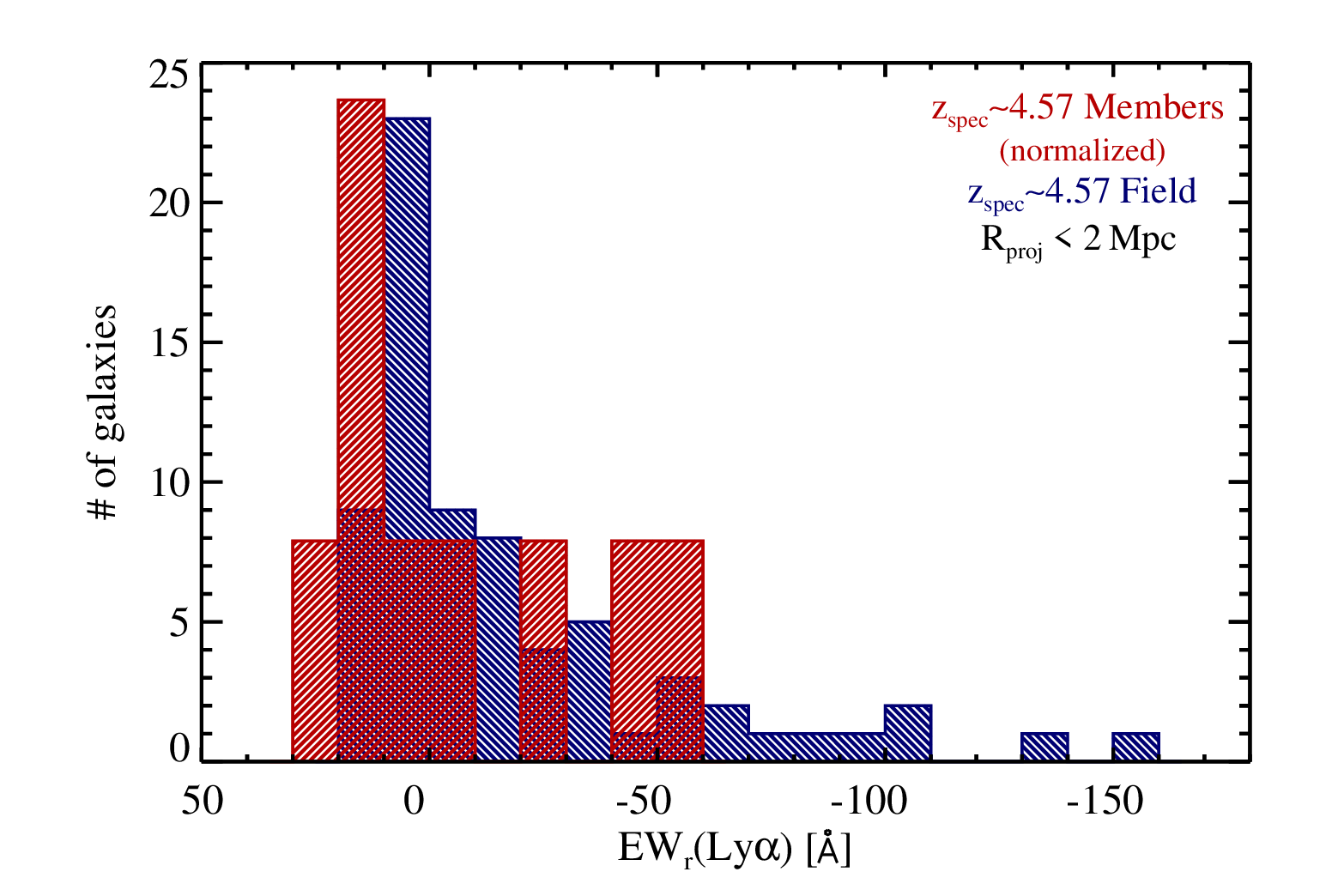}
\caption{\emph{Left:} Unit-weighted (normalized) mean rest-frame VUDS spectra of the nine spec-$z$ members of the PCl J1001+0220 proto-cluster plotted against that of the 53 coeval
field galaxies (one galaxy was removed due to the presence of severe reduction artifacts). The requisite stellar mass limit for each coaddition is given in the figure panel. 
Each spectrum is smoothed with a Gaussian of FWHM=1.5 pixels (1.45\AA).
Important absorption and emission features frequently observed in VUDS spectra are marked by vertical lines. Clear differences can be observed between the two spectra, with
the proto-cluster members exhibiting, on average, reduced Ly$\alpha$ emission and exaggerated absorption in Hydrogen and various metal lines relative to the average
coeval field galaxy. In both cases, a significant ($\sim$300 km s$^{-1}$) offset to the red can be seen for the Ly$\alpha$ line potentially indicating that large-scale outflows are
pervasive in galaxies at this redshift. \emph{Right:} Distribution of rest-frame Ly$\alpha$ equivalent widths (EWs) for the proto-cluster (red histogram) and coeval field sample
(blue histogram) measured on individual spectra. Negative or positive EW indicates the line is seen in emission or absorption, respectively. The two distributions are inconsistent at $>3\sigma$,
the Ly$\alpha$ feature being more likely to be seen in absorption for the proto-cluster members.}
\label{fig:coaddnEWs}
\end{figure*}

Here we attempt two basic comparisons. In the first comparison we plot in the left panel of Figure \ref{fig:coaddnEWs} the coadded spectrum of all nine PCl J1001+0220
members against that of the coeval field sample. These coadded spectra are generated following the methodology of \citet{lem14a} and show the 
unit-weighted (normalized) average (mean) spectra of the two populations. The wavelength region in the left panel of Figure \ref{fig:coaddnEWs} is tailored
such that only those wavelengths where all galaxies contribute the coadded spectra for each population are shown. Immediately it can be seen that, while the overall continuum
shape is similar between the two coadded spectra, a fact which bolsters the $\beta$-slope analysis presented in \S\ref{mainproperties}, there appears a noticeable difference 
in the strength of the emission component of the Ly$\alpha$ line. We use a custom bandpass \textsc{IDL} routine to measure 
the EW(Ly$\alpha$) of both coadded spectra, which, similarly to \citet{paolo15}, uses the continuum just redward of Ly$\alpha$ to set the continuum level. For individual 
measurements of EW(Ly$\alpha$), this routine returns measurements which are broadly consistent with those of \citet{paolo15}. Using this method we measure 
EW(Ly$\alpha$)=-4.7$\pm$0.6 and -15.4$\pm$0.3\AA\ for the proto-cluster member and coeval field galaxies, respectively. We note that these cannot be directly compared to 
the mean values of the individual measurements due to the fact that each spectrum has its average flux density scaled to unity prior to coadding. Uncertainties on 
mean EWs are derived from a combination of the covariance matrix of the linear continuum fit and the error spectrum of the coadded spectra constructed from the 
individual \textsc{VIPGI} error spectra 
of the input galaxies. Though there are some known issues with the latter, neither EWs or their associated uncertainties are found to change significantly if
a running measure of the root mean square fluctuations is instead used. The suppression of Ly$\alpha$ relative to the level of the UV continuum amongst the proto-cluster
galaxies is clearly significant, a difference which persists in significance and directionality if we instead perform unit-weighted median coadding. 
% EW(Ly$\alpha$)=-0.0$\pm$0.7 and -5.4$\pm$0.3\AA\ for the proto-cluster member and coeval field samples, respectively. 
In the right panel of Figure \ref{fig:coaddnEWs} we plot the distributions of the individual measurements of EW(Ly$\alpha$) as measured by 
\citet{paolo15} for the proto-cluster member and coeval field samples used to construct the coadded spectra. By eye it appears that the proto-cluster
members are skewed to less negative EW(Ly$\alpha$) values with respect to those of the coeval field sample, an observation which is borne out both by a KS test (rejection
of null hypothesis at $\sim$99\%) and by a comparison of the mean values (-5.5 vs. -37.6\AA, respectively).  

It is important to note, though, that, as we do not preserve flux density in our coaddition method, this suppression cannot necessarily be interpreted as a suppression of Ly$\alpha$ line 
luminosity. For this to also be true, the dust-uncorrected FUV luminosity between the two samples must be similar. Both a KS test and a comparison of the average 
of the $M_{FUV}$ values of the proto-cluster member and coeval field galaxies ($\tilde{M_{FUV}}=-21.77\pm0.18$ and $-21.60\pm0.06$, respectively) show no 
significant offset between the two samples. Echoing this result, a KS test performed directly on the Ly$\alpha$ line luminosity measurements, made following 
the method of \citet{paolo15}, rejects the hypothesis that the two distributions are drawn from the same underlying parent sample at $>99$\%. %We preferred, however, not to 
% rely solely on these measurements because they require good knowledge 
%of the systematics associated with the spectrophotometric calibration. Regardless,
It appears that the suppression of Ly$\alpha$ relative to the UV continuum observed in the proto-cluster members relative to the coeval field galaxies extends to an 
absolute suppression of Ly$\alpha$ luminosity. Given that the two populations appear to have similar dust properties and UV-derived SFRs, and because the absolute brightness
of Ly$\alpha$ in the absence of resonant scattering and dust is directly proportional to the amount of star formation, this decrease can be translated directly into 
the fraction of Ly$\alpha$ photons which escape the average galaxy in the two populations. Adopting the Ly$\alpha$ SFR conversion of \citet{lem09} with no correction for 
intergalactic medium absorption, converting to a \citet{chab03} IMF, and replacing L(Ly$\alpha$) with a combination of EW(Ly$\alpha$) and $M_{FUV}$ gives: 

\begin{equation}
SFR(L_{\rm{Ly}\alpha}) = -8.6\times10^{9} (10^{\frac{-(\langle M_{FUV} \rangle +48.6)}{2.5}}) \langle EW(\rm{Ly}\alpha) \rangle \, \mathcal{M}_{\odot} \, \rm{yr}^{-1}\\
\label{eqn:SFRLya}
,\end{equation}

\noindent where $M_{FUV}$ is the (linear) average $M_{FUV}$ of the sample, $\langle$EW(Ly$\alpha$)$\rangle$ is measured in the coadded spectra, and the constant of proportionality comes from 
a combination of that given in \citet{lem09}, 4$\pi d_{L}^2$, where $d_{L}$ is 10 pc in cm, c is cm s$^{-1}$, and the $\lambda_{eff}$ of the FUV filter 
($1539\times10^{-8}$ cm, \citealt{ciao11}). We then define the escape fraction of Ly$\alpha$ as:

\begin{equation}
f_{esc, \, Ly\alpha}= SFR(L_{\rm{Ly}\alpha})/SFR_{SED}\\
\label{eqn:fesclya}
,\end{equation}

\noindent where the SFR$_{SED}$ are the average values given in \S\ref{mainproperties}. For the full VUDS sample in COSMOS at $2 \le z \le 5$ with flags $\ge$ X2 we have 
confirmed that, as EW(Ly$\alpha$) becomes more negative, that is, more Ly$\alpha$ flux for a given UV continuum level, $f_{esc, \, \rm{Ly}\alpha}$ approaches 100\% using the 
above formalism. The resulting values for the proto-cluster member and coeval field sample are $f_{esc, \, \rm{Ly}\alpha} = 1.8^{+0.3}_{-1.7}$\% and $4.0^{+1.0}_{-0.8}$\%, 
respectively. The coeval field value is broadly consistent with the global $f_{esc, \, \rm{Ly}\alpha}$ at $4 < z < 5$ as derived through comparisons between Ly$\alpha$ and 
UV luminosity functions \citep{hayes11, zheng13} or other measures (e.g., \citealt{wardlow14, paolo15} though these measurements were limited to LAEs only), especially 
considering the unknown uncertainties of differing sample selections and measurement schemes. The average $f_{esc, \, \rm{Ly}\alpha}$ for proto-cluster members appears, 
however, at the extreme low end of measured values at these redshifts. 

A variety of factors can increase the chance of the escape of Ly$\alpha$ photons or the appearance thereof: geometric effects, observational bias, dust content and geometry, 
strong outflows, and decreased column densities of neutral hydrogen along the line of sight. Because we are averaging over a reasonably large number of galaxies in each 
sample, geometrical effects (see, e.g., \citealt{finkel08}), are likely not the cause of the observed difference. Similarly, because we are comparing galaxy samples 
with the same mean redshift observed in an identical manner, 
it is unlikely that differing measurement apertures, a pernicious effect when measuring $f_{esc, \, \rm{Ly}\alpha}$ (e.g., \citealt{mathee16}), affect our measurements here.
Further, we have shown through a variety of methods that the dust content between the two samples
is likely to be the same. Therefore,
we are left with only two possibilities: increased prevalence and strength of outflows amongst the coeval sample or decreased
\ion{H}{I} column density. It has been shown that the presence of outflows and their strength depend, in the absence of strong AGN activity, on the instantaneous
SFR (e.g., \citealt{heckman01, weiner09, erb15}), a quantity which, as far as we can tell, is the same between the two samples. In addition, we see no difference 
in the velocity offset of the Ly$\alpha$ line in the coadded spectra with respect to the ISM lines, 310 versus 270 km s$^{-1}$ for the coeval field and proto-cluster 
member coadded spectra, respectively, which suggests similar \ion{H}{I} kinematics between the two galaxies. However, this test is far from definitive as the relationship between 
these components is not necessarily straightforward (e.g., \citealt{alice03, loopylucia17}). 

What remains, then, is the intriguing possibility that the proto-cluster galaxies see a denser medium in their immediate vicinity or a more isotropic medium 
resulting in a higher covering fraction of \ion{H}{I}. Such conditions would cause a subsequent drop of Ly$\alpha$ photons able to escape (e.g., \citealt{reddy16}) 
due to their increased possibility of interacting with a dust grain for a given dust content and geometry. The presence of a large and dense neutral medium 
is a possibility which has been raised previously with the observation of large scale Ly$\alpha$ absorption in another VUDS proto-cluster \citep{olga14} and 
in Ly$\alpha$ tomography studies (e.g., \citealt{lee16}) and is consistent with the idea of the assembly of a nascent proto-ICM. Assuming a one-to-one correspondence between increased \ion{H}{I} and an increase in heavy elements, the possibility of denser medium is also broadly consistent with
the relative strength ISM absorption features, as the average ratio between EWs measured on the proto-cluster member and coeval field coadded spectra for the six ISM absorption
features labeled in Figure \ref{fig:coaddnEWs} is 1.7$\pm$0.2, measured using the bandpass method of \citet{lem10} with custom bandpasses. From comparing Gaussian
fits to each of the features in the two coadditions we find that the increase in the average EW is driven almost entirely by an increased (relative) depth of the 
absorption feature rather than by an increase in the feature width. Since all ISM features should be unresolved in VUDS data, this lack of appreciable difference in the widths
measured in the ISM features of the two coadded spectra strongly indicates that the increased absorption strength in the cluster member coadded spectrum is real. However, the relationship 
between ionic column density and ISM absorption strength is a complicated one (see \citealt{marghe12} and references therein) and, as such, while suggestive, these measurements 
do not offer definitive proof that the galaxies inhabiting the proto-cluster are experiencing a denser medium. Such lines of inquiry will continue to be followed 
for the entire sample of VUDS proto-structures. It may be, though, that observations of cold gas from, for example, the Atacama Large Millimeter/submillimeter 
Array will be needed to definitively confirm or deny this assembly, a prospect which may have already begun to be realized \citep{ginolfi16, helpfulhelmut17}. 

\section{Summary and conclusions}
\label{conclusions}

In this study we reported on the detection, spectroscopic confirmation, and the mapping of the PCl J1001+0220 proto-cluster at $z\sim4.57$ in 
the COSMOS field. We undertook a preliminary investigation of various properties of the spectroscopic and photometric proto-cluster member galaxies
and contrast these with matched field samples. The following were the main conclusions of these investigations.

\begin{itemize}
\renewcommand{\labelitemi}{$\bullet$}

\item PCl J1001+0220 was discovered in a blind spectroscopic survey without the use of traditional beacons of proto-structure or proto-cluster activity
such as quasars or other prodigious radio emitters. Its large extent on the plane of the sky is consistent with both predicted and observed sizes 
of high-redshift proto-clusters. 

\item The galaxy overdensity of the proto-cluster was measured using three methods: spectroscopic overdensity, photometric redshift overdensity, 
and a new method, Voronoi Monte-Carlo mapping, which statistically combines both the spectroscopic and photometric redshifts. While the level of 
overdensity was found to be dependent on the method, in all three methods PCl J1001+0220 was significantly detected. The overdensity derived from the
Voronoi Monte-Carlo mapping was considered the most secure, with PCl J1001+0220 exhibiting an average $\delta_{gal}=3.30\pm0.32$ over a sky
area of 7.58 Mpc$^{2}$ and a line-of-sight distance of 7.49 Mpc.  

\item Several methods were attempted to measure the total mass associated with PCl J1001+0220. Though subject to a large number of assumptions and uncertainties, 
this analysis provided a general picture that PCl J1001+0220 will evolve into a cluster of mass $\sim3-10\times10^{14}$ $\mathcal{M}_{\odot}$ by the present 
day, with the mass estimated from the overdensity derived from the Voronoi Monte Carlo approach falling at the low end of these estimates.
 
\item When comparing various member populations to similarly selected coeval field samples we found tentative evidence for a fractional excess of older and
more massive galaxies within the proto-cluster bounds suggesting that the SFHs of galaxies in the proto-cluster environment differ
appreciably, on average, from those in the field. 

\item A variety of methods were used to attempt to quantify the level of AGN activity amongst the various member samples using some of the deepest 
multiwavelength data ever compiled. In contrast with studies of many other high-redshift proto-clusters, no indication was found for the presence of 
any type of AGN within the proto-cluster.

\item No evidence was found for either suppression or enhancement of star formation activity within the proto-cluster bounds through comparison of 
dust-corrected rest-frame UV star-formation indicators. Through cross-matching and stacking newly obtained deep JVLA 3 GHz imaging of the COSMOS 
field for both the proto-cluster member and coeval field samples it was demonstrated that neither environment contained a large number of dusty
starbursting galaxies. Limits on the stacking analysis and the distribution of measured $\beta$ slopes precluded the possibility of copious amounts 
of dust in the average galaxy in both populations.

\item The rest-frame FUV morphologies from the \emph{HST}/$F814W$ mosaic of the COSMOS field for the proto-cluster member and coeval field population 
were compared visually, parametrically, and non-parametrically. The nine $z_{spec}$ members spanned the entirety of all quantifiable morphological 
measurements of the corresponding coeval field sample which suggested, in tandem with a possible increase in the average number of background-corrected 
close pairs in both the $z_{spec}$ and $z_{phot}$ member sample, a potential enhancement of galaxy-galaxy interactions within the proto-cluster. 

\item The individual and stacked spectral properties of the member and coeval field galaxies were contrasted. The member population exhibited a decided
suppression in the strength of the Ly$\alpha$ emission feature relative to the coeval field as well as a decrease in the estimated average 
Ly$\alpha$ escape fraction ($1.8^{+0.3}_{-1.7}$\% vs. $4.0^{+1.0}_{-0.8}$\%, respectively). Additionally, the average absorption of ISM features in 
the stacked spectra was markedly stronger for the proto-cluster members. These lines of evidence were used to suggest the possible presence of a 
large and dense medium housed in the proto-cluster. 

\end{itemize} 

Despite the relatively small sample assembled here, the observed properties of galaxies in PCl J1001+0220 tentatively demonstrate that local environment 
may already impact galaxy evolution and that the effects which give rise to the massive quiescent galaxies that dominate local galaxy clusters 
may begin to be felt as early as just 1.3 Gyr after the Big Bang. %at the edge of the reionization era. 
%early cosmic times, just 1.3 Gyr after the Big Bang. While the effects of environment in rich clusters are well documented in the local universe, Cl J1001+0220 
%provides an example that environmentally-driven evolution has possibly already started at the edge of the reionization era. 
Further, the simple presence of PCl J1001+0220 and the inference of the large amount of mass assembled in the proto-cluster by $z\sim4.57$ is a clear indication 
that such overdensities were able to begin to distinguish themselves from average density regions of the universe well before the end of the epoch of reionization.
While the existence and nature of PCl J1001+0220 was proven definitively in this study, many of the results pertaining to galaxy evolution within
PCl J1001+0220 presented in this work do not have the statistical weight to be anything other than suggestive. That we were able to even attempt 
such lines of analysis in some sort of controlled manner at these redshifts already constitutes a triumph. It will be up to further spectroscopic 
observations of PCl J1001+0220, followup observations at other wavelengths, and future works using the full power of
the VUDS survey or other large-scale spectroscopic surveys, such as the Subaru Prime Focus Spectrograph 
Galaxy Evolution Survey, to determine whether the tentative signs of early-onset environmentally-driven evolution seen here are legitimate 
hallmarks of the general population of high-redshift proto-clusters.
 
%The average number of background-corrected 
%close pairs were counted for both the $z_{spec}$ member and coeval field samples (0.4$\pm$0.3 and 0.1$\pm$0.1, respectively) and the $z_{phot}$ member 
%and coeval field samples (0.17$\pm$0.08 vs. 0.08$\pm$0.02, respectively)  

% Thanks. For listening. Thank Thibaud and Iary and AA. And the EZGal people. And Pol and Matevey for their fine company. 
\begin{acknowledgements}
%Thanks. For listening.
This work was supported by funding from the European Research Council Advanced Grant ERC-2010-AdG-268107-EARLY and by INAF Grants PRIN 2010, PRIN 2012 and PICS 2013.
This work was additionally supported by the National Science Foundation under Grant No. 1411943 and NASA Grant Number NNX15AK92G.
VS, ID and MN acknowledge the European Union's Seventh Framework programme under grant agreement 337595 (ERC Starting Grant, "CoSMass").
We thank Thibaud Moutard, Iary Davidzon, Lori M.\ Lubin, and Adam R. Tomczak for their insight relevant to the various analyses presented in this paper and their willingness 
to impart it through conversation. We also thank Conor Mancone and Anthony Gonzalez for their support with {\sc EZGal} under circumstances that made such support 
difficult and Clotilde Laigle for generously providing access to the COSMOS2015 catalog at an early stage of this work, a gesture which immensely helped its development. 
BCL thanks Pol Mollitor and Matvey Sapunov for providing the finest of company during the entirety of this work and for continually helping to create the 
destruction necessary, amongst other things, to write. This work is based, in part, on observations made with the Spitzer Space Telescope, which is operated by the 
Jet Propulsion Laboratory, California Institute of Technology under a contract with NASA. The scientific results reported in this article are also based in part on 
observations made by the Chandra X-ray 
Observatory and data obtained from the Chandra Data Archive. We thank ESO staff for their continuous support for the VUDS survey, particularly the Paranal staff 
conducting the observations and Marina Rejkuba and the ESO user support group in Garching.
\end{acknowledgements}

\vskip6pt\noindent{{\it\large Facilities}: CFHT, ESO:VISTA, GALEX, Herschel, HST (ACS), Spitzer, Subaru, VLA, VLT:Melipal, XMM}

\bibliographystyle{aa} % style aa.bst
\bibliography{vuds_COSMOS_LSS_morecompetitive} % your references Yourfile.bib

\newcommand{\noop}[1]{}
\begin{thebibliography}{259}
\expandafter\ifx\csname natexlab\endcsname\relax\def\natexlab#1{#1}\fi

\bibitem[{{Abell}(1958)}]{abell58}
{Abell}, G.~O. 1958, \apjs, 3, 211

\bibitem[{{Allen} {et~al.}(2015){Allen}, {Kacprzak}, {Spitler}, {Glazebrook},
  {Labb{\'e}}, {Tran}, {Straatman}, {Nanayakkara}, {Brammer}, {Quadri},
  {Cowley}, {Monson}, {Papovich}, {Persson}, {Rees}, {Tilvi}, \&
  {Tomczak}}]{allen15}
{Allen}, R.~J., {Kacprzak}, G.~G., {Spitler}, L.~R., {et~al.} 2015, \apj, 806,
  3

\bibitem[{{{\'A}lvarez-M{\'a}rquez} {et~al.}(2016){{\'A}lvarez-M{\'a}rquez},
  {Burgarella}, {Heinis}, {Buat}, {Lo Faro}, {B{\'e}thermin},
  {L{\'o}pez-Fort{\'{\i}}n}, {Cooray}, {Farrah}, {Hurley}, {Ibar}, {Ilbert},
  {Koekemoer}, {Lemaux}, {P{\'e}rez-Fournon}, {Rodighiero}, {Salvato}, {Scott},
  {Taniguchi}, {Vieira}, \& {Wang}}]{javier16}
{{\'A}lvarez-M{\'a}rquez}, J., {Burgarella}, D., {Heinis}, S., {et~al.} 2016,
  \aap, 587, A122

\bibitem[{{Andreon}(2012)}]{andreon12}
{Andreon}, S. 2012, \aap, 548, A83

\bibitem[{{Aragon-Calvo} {et~al.}(2015){Aragon-Calvo}, {Weygaert}, {Jones}, \&
  {Mobasher}}]{aragon-calvo15}
{Aragon-Calvo}, M.~A., {Weygaert}, R.~v.~d., {Jones}, B.~J.~T., \& {Mobasher},
  B. 2015, \mnras, 454, 463

\bibitem[{{Aravena} {et~al.}(2010){Aravena}, {Bertoldi}, {Carilli},
  {Schinnerer}, {McCracken}, {Salvato}, {Riechers}, {Sheth}, {Sm{\v
  o}lci{\'c}}, {Capak}, {Koekemoer}, \& {Menten}}]{aravena10}
{Aravena}, M., {Bertoldi}, F., {Carilli}, C., {et~al.} 2010, \apjl, 708, L36

\bibitem[{{Aretxaga} {et~al.}(2011){Aretxaga}, {Wilson}, {Aguilar}, {Alberts},
  {Scott}, {Scoville}, {Yun}, {Austermann}, {Downes}, {Ezawa}, {Hatsukade},
  {Hughes}, {Kawabe}, {Kohno}, {Oshima}, {Perera}, {Tamura}, \&
  {Zeballos}}]{Aretxaga11}
{Aretxaga}, I., {Wilson}, G.~W., {Aguilar}, E., {et~al.} 2011, \mnras, 415,
  3831

\bibitem[{{Arnouts} {et~al.}(1999){Arnouts}, {Cristiani}, {Moscardini},
  {Matarrese}, {Lucchin}, {Fontana}, \& {Giallongo}}]{stephane99}
{Arnouts}, S., {Cristiani}, S., {Moscardini}, L., {et~al.} 1999, \mnras, 310,
  540

\bibitem[{{Arnouts} {et~al.}(2013){Arnouts}, {Le Floc'h}, {Chevallard},
  {Johnson}, {Ilbert}, {Treyer}, {Aussel}, {Capak}, {Sanders}, {Scoville},
  {McCracken}, {Milliard}, {Pozzetti}, \& {Salvato}}]{stephane13}
{Arnouts}, S., {Le Floc'h}, E., {Chevallard}, J., {et~al.} 2013, \aap, 558, A67

\bibitem[{{Ascaso} {et~al.}(2015){Ascaso}, {Ben{\'{\i}}tez},
  {Fern{\'a}ndez-Soto}, {Arnalte-Mur}, {L{\'o}pez-Sanjuan}, {Molino},
  {Schoenell}, {Jim{\'e}nez-Teja}, {Merson}, {Huertas-Company},
  {D{\'{\i}}az-Garc{\'{\i}}a}, {Mart{\'{\i}}nez}, {Cenarro}, {Dupke},
  {M{\'a}rquez}, {Masegosa}, {Nieves-Seoane}, {Povi{\'c}}, {Varela},
  {Viironen}, {Aguerri}, {Olmo}, {Moles}, {Perea}, {Alfaro},
  {Aparicio-Villegas}, {Broadhurst}, {Cabrera-Ca{\~n}o}, {Castander}, {Cepa},
  {Cervi{\~n}o}, {Delgado}, {Crist{\'o}bal-Hornillos}, {Hurtado-Gil},
  {Husillos}, {Infante}, {Prada}, \& {Quintana}}]{superB15}
{Ascaso}, B., {Ben{\'{\i}}tez}, N., {Fern{\'a}ndez-Soto}, A., {et~al.} 2015,
  \mnras, 452, 549

\bibitem[{{Ascaso} {et~al.}(2014){Ascaso}, {Lemaux}, {Lubin}, {Gal},
  {Kocevski}, {Rumbaugh}, \& {Squires}}]{superB14}
{Ascaso}, B., {Lemaux}, B.~C., {Lubin}, L.~M., {et~al.} 2014, \mnras, 442, 589

\bibitem[{{Balogh} {et~al.}(2016){Balogh}, {McGee}, {Mok}, {Muzzin}, {van der
  Burg}, {Bower}, {Finoguenov}, {Hoekstra}, {Lidman}, {Mulchaey}, {Noble},
  {Parker}, {Tanaka}, {Wilman}, {Webb}, {Wilson}, \& {Yee}}]{balogh16}
{Balogh}, M.~L., {McGee}, S.~L., {Mok}, A., {et~al.} 2016, \mnras, 456, 4364

\bibitem[{{Becker} {et~al.}(2015){Becker}, {Bolton}, \& {Lidz}}]{becker15}
{Becker}, G.~D., {Bolton}, J.~S., \& {Lidz}, A. 2015, \pasa, 32, e045

\bibitem[{{Becker} {et~al.}(2001){Becker}, {Fan}, {White}, {Strauss},
  {Narayanan}, {Lupton}, {Gunn}, {Annis}, {Bahcall}, {Brinkmann}, {Connolly},
  {Csabai}, {Czarapata}, {Doi}, {Heckman}, {Hennessy}, {Ivezi{\'c}}, {Knapp},
  {Lamb}, {McKay}, {Munn}, {Nash}, {Nichol}, {Pier}, {Richards}, {Schneider},
  {Stoughton}, {Szalay}, {Thakar}, \& {York}}]{thebob01}
{Becker}, R.~H., {Fan}, X., {White}, R.~L., {et~al.} 2001, \aj, 122, 2850

\bibitem[{{Bell}(2003)}]{bell03}
{Bell}, E.~F. 2003, \apj, 586, 794

\bibitem[{{Berta} {et~al.}(2013){Berta}, {Lutz}, {Santini}, {Wuyts}, {Rosario},
  {Brisbin}, {Cooray}, {Franceschini}, {Gruppioni}, {Hatziminaoglou}, {Hwang},
  {Le Floc'h}, {Magnelli}, {Nordon}, {Oliver}, {Page}, {Popesso}, {Pozzetti},
  {Pozzi}, {Riguccini}, {Rodighiero}, {Roseboom}, {Scott}, {Symeonidis},
  {Valtchanov}, {Viero}, \& {Wang}}]{berta13}
{Berta}, S., {Lutz}, D., {Santini}, P., {et~al.} 2013, \aap, 551, A100

\bibitem[{{Bertin} \& {Arnouts}(1996)}]{BertinArn96}
{Bertin}, E. \& {Arnouts}, S. 1996, \aaps, 117, 393

\bibitem[{{Bielby} {et~al.}(2016){Bielby}, {Tummuangpak}, {Shanks}, {Francke},
  {Crighton}, {Ba{\~n}ados}, {Gonz{\'a}lez-L{\'o}pez}, {Infante}, \&
  {Orsi}}]{bielby16}
{Bielby}, R.~M., {Tummuangpak}, P., {Shanks}, T., {et~al.} 2016, \mnras, 456,
  4061

\bibitem[{{Bleem} {et~al.}(2015){Bleem}, {Stalder}, {de Haan}, {Aird}, {Allen},
  {Applegate}, {Ashby}, {Bautz}, {Bayliss}, {Benson}, {Bocquet}, {Brodwin},
  {Carlstrom}, {Chang}, {Chiu}, {Cho}, {Clocchiatti}, {Crawford}, {Crites},
  {Desai}, {Dietrich}, {Dobbs}, {Foley}, {Forman}, {George}, {Gladders},
  {Gonzalez}, {Halverson}, {Hennig}, {Hoekstra}, {Holder}, {Holzapfel},
  {Hrubes}, {Jones}, {Keisler}, {Knox}, {Lee}, {Leitch}, {Liu}, {Lueker},
  {Luong-Van}, {Mantz}, {Marrone}, {McDonald}, {McMahon}, {Meyer}, {Mocanu},
  {Mohr}, {Murray}, {Padin}, {Pryke}, {Reichardt}, {Rest}, {Ruel}, {Ruhl},
  {Saliwanchik}, {Saro}, {Sayre}, {Schaffer}, {Schrabback}, {Shirokoff},
  {Song}, {Spieler}, {Stanford}, {Staniszewski}, {Stark}, {Story}, {Stubbs},
  {Vanderlinde}, {Vieira}, {Vikhlinin}, {Williamson}, {Zahn}, \&
  {Zenteno}}]{breem15}
{Bleem}, L.~E., {Stalder}, B., {de Haan}, T., {et~al.} 2015, \apjs, 216, 27

\bibitem[{{Boselli} {et~al.}(2011){Boselli}, {Boissier}, {Heinis}, {Cortese},
  {Ilbert}, {Hughes}, {Cucciati}, {Davies}, {Ferrarese}, {Giovanelli},
  {Haynes}, {Baes}, {Balkowski}, {Brosch}, {Chapman}, {Charmandaris},
  {Clemens}, {Dariush}, {De Looze}, {di Serego Alighieri}, {Duc}, {Durrell},
  {Emsellem}, {Erben}, {Fritz}, {Garcia-Appadoo}, {Gavazzi}, {Grossi},
  {Jord{\'a}n}, {Hess}, {Huertas-Company}, {Hunt}, {Kent}, {Lambas}, {Lan{\c
  c}on}, {MacArthur}, {Madden}, {Magrini}, {Mei}, {Momjian}, {Olowin},
  {Papastergis}, {Smith}, {Solanes}, {Spector}, {Spekkens}, {Taylor},
  {Valotto}, {van Driel}, {Verstappen}, {Vlahakis}, {Vollmer}, \&
  {Xilouris}}]{ciao11}
{Boselli}, A., {Boissier}, S., {Heinis}, S., {et~al.} 2011, \aap, 528, A107

\bibitem[{{Boulade} {et~al.}(2003){Boulade}, {Charlot}, {Abbon}, {Aune},
  {Borgeaud}, {Carton}, {Carty}, {Da Costa}, {Deschamps}, {Desforge},
  {Eppell{\'e}}, {Gallais}, {Gosset}, {Granelli}, {Gros}, {de Kat}, {Loiseau},
  {Ritou}, {Rouss{\'e}}, {Starzynski}, {Vignal}, \& {Vigroux}}]{boulade03}
{Boulade}, O., {Charlot}, X., {Abbon}, P., {et~al.} 2003, in \procspie, Vol.
  4841, Instrument Design and Performance for Optical/Infrared Ground-based
  Telescopes, ed. M.~{Iye} \& A.~F.~M. {Moorwood}, 72--81

\bibitem[{{Bouwens} {et~al.}(2015){Bouwens}, {Illingworth}, {Oesch}, {Trenti},
  {Labb{\'e}}, {Bradley}, {Carollo}, {van Dokkum}, {Gonzalez}, {Holwerda},
  {Franx}, {Spitler}, {Smit}, \& {Magee}}]{anykey15}
{Bouwens}, R.~J., {Illingworth}, G.~D., {Oesch}, P.~A., {et~al.} 2015, \apj,
  803, 34

\bibitem[{{Brusa} {et~al.}(2010){Brusa}, {Civano}, {Comastri}, {Miyaji},
  {Salvato}, {Zamorani}, {Cappelluti}, {Fiore}, {Hasinger}, {Mainieri},
  {Merloni}, {Bongiorno}, {Capak}, {Elvis}, {Gilli}, {Hao}, {Jahnke},
  {Koekemoer}, {Ilbert}, {Le Floc'h}, {Lusso}, {Mignoli}, {Schinnerer},
  {Silverman}, {Treister}, {Trump}, {Vignali}, {Zamojski}, {Aldcroft},
  {Aussel}, {Bardelli}, {Bolzonella}, {Cappi}, {Caputi}, {Contini},
  {Finoguenov}, {Fruscione}, {Garilli}, {Impey}, {Iovino}, {Iwasawa},
  {Kampczyk}, {Kartaltepe}, {Kneib}, {Knobel}, {Kovac}, {Lamareille},
  {Leborgne}, {Le Brun}, {Le Fevre}, {Lilly}, {Maier}, {McCracken}, {Pello},
  {Peng}, {Perez-Montero}, {de Ravel}, {Sanders}, {Scodeggio}, {Scoville},
  {Tanaka}, {Taniguchi}, {Tasca}, {de la Torre}, {Tresse}, {Vergani}, \&
  {Zucca}}]{brusa10}
{Brusa}, M., {Civano}, F., {Comastri}, A., {et~al.} 2010, \apj, 716, 348

\bibitem[{{Bruzual} \& {Charlot}(2003)}]{bc03}
{Bruzual}, G. \& {Charlot}, S. 2003, \mnras, 344, 1000

\bibitem[{{Cai} {et~al.}(2017){Cai}, {Fan}, {Bian}, {Zabludoff}, {Yang},
  {Prochaska}, {McGreer}, {Zheng}, {Kashikawa}, {Wang}, {Frye}, {Green}, \&
  {Jiang}}]{whathappensifIdonthavemyticket17}
{Cai}, Z., {Fan}, X., {Bian}, F., {et~al.} 2017, \apj, 839, 131

\bibitem[{{Calzetti} {et~al.}(1994){Calzetti}, {Kinney}, \&
  {Storchi-Bergmann}}]{calzetti94}
{Calzetti}, D., {Kinney}, A.~L., \& {Storchi-Bergmann}, T. 1994, \apj, 429, 582

\bibitem[{{Campanelli} {et~al.}(2012){Campanelli}, {Fogli}, {Kahniashvili},
  {Marrone}, \& {Ratra}}]{campanelli12}
{Campanelli}, L., {Fogli}, G.~L., {Kahniashvili}, T., {Marrone}, A., \&
  {Ratra}, B. 2012, European Physical Journal C, 72, 2218

\bibitem[{{Capak} {et~al.}(2007){Capak}, {Aussel}, {Ajiki}, {McCracken},
  {Mobasher}, {Scoville}, {Shopbell}, {Taniguchi}, {Thompson}, {Tribiano},
  {Sasaki}, {Blain}, {Brusa}, {Carilli}, {Comastri}, {Carollo}, {Cassata},
  {Colbert}, {Ellis}, {Elvis}, {Giavalisco}, {Green}, {Guzzo}, {Hasinger},
  {Ilbert}, {Impey}, {Jahnke}, {Kartaltepe}, {Kneib}, {Koda}, {Koekemoer},
  {Komiyama}, {Leauthaud}, {Le Fevre}, {Lilly}, {Liu}, {Massey}, {Miyazaki},
  {Murayama}, {Nagao}, {Peacock}, {Pickles}, {Porciani}, {Renzini}, {Rhodes},
  {Rich}, {Salvato}, {Sanders}, {Scarlata}, {Schiminovich}, {Schinnerer},
  {Scodeggio}, {Sheth}, {Shioya}, {Tasca}, {Taylor}, {Yan}, \&
  {Zamorani}}]{capak07}
{Capak}, P., {Aussel}, H., {Ajiki}, M., {et~al.} 2007, \apjs, 172, 99

\bibitem[{{Capak} {et~al.}(2008){Capak}, {Carilli}, {Lee}, {Aldcroft},
  {Aussel}, {Schinnerer}, {Wilson}, {Yun}, {Blain}, {Giavalisco}, {Ilbert},
  {Kartaltepe}, {Lee}, {McCracken}, {Mobasher}, {Salvato}, {Sasaki}, {Scott},
  {Sheth}, {Shioya}, {Thompson}, {Elvis}, {Sanders}, {Scoville}, \&
  {Tanaguchi}}]{capak08}
{Capak}, P., {Carilli}, C.~L., {Lee}, N., {et~al.} 2008, \apjl, 681, L53

\bibitem[{{Capak} {et~al.}(2011{\natexlab{a}}){Capak}, {Riechers}, {Scoville},
  {Carilli}, {Cox}, {Neri}, {Salvato}, {Schinnerer}, {Yan}, {Wilson}, {Yun},
  {Civano}, {Elvis}, {Karim}, {Mobasher}, \& {Staguhn}}]{pcap11}
{Capak}, P.~L., {Riechers}, D., {Scoville}, N., {et~al.} 2011{\natexlab{a}}, in
  Bulletin of the American Astronomical Society, Vol.~43, American Astronomical
  Society Meeting Abstracts \#217, 430.23

\bibitem[{{Capak} {et~al.}(2011{\natexlab{b}}){Capak}, {Riechers}, {Scoville},
  {Carilli}, {Cox}, {Neri}, {Robertson}, {Salvato}, {Schinnerer}, {Yan},
  {Wilson}, {Yun}, {Civano}, {Elvis}, {Karim}, {Mobasher}, \&
  {Staguhn}}]{capak11}
{Capak}, P.~L., {Riechers}, D., {Scoville}, N.~Z., {et~al.} 2011{\natexlab{b}},
  \nat, 470, 233

\bibitem[{{Cappelluti} {et~al.}(2009){Cappelluti}, {Brusa}, {Hasinger},
  {Comastri}, {Zamorani}, {Finoguenov}, {Gilli}, {Puccetti}, {Miyaji},
  {Salvato}, {Vignali}, {Aldcroft}, {B{\"o}hringer}, {Brunner}, {Civano},
  {Elvis}, {Fiore}, {Fruscione}, {Griffiths}, {Guzzo}, {Iovino}, {Koekemoer},
  {Mainieri}, {Scoville}, {Shopbell}, {Silverman}, \& {Urry}}]{cappelluti09}
{Cappelluti}, N., {Brusa}, M., {Hasinger}, G., {et~al.} 2009, \aap, 497, 635

\bibitem[{{Caputi} {et~al.}(2015){Caputi}, {Ilbert}, {Laigle}, {McCracken}, {Le
  F{\`e}vre}, {Fynbo}, {Milvang-Jensen}, {Capak}, {Salvato}, \&
  {Taniguchi}}]{caputi15}
{Caputi}, K.~I., {Ilbert}, O., {Laigle}, C., {et~al.} 2015, \apj, 810, 73

\bibitem[{{Casey} {et~al.}(2013){Casey}, {Chen}, {Cowie}, {Barger}, {Capak},
  {Ilbert}, {Koss}, {Lee}, {Le Floc'h}, {Sanders}, \&
  {Williams}}]{connivingcasey13}
{Casey}, C.~M., {Chen}, C.-C., {Cowie}, L.~L., {et~al.} 2013, \mnras, 436, 1919

\bibitem[{{Casey} {et~al.}(2015){Casey}, {Cooray}, {Capak}, {Fu}, {Kovac},
  {Lilly}, {Sanders}, {Scoville}, \& {Treister}}]{conivingcasey15}
{Casey}, C.~M., {Cooray}, A., {Capak}, P., {et~al.} 2015, \apjl, 808, L33

\bibitem[{{Cassata} {et~al.}(2015){Cassata}, {Tasca}, {Le F{\`e}vre}, {Lemaux},
  {Garilli}, {Le Brun}, {Maccagni}, {Pentericci}, {Thomas}, {Vanzella},
  {Zamorani}, {Zucca}, {Amorin}, {Bardelli}, {Capak}, {Cassar{\`a}},
  {Castellano}, {Cimatti}, {Cuby}, {Cucciati}, {de la Torre}, {Durkalec},
  {Fontana}, {Giavalisco}, {Grazian}, {Hathi}, {Ilbert}, {Moreau}, {Paltani},
  {Ribeiro}, {Salvato}, {Schaerer}, {Scodeggio}, {Sommariva}, {Talia},
  {Taniguchi}, {Tresse}, {Vergani}, {Wang}, {Charlot}, {Contini}, {Fotopoulou},
  {Koekemoer}, {L{\'o}pez-Sanjuan}, {Mellier}, \& {Scoville}}]{paolo15}
{Cassata}, P., {Tasca}, L.~A.~M., {Le F{\`e}vre}, O., {et~al.} 2015, \aap, 573,
  A24

\bibitem[{{Castellano} {et~al.}(2014){Castellano}, {Sommariva}, {Fontana},
  {Pentericci}, {Santini}, {Grazian}, {Amorin}, {Donley}, {Dunlop}, {Ferguson},
  {Fiore}, {Galametz}, {Giallongo}, {Guo}, {Huang}, {Koekemoer}, {Maiolino},
  {McLure}, {Paris}, {Schaerer}, {Troncoso}, \& {Vanzella}}]{castellano14}
{Castellano}, M., {Sommariva}, V., {Fontana}, A., {et~al.} 2014, \aap, 566, A19

\bibitem[{{Castellano} {et~al.}(2016){Castellano}, {Yue}, {Ferrara}, {Merlin},
  {Fontana}, {Amor{\'{\i}}n}, {Grazian}, {M{\'a}rmol-Queralto},
  {Micha{\l}owski}, {Mortlock}, {Paris}, {Parsa}, {Pilo}, \&
  {Santini}}]{castellano16}
{Castellano}, M., {Yue}, B., {Ferrara}, A., {et~al.} 2016, \apjl, 823, L40

\bibitem[{{Chabrier}(2003)}]{chab03}
{Chabrier}, G. 2003, \pasp, 115, 763

\bibitem[{{Charlot} \& {Fall}(2000)}]{charlotnfall00}
{Charlot}, S. \& {Fall}, S.~M. 2000, \apj, 539, 718

\bibitem[{{Chiang} {et~al.}(2013){Chiang}, {Overzier}, \&
  {Gebhardt}}]{coolchiang13}
{Chiang}, Y.-K., {Overzier}, R., \& {Gebhardt}, K. 2013, \apj, 779, 127

\bibitem[{{Chiang} {et~al.}(2014){Chiang}, {Overzier}, \&
  {Gebhardt}}]{coolchiang14}
{Chiang}, Y.-K., {Overzier}, R., \& {Gebhardt}, K. 2014, \apjl, 782, L3

\bibitem[{{Chiang} {et~al.}(2015){Chiang}, {Overzier}, {Gebhardt},
  {Finkelstein}, {Chiang}, {Hill}, {Blanc}, {Drory}, {Chonis}, {Zeimann},
  {Hagen}, {Schneider}, {Jogee}, {Ciardullo}, \& {Gronwall}}]{coolchiang15}
{Chiang}, Y.-K., {Overzier}, R.~A., {Gebhardt}, K., {et~al.} 2015, \apj, 808,
  37

\bibitem[{{Ciardi} {et~al.}(2003){Ciardi}, {Stoehr}, \& {White}}]{ciardi03}
{Ciardi}, B., {Stoehr}, F., \& {White}, S.~D.~M. 2003, \mnras, 343, 1101

\bibitem[{{Civano} {et~al.}(2016){Civano}, {Marchesi}, {Comastri}, {Urry},
  {Elvis}, {Cappelluti}, {Puccetti}, {Brusa}, {Zamorani}, {Hasinger},
  {Aldcroft}, {Alexander}, {Allevato}, {Brunner}, {Capak}, {Finoguenov},
  {Fiore}, {Fruscione}, {Gilli}, {Glotfelty}, {Griffiths}, {Hao}, {Harrison},
  {Jahnke}, {Kartaltepe}, {Karim}, {LaMassa}, {Lanzuisi}, {Miyaji}, {Ranalli},
  {Salvato}, {Sargent}, {Scoville}, {Schawinski}, {Schinnerer}, {Silverman},
  {Smolcic}, {Stern}, {Toft}, {Trakhenbrot}, {Treister}, \&
  {Vignali}}]{civano16}
{Civano}, F., {Marchesi}, S., {Comastri}, A., {et~al.} 2016, \apj, 819, 62

\bibitem[{{Clerc} {et~al.}(2014){Clerc}, {Adami}, {Lieu}, {Maughan}, {Pacaud},
  {Pierre}, {Sadibekova}, {Smith}, {Valageas}, {Altieri}, {Benoist},
  {Maurogordato}, \& {Willis}}]{clerc14}
{Clerc}, N., {Adami}, C., {Lieu}, M., {et~al.} 2014, \mnras, 444, 2723

\bibitem[{{Clowe} {et~al.}(1998){Clowe}, {Luppino}, {Kaiser}, {Henry}, \&
  {Gioia}}]{clowe98}
{Clowe}, D., {Luppino}, G.~A., {Kaiser}, N., {Henry}, J.~P., \& {Gioia}, I.~M.
  1998, \apjl, 497, L61

\bibitem[{{Colless} \& {Dunn}(1996)}]{collessndunn96}
{Colless}, M. \& {Dunn}, A.~M. 1996, \apj, 458, 435

\bibitem[{{Condon}(1992)}]{condon92}
{Condon}, J.~J. 1992, \araa, 30, 575

\bibitem[{{Contini} {et~al.}(2016){Contini}, {De Lucia}, {Hatch}, {Borgani}, \&
  {Kang}}]{contini16}
{Contini}, E., {De Lucia}, G., {Hatch}, N., {Borgani}, S., \& {Kang}, X. 2016,
  \mnras, 456, 1924

\bibitem[{{Cooke} {et~al.}(2016){Cooke}, {Hatch}, {Stern}, {Rettura},
  {Brodwin}, {Galametz}, {Wylezalek}, {Bridge}, {Conselice}, {De Breuck},
  {Gonzalez}, \& {Jarvis}}]{cooke16}
{Cooke}, E.~A., {Hatch}, N.~A., {Stern}, D., {et~al.} 2016, \apj, 816, 83

\bibitem[{{Cooper} {et~al.}(2007){Cooper}, {Newman}, {Coil}, {Croton}, {Gerke},
  {Yan}, {Davis}, {Faber}, {Guhathakurta}, {Koo}, {Weiner}, \&
  {Willmer}}]{mcoopz07}
{Cooper}, M.~C., {Newman}, J.~A., {Coil}, A.~L., {et~al.} 2007, \mnras, 376,
  1445

\bibitem[{{Cucciati} {et~al.}(2014){Cucciati}, {Zamorani}, {Lemaux},
  {Bardelli}, {Cimatti}, {Le F{\`e}vre}, {Cassata}, {Garilli}, {Le Brun},
  {Maccagni}, {Pentericci}, {Tasca}, {Thomas}, {Vanzella}, {Zucca}, {Amorin},
  {Capak}, {Cassar{\`a}}, {Castellano}, {Cuby}, {de la Torre}, {Durkalec},
  {Fontana}, {Giavalisco}, {Grazian}, {Hathi}, {Ilbert}, {Moreau}, {Paltani},
  {Ribeiro}, {Salvato}, {Schaerer}, {Scodeggio}, {Sommariva}, {Talia},
  {Taniguchi}, {Tresse}, {Vergani}, {Wang}, {Charlot}, {Contini}, {Fotopoulou},
  {L{\'o}pez-Sanjuan}, {Mellier}, \& {Scoville}}]{olga14}
{Cucciati}, O., {Zamorani}, G., {Lemaux}, B.~C., {et~al.} 2014, \aap, 570, A16

\bibitem[{{da Cunha} {et~al.}(2008){da Cunha}, {Charlot}, \&
  {Elbaz}}]{dacunha08}
{da Cunha}, E., {Charlot}, S., \& {Elbaz}, D. 2008, \mnras, 388, 1595

\bibitem[{{Daddi} {et~al.}(2009){Daddi}, {Dannerbauer}, {Stern}, {Dickinson},
  {Morrison}, {Elbaz}, {Giavalisco}, {Mancini}, {Pope}, \& {Spinrad}}]{daddi09}
{Daddi}, E., {Dannerbauer}, H., {Stern}, D., {et~al.} 2009, \apj, 694, 1517

\bibitem[{{D'Aloisio} {et~al.}(2015){D'Aloisio}, {McQuinn}, \&
  {Trac}}]{d'aloisio15}
{D'Aloisio}, A., {McQuinn}, M., \& {Trac}, H. 2015, \apjl, 813, L38

\bibitem[{{Dannerbauer} {et~al.}(2017){Dannerbauer}, {Lehnert}, {Emonts},
  {Ziegler}, {Altieri}, {De Breuck}, {Hatch}, {Kodama}, {Koyama}, {Kurk},
  {Matiz}, {Miley}, {Narayanan}, {Norris}, {Overzier}, {Roettgering},
  {Sargent}, {Seymour}, {Tanaka}, {Valtchanov}, \&
  {Wylezalek}}]{helpfulhelmut17}
{Dannerbauer}, H., {Lehnert}, M.~D., {Emonts}, B.~H.~C., {et~al.} 2017, ArXiv
  e-prints [\eprint[arXiv]{1701.05250}]

\bibitem[{{Darvish} {et~al.}(2015){Darvish}, {Mobasher}, {Sobral}, {Scoville},
  \& {Aragon-Calvo}}]{darvish15}
{Darvish}, B., {Mobasher}, B., {Sobral}, D., {Scoville}, N., \& {Aragon-Calvo},
  M. 2015, \apj, 805, 121

\bibitem[{{Davidzon} {et~al.}(2016){Davidzon}, {Cucciati}, {Bolzonella}, {De
  Lucia}, {Zamorani}, {Arnouts}, {Moutard}, {Ilbert}, {Garilli}, {Scodeggio},
  {Guzzo}, {Abbas}, {Adami}, {Bel}, {Bottini}, {Branchini}, {Cappi}, {Coupon},
  {de la Torre}, {Di Porto}, {Fritz}, {Franzetti}, {Fumana}, {Granett},
  {Guennou}, {Iovino}, {Krywult}, {Le Brun}, {Le F{\`e}vre}, {Maccagni},
  {Ma{\l}ek}, {Marulli}, {McCracken}, {Mellier}, {Moscardini}, {Polletta},
  {Pollo}, {Tasca}, {Tojeiro}, {Vergani}, \& {Zanichelli}}]{iary16}
{Davidzon}, I., {Cucciati}, O., {Bolzonella}, M., {et~al.} 2016, \aap, 586, A23

\bibitem[{{Davidzon} {et~al.}(2017){Davidzon}, {Ilbert}, {Laigle}, {Coupon},
  {McCracken}, {Delvecchio}, {Masters}, {Capak}, {Hsieh}, {Tresse}, {Le Fevre},
  {Bethermin}, {Chang}, {Faisst}, {Le Floc'h}, {Steinhardt}, {Toft}, {Aussel},
  {Dubois}, {Hasinger}, {Salvato}, {Sanders}, {Scoville}, \&
  {Silverman}}]{idolatrousiary17}
{Davidzon}, I., {Ilbert}, O., {Laigle}, C., {et~al.} 2017, ArXiv e-prints
  [\eprint[arXiv]{1701.02734}]

\bibitem[{{Davies} \& {Furlanetto}(2016)}]{davies16}
{Davies}, F.~B. \& {Furlanetto}, S.~R. 2016, \mnras, 460, 1328

\bibitem[{{Davis} \& {Wilkinson}(1974)}]{davisnwilkinson74}
{Davis}, M. \& {Wilkinson}, D.~T. 1974, \apj, 192, 251

\bibitem[{{De Propris} {et~al.}(2013){De Propris}, {Phillipps}, \&
  {Bremer}}]{depropris13}
{De Propris}, R., {Phillipps}, S., \& {Bremer}, M.~N. 2013, \mnras, 434, 3469

\bibitem[{{de Ravel} {et~al.}(2009){de Ravel}, {Le F{\`e}vre}, {Tresse},
  {Bottini}, {Garilli}, {Le Brun}, {Maccagni}, {Scaramella}, {Scodeggio},
  {Vettolani}, {Zanichelli}, {Adami}, {Arnouts}, {Bardelli}, {Bolzonella},
  {Cappi}, {Charlot}, {Ciliegi}, {Contini}, {Foucaud}, {Franzetti},
  {Gavignaud}, {Guzzo}, {Ilbert}, {Iovino}, {Lamareille}, {McCracken},
  {Marano}, {Marinoni}, {Mazure}, {Meneux}, {Merighi}, {Paltani}, {Pell{\`o}},
  {Pollo}, {Pozzetti}, {Radovich}, {Vergani}, {Zamorani}, {Zucca}, {Bondi},
  {Bongiorno}, {Brinchmann}, {Cucciati}, {de La Torre}, {Gregorini}, {Memeo},
  {Perez-Montero}, {Mellier}, {Merluzzi}, \& {Temporin}}]{dirtydeR09}
{de Ravel}, L., {Le F{\`e}vre}, O., {Tresse}, L., {et~al.} 2009, \aap, 498, 379

\bibitem[{{Delvecchio} {et~al.}(2014){Delvecchio}, {Gruppioni}, {Pozzi},
  {Berta}, {Zamorani}, {Cimatti}, {Lutz}, {Scott}, {Vignali}, {Cresci},
  {Feltre}, {Cooray}, {Vaccari}, {Fritz}, {Le Floc'h}, {Magnelli}, {Popesso},
  {Oliver}, {Bock}, {Carollo}, {Contini}, {Le F{\'e}vre}, {Lilly}, {Mainieri},
  {Renzini}, \& {Scodeggio}}]{ivan14}
{Delvecchio}, I., {Gruppioni}, C., {Pozzi}, F., {et~al.} 2014, \mnras, 439,
  2736

\bibitem[{{Dey} {et~al.}(2016){Dey}, {Lee}, {Reddy}, {Cooper}, {Inami}, {Hong},
  {Gonzalez}, \& {Jannuzi}}]{dey16}
{Dey}, A., {Lee}, K.-S., {Reddy}, N., {et~al.} 2016, \apj, 823, 11

\bibitem[{{Diener} {et~al.}(2013){Diener}, {Lilly}, {Knobel}, {Zamorani},
  {Lemson}, {Kampczyk}, {Scoville}, {Carollo}, {Contini}, {Kneib}, {Le Fevre},
  {Mainieri}, {Renzini}, {Scodeggio}, {Bardelli}, {Bolzonella}, {Bongiorno},
  {Caputi}, {Cucciati}, {de la Torre}, {de Ravel}, {Franzetti}, {Garilli},
  {Iovino}, {Kova{\v c}}, {Lamareille}, {Le Borgne}, {Le Brun}, {Maier},
  {Mignoli}, {Pello}, {Peng}, {Perez Montero}, {Presotto}, {Silverman},
  {Tanaka}, {Tasca}, {Tresse}, {Vergani}, {Zucca}, {Bordoloi}, {Cappi},
  {Cimatti}, {Coppa}, {Koekemoer}, {L{\'o}pez-Sanjuan}, {McCracken}, {Moresco},
  {Nair}, {Pozzetti}, \& {Welikala}}]{dutifuldiener13}
{Diener}, C., {Lilly}, S.~J., {Knobel}, C., {et~al.} 2013, \apj, 765, 109

\bibitem[{{Diener} {et~al.}(2015){Diener}, {Lilly}, {Ledoux}, {Zamorani},
  {Bolzonella}, {Murphy}, {Capak}, {Ilbert}, \& {McCracken}}]{dutifuldiener15}
{Diener}, C., {Lilly}, S.~J., {Ledoux}, C., {et~al.} 2015, \apj, 802, 31

\bibitem[{{Digby-North} {et~al.}(2010){Digby-North}, {Nandra}, {Laird},
  {Steidel}, {Georgakakis}, {Bogosavljevi{\'c}}, {Erb}, {Shapley}, {Reddy}, \&
  {Aird}}]{dig10}
{Digby-North}, J.~A., {Nandra}, K., {Laird}, E.~S., {et~al.} 2010, \mnras, 407,
  846

\bibitem[{{Donley} {et~al.}(2012){Donley}, {Koekemoer}, {Brusa}, {Capak},
  {Cardamone}, {Civano}, {Ilbert}, {Impey}, {Kartaltepe}, {Miyaji}, {Salvato},
  {Sanders}, {Trump}, \& {Zamorani}}]{donley12}
{Donley}, J.~L., {Koekemoer}, A.~M., {Brusa}, M., {et~al.} 2012, \apj, 748, 142

\bibitem[{{Dressler} {et~al.}(2013){Dressler}, {Oemler}, {Poggianti},
  {Gladders}, {Abramson}, \& {Vulcani}}]{dressler13}
{Dressler}, A., {Oemler}, Jr., A., {Poggianti}, B.~M., {et~al.} 2013, \apj,
  770, 62

\bibitem[{{Dressler} {et~al.}(1999){Dressler}, {Smail}, {Poggianti}, {Butcher},
  {Couch}, {Ellis}, \& {Oemler}}]{dressler99}
{Dressler}, A., {Smail}, I., {Poggianti}, B.~M., {et~al.} 1999, \apjs, 122, 51

\bibitem[{{Durkalec} {et~al.}(2017){Durkalec}, {Le F{\'e}vre}, {Pollo},
  {Zamorani}, {Lemaux}, {Garilli}, {Bardelli}, {Hathi}, {Koekemoer}, {Pforr},
  \& {Zucca}}]{ania17}
{Durkalec}, A., {Le F{\'e}vre}, O., {Pollo}, A., {et~al.} 2017, ArXiv e-prints
  [\eprint[arXiv]{1703.02049}]

\bibitem[{{Elvis} {et~al.}(2009){Elvis}, {Civano}, {Vignali}, {Puccetti},
  {Fiore}, {Cappelluti}, {Aldcroft}, {Fruscione}, {Zamorani}, {Comastri},
  {Brusa}, {Gilli}, {Miyaji}, {Damiani}, {Koekemoer}, {Finoguenov}, {Brunner},
  {Urry}, {Silverman}, {Mainieri}, {Hasinger}, {Griffiths}, {Carollo}, {Hao},
  {Guzzo}, {Blain}, {Calzetti}, {Carilli}, {Capak}, {Ettori}, {Fabbiano},
  {Impey}, {Lilly}, {Mobasher}, {Rich}, {Salvato}, {Sanders}, {Schinnerer},
  {Scoville}, {Shopbell}, {Taylor}, {Taniguchi}, \& {Volonteri}}]{elvis09}
{Elvis}, M., {Civano}, F., {Vignali}, C., {et~al.} 2009, \apjs, 184, 158

\bibitem[{{Erb}(2015)}]{erb15}
{Erb}, D.~K. 2015, \nat, 523, 169

\bibitem[{{Faber} {et~al.}(2003){Faber}, {Phillips}, {Kibrick}, {Alcott},
  {Allen}, {Burrous}, {Cantrall}, {Clarke}, {Coil}, {Cowley}, {Davis}, {Deich},
  {Dietsch}, {Gilmore}, {Harper}, {Hilyard}, {Lewis}, {McVeigh}, {Newman},
  {Osborne}, {Schiavon}, {Stover}, {Tucker}, {Wallace}, {Wei}, {Wirth}, \&
  {Wright}}]{fab03}
{Faber}, S.~M., {Phillips}, A.~C., {Kibrick}, R.~I., {et~al.} 2003, in Society
  of Photo-Optical Instrumentation Engineers (SPIE) Conference Series, Vol.
  4841, Instrument Design and Performance for Optical/Infrared Ground-based
  Telescopes, ed. M.~{Iye} \& A.~F.~M. {Moorwood}, 1657--1669

\bibitem[{{Fakhouri} {et~al.}(2010){Fakhouri}, {Ma}, \&
  {Boylan-Kolchin}}]{FakhourinMa10}
{Fakhouri}, O., {Ma}, C.-P., \& {Boylan-Kolchin}, M. 2010, \mnras, 406, 2267

\bibitem[{{Fazio} {et~al.}(2004){Fazio}, {Hora}, {Allen}, {Ashby}, {Barmby},
  {Deutsch}, {Huang}, {Kleiner}, {Marengo}, {Megeath}, {Melnick}, {Pahre},
  {Patten}, {Polizotti}, {Smith}, {Taylor}, {Wang}, {Willner}, {Hoffmann},
  {Pipher}, {Forrest}, {McMurty}, {McCreight}, {McKelvey}, {McMurray}, {Koch},
  {Moseley}, {Arendt}, {Mentzell}, {Marx}, {Losch}, {Mayman}, {Eichhorn},
  {Krebs}, {Jhabvala}, {Gezari}, {Fixsen}, {Flores}, {Shakoorzadeh}, {Jungo},
  {Hakun}, {Workman}, {Karpati}, {Kichak}, {Whitley}, {Mann}, {Tollestrup},
  {Eisenhardt}, {Stern}, {Gorjian}, {Bhattacharya}, {Carey}, {Nelson},
  {Glaccum}, {Lacy}, {Lowrance}, {Laine}, {Reach}, {Stauffer}, {Surace},
  {Wilson}, {Wright}, {Hoffman}, {Domingo}, \& {Cohen}}]{fazio04}
{Fazio}, G.~G., {Hora}, J.~L., {Allen}, L.~E., {et~al.} 2004, \apjs, 154, 10

\bibitem[{{Feltre} {et~al.}(2012){Feltre}, {Hatziminaoglou}, {Fritz}, \&
  {Franceschini}}]{feltre12}
{Feltre}, A., {Hatziminaoglou}, E., {Fritz}, J., \& {Franceschini}, A. 2012,
  \mnras, 426, 120

\bibitem[{{Finkelstein} {et~al.}(2012){Finkelstein}, {Papovich}, {Salmon},
  {Finlator}, {Dickinson}, {Ferguson}, {Giavalisco}, {Koekemoer}, {Reddy},
  {Bassett}, {Conselice}, {Dunlop}, {Faber}, {Grogin}, {Hathi}, {Kocevski},
  {Lai}, {Lee}, {McLure}, {Mobasher}, \& {Newman}}]{finkel12}
{Finkelstein}, S.~L., {Papovich}, C., {Salmon}, B., {et~al.} 2012, \apj, 756,
  164

\bibitem[{{Finkelstein} {et~al.}(2008){Finkelstein}, {Rhoads}, {Malhotra},
  {Grogin}, \& {Wang}}]{finkel08}
{Finkelstein}, S.~L., {Rhoads}, J.~E., {Malhotra}, S., {Grogin}, N., \& {Wang},
  J. 2008, \apj, 678, 655

\bibitem[{{Ford} {et~al.}(1998){Ford}, {Bartko}, {Bely}, {Broadhurst},
  {Burrows}, {Cheng}, {Clampin}, {Crocker}, {Feldman}, {Golimowski}, {Hartig},
  {Illingworth}, {Kimble}, {Lesser}, {Miley}, {Neff}, {Postman}, {Sparks},
  {Tsvetanov}, {White}, {Sullivan}, {Krebs}, {Leviton}, {La Jeunesse},
  {Burmester}, {Fike}, {Johnson}, {Slusher}, {Volmer}, \& {Woodruff}}]{ford98}
{Ford}, H.~C., {Bartko}, F., {Bely}, P.~Y., {et~al.} 1998, in \procspie, Vol.
  3356, Space Telescopes and Instruments V, ed. P.~Y. {Bely} \& J.~B.
  {Breckinridge}, 234--248

\bibitem[{{Fort} \& {Mellier}(1994)}]{fort94}
{Fort}, B. \& {Mellier}, Y. 1994, \aapr, 5, 239

\bibitem[{{Franck} \& {McGaugh}(2016{\natexlab{a}})}]{francknmcgaugh16b}
{Franck}, J.~R. \& {McGaugh}, S.~S. 2016{\natexlab{a}}, \apj, 833, 15

\bibitem[{{Franck} \& {McGaugh}(2016{\natexlab{b}})}]{faithlessFranck16b}
{Franck}, J.~R. \& {McGaugh}, S.~S. 2016{\natexlab{b}}, \apj, 833, 15

\bibitem[{{Franck} \& {McGaugh}(2016{\natexlab{c}})}]{francknmcgaugh16a}
{Franck}, J.~R. \& {McGaugh}, S.~S. 2016{\natexlab{c}}, \apj, 817, 158

\bibitem[{{Franzetti} {et~al.}(2008){Franzetti}, {Scodeggio}, {Garilli},
  {Fumana}, \& {Paioro}}]{franzetti08}
{Franzetti}, P., {Scodeggio}, M., {Garilli}, B., {Fumana}, M., \& {Paioro}, L.
  2008, in Astronomical Society of the Pacific Conference Series, Vol. 394,
  Astronomical Data Analysis Software and Systems XVII, ed. R.~W. {Argyle},
  P.~S. {Bunclark}, \& J.~R. {Lewis}, 642

\bibitem[{{Fritz} {et~al.}(2006){Fritz}, {Franceschini}, \&
  {Hatziminaoglou}}]{fritz06}
{Fritz}, J., {Franceschini}, A., \& {Hatziminaoglou}, E. 2006, \mnras, 366, 767

\bibitem[{{Fukugita} {et~al.}(1996){Fukugita}, {Ichikawa}, {Gunn}, {Doi},
  {Shimasaku}, \& {Schneider}}]{fukugita96}
{Fukugita}, M., {Ichikawa}, T., {Gunn}, J.~E., {et~al.} 1996, \aj, 111, 1748

\bibitem[{{Gal} {et~al.}(2008){Gal}, {Lemaux}, {Lubin}, {Kocevski}, \&
  {Squires}}]{gal08}
{Gal}, R.~R., {Lemaux}, B.~C., {Lubin}, L.~M., {Kocevski}, D., \& {Squires},
  G.~K. 2008, \apj, 684, 933

\bibitem[{{Garmire} {et~al.}(2003){Garmire}, {Bautz}, {Ford}, {Nousek}, \&
  {Ricker}}]{garmire03}
{Garmire}, G.~P., {Bautz}, M.~W., {Ford}, P.~G., {Nousek}, J.~A., \& {Ricker},
  Jr., G.~R. 2003, in \procspie, Vol. 4851, X-Ray and Gamma-Ray Telescopes and
  Instruments for Astronomy., ed. J.~E. {Truemper} \& H.~D. {Tananbaum}, 28--44

\bibitem[{{Gawiser} {et~al.}(2006){Gawiser}, {van Dokkum}, {Gronwall},
  {Ciardullo}, {Blanc}, {Castander}, {Feldmeier}, {Francke}, {Franx},
  {Haberzettl}, {Herrera}, {Hickey}, {Infante}, {Lira}, {Maza}, {Quadri},
  {Richardson}, {Schawinski}, {Schirmer}, {Taylor}, {Treister}, {Urry}, \&
  {Virani}}]{imfukingericgawiser06}
{Gawiser}, E., {van Dokkum}, P.~G., {Gronwall}, C., {et~al.} 2006, \apjl, 642,
  L13

\bibitem[{{Geach} {et~al.}(2017){Geach}, {Dunlop}, {Halpern}, {Smail}, {van der
  Werf}, {Alexander}, {Almaini}, {Aretxaga}, {Arumugam}, {Asboth}, {Banerji},
  {Beanlands}, {Best}, {Blain}, {Birkinshaw}, {Chapin}, {Chapman}, {Chen},
  {Chrysostomou}, {Clarke}, {Clements}, {Conselice}, {Coppin}, {Cowley},
  {Danielson}, {Eales}, {Edge}, {Farrah}, {Gibb}, {Harrison}, {Hine}, {Hughes},
  {Ivison}, {Jarvis}, {Jenness}, {Jones}, {Karim}, {Koprowski}, {Knudsen},
  {Lacey}, {Mackenzie}, {Marsden}, {McAlpine}, {McMahon}, {Meijerink},
  {Micha{\l}owski}, {Oliver}, {Page}, {Peacock}, {Rigopoulou}, {Robson},
  {Roseboom}, {Rotermund}, {Scott}, {Serjeant}, {Simpson}, {Simpson}, {Smith},
  {Spaans}, {Stanley}, {Stevens}, {Swinbank}, {Targett}, {Thomson}, {Valiante},
  {Wake}, {Webb}, {Willott}, {Zavala}, \& {Zemcov}}]{geach17}
{Geach}, J.~E., {Dunlop}, J.~S., {Halpern}, M., {et~al.} 2017, \mnras, 465,
  1789

\bibitem[{{Giavalisco} {et~al.}(2004){Giavalisco}, {Dickinson}, {Ferguson},
  {Ravindranath}, {Kretchmer}, {Moustakas}, {Madau}, {Fall}, {Gardner},
  {Livio}, {Papovich}, {Renzini}, {Spinrad}, {Stern}, \&
  {Riess}}]{giavalisco04}
{Giavalisco}, M., {Dickinson}, M., {Ferguson}, H.~C., {et~al.} 2004, \apjl,
  600, L103

\bibitem[{{Ginolfi} {et~al.}(2016){Ginolfi}, {Maiolino}, {Nagao}, {Carniani},
  {Belfiore}, {Cresci}, {Hatsukade}, {Mannucci}, {Marconi}, {Pallottini},
  {Schneider}, \& {Santini}}]{ginolfi16}
{Ginolfi}, M., {Maiolino}, R., {Nagao}, T., {et~al.} 2016, ArXiv e-prints
  [\eprint[arXiv]{1611.07026}]

\bibitem[{{Gladders} {et~al.}(2003){Gladders}, {Hoekstra}, {Yee}, {Hall}, \&
  {Barrientos}}]{gladders03}
{Gladders}, M.~D., {Hoekstra}, H., {Yee}, H.~K.~C., {Hall}, P.~B., \&
  {Barrientos}, L.~F. 2003, \apj, 593, 48

\bibitem[{{Gladders} \& {Yee}(2005)}]{gladdersnyee05}
{Gladders}, M.~D. \& {Yee}, H.~K.~C. 2005, \apjs, 157, 1

\bibitem[{{Gobat} {et~al.}(2011){Gobat}, {Daddi}, {Onodera}, {Finoguenov},
  {Renzini}, {Arimoto}, {Bouwens}, {Brusa}, {Chary}, {Cimatti}, {Dickinson},
  {Kong}, \& {Mignoli}}]{gobat11}
{Gobat}, R., {Daddi}, E., {Onodera}, M., {et~al.} 2011, \aap, 526, A133

\bibitem[{{Goto} {et~al.}(2003){Goto}, {Yamauchi}, {Fujita}, {Okamura},
  {Sekiguchi}, {Smail}, {Bernardi}, \& {Gomez}}]{goto03a}
{Goto}, T., {Yamauchi}, C., {Fujita}, Y., {et~al.} 2003, \mnras, 346, 601

\bibitem[{{Griffin} {et~al.}(2010){Griffin}, {Abergel}, {Abreu}, {Ade},
  {Andr{\'e}}, {Augueres}, {Babbedge}, {Bae}, {Baillie}, {Baluteau}, {Barlow},
  {Bendo}, {Benielli}, {Bock}, {Bonhomme}, {Brisbin}, {Brockley-Blatt},
  {Caldwell}, {Cara}, {Castro-Rodriguez}, {Cerulli}, {Chanial}, {Chen},
  {Clark}, {Clements}, {Clerc}, {Coker}, {Communal}, {Conversi}, {Cox},
  {Crumb}, {Cunningham}, {Daly}, {Davis}, {de Antoni}, {Delderfield}, {Devin},
  {di Giorgio}, {Didschuns}, {Dohlen}, {Donati}, {Dowell}, {Dowell}, {Duband},
  {Dumaye}, {Emery}, {Ferlet}, {Ferrand}, {Fontignie}, {Fox}, {Franceschini},
  {Frerking}, {Fulton}, {Garcia}, {Gastaud}, {Gear}, {Glenn}, {Goizel},
  {Griffin}, {Grundy}, {Guest}, {Guillemet}, {Hargrave}, {Harwit}, {Hastings},
  {Hatziminaoglou}, {Herman}, {Hinde}, {Hristov}, {Huang}, {Imhof}, {Isaak},
  {Israelsson}, {Ivison}, {Jennings}, {Kiernan}, {King}, {Lange}, {Latter},
  {Laurent}, {Laurent}, {Leeks}, {Lellouch}, {Levenson}, {Li}, {Li},
  {Lilienthal}, {Lim}, {Liu}, {Lu}, {Madden}, {Mainetti}, {Marliani}, {McKay},
  {Mercier}, {Molinari}, {Morris}, {Moseley}, {Mulder}, {Mur}, {Naylor},
  {Nguyen}, {O'Halloran}, {Oliver}, {Olofsson}, {Olofsson}, {Orfei}, {Page},
  {Pain}, {Panuzzo}, {Papageorgiou}, {Parks}, {Parr-Burman}, {Pearce},
  {Pearson}, {P{\'e}rez-Fournon}, {Pinsard}, {Pisano}, {Podosek}, {Pohlen},
  {Polehampton}, {Pouliquen}, {Rigopoulou}, {Rizzo}, {Roseboom}, {Roussel},
  {Rowan-Robinson}, {Rownd}, {Saraceno}, {Sauvage}, {Savage}, {Savini},
  {Sawyer}, {Scharmberg}, {Schmitt}, {Schneider}, {Schulz}, {Schwartz},
  {Shafer}, {Shupe}, {Sibthorpe}, {Sidher}, {Smith}, {Smith}, {Smith},
  {Spencer}, {Stobie}, {Sudiwala}, {Sukhatme}, {Surace}, {Stevens}, {Swinyard},
  {Trichas}, {Tourette}, {Triou}, {Tseng}, {Tucker}, {Turner}, {Vaccari},
  {Valtchanov}, {Vigroux}, {Virique}, {Voellmer}, {Walker}, {Ward}, {Waskett},
  {Weilert}, {Wesson}, {White}, {Whitehouse}, {Wilson}, {Winter}, {Woodcraft},
  {Wright}, {Xu}, {Zavagno}, {Zemcov}, {Zhang}, \& {Zonca}}]{griffin10}
{Griffin}, M.~J., {Abergel}, A., {Abreu}, A., {et~al.} 2010, \aap, 518, L3

\bibitem[{{Grogin} {et~al.}(2011){Grogin}, {Kocevski}, {Faber}, {Ferguson},
  {Koekemoer}, {Riess}, {Acquaviva}, {Alexander}, {Almaini}, {Ashby}, {Barden},
  {Bell}, {Bournaud}, {Brown}, {Caputi}, {Casertano}, {Cassata}, {Castellano},
  {Challis}, {Chary}, {Cheung}, {Cirasuolo}, {Conselice}, {Roshan Cooray},
  {Croton}, {Daddi}, {Dahlen}, {Dav{\'e}}, {de Mello}, {Dekel}, {Dickinson},
  {Dolch}, {Donley}, {Dunlop}, {Dutton}, {Elbaz}, {Fazio}, {Filippenko},
  {Finkelstein}, {Fontana}, {Gardner}, {Garnavich}, {Gawiser}, {Giavalisco},
  {Grazian}, {Guo}, {Hathi}, {H{\"a}ussler}, {Hopkins}, {Huang}, {Huang},
  {Jha}, {Kartaltepe}, {Kirshner}, {Koo}, {Lai}, {Lee}, {Li}, {Lotz}, {Lucas},
  {Madau}, {McCarthy}, {McGrath}, {McIntosh}, {McLure}, {Mobasher},
  {Moustakas}, {Mozena}, {Nandra}, {Newman}, {Niemi}, {Noeske}, {Papovich},
  {Pentericci}, {Pope}, {Primack}, {Rajan}, {Ravindranath}, {Reddy}, {Renzini},
  {Rix}, {Robaina}, {Rodney}, {Rosario}, {Rosati}, {Salimbeni}, {Scarlata},
  {Siana}, {Simard}, {Smidt}, {Somerville}, {Spinrad}, {Straughn}, {Strolger},
  {Telford}, {Teplitz}, {Trump}, {van der Wel}, {Villforth}, {Wechsler},
  {Weiner}, {Wiklind}, {Wild}, {Wilson}, {Wuyts}, {Yan}, \& {Yun}}]{grogin11}
{Grogin}, N.~A., {Kocevski}, D.~D., {Faber}, S.~M., {et~al.} 2011, \apjs, 197,
  35

\bibitem[{{Guaita} {et~al.}(2017){Guaita}, {Talia}, {Pentericci}, {Verhamme},
  {Cassata}, {Lemaux}, {Orlitova}, {Ribeiro}, {Schaerer}, {Zamorani},
  {Garilli}, {Le Brun}, {Le F{\`e}vre}, {Maccagni}, {Tasca}, {Thomas},
  {Vanzella}, {Zucca}, {Amorin}, {Bardelli}, {Castellano}, {Grazian}, {Hathi},
  {Koekemoer}, \& {Marchi}}]{loopylucia17}
{Guaita}, L., {Talia}, M., {Pentericci}, L., {et~al.} 2017, \aap, 606, A19

\bibitem[{{Guo} {et~al.}(2015){Guo}, {Ferguson}, {Bell}, {Koo}, {Conselice},
  {Giavalisco}, {Kassin}, {Lu}, {Lucas}, {Mandelker}, {McIntosh}, {Primack},
  {Ravindranath}, {Barro}, {Ceverino}, {Dekel}, {Faber}, {Fang}, {Koekemoer},
  {Noeske}, {Rafelski}, \& {Straughn}}]{guo15}
{Guo}, Y., {Ferguson}, H.~C., {Bell}, E.~F., {et~al.} 2015, \apj, 800, 39

\bibitem[{{Gutermuth} {et~al.}(2005){Gutermuth}, {Megeath}, {Pipher},
  {Williams}, {Allen}, {Myers}, \& {Raines}}]{guter05}
{Gutermuth}, R.~A., {Megeath}, S.~T., {Pipher}, J.~L., {et~al.} 2005, \apj,
  632, 397

\bibitem[{{Halliday} {et~al.}(2004){Halliday}, {Milvang-Jensen}, {Poirier},
  {Poggianti}, {Jablonka}, {Arag{\'o}n-Salamanca}, {Saglia}, {De Lucia},
  {Pell{\'o}}, {Simard}, {Clowe}, {Rudnick}, {Dalcanton}, {White}, \&
  {Zaritsky}}]{halliday04}
{Halliday}, C., {Milvang-Jensen}, B., {Poirier}, S., {et~al.} 2004, \aap, 427,
  397

\bibitem[{{Hansen} {et~al.}(2009){Hansen}, {Sheldon}, {Wechsler}, \&
  {Koester}}]{hansen09}
{Hansen}, S.~M., {Sheldon}, E.~S., {Wechsler}, R.~H., \& {Koester}, B.~P. 2009,
  \apj, 699, 1333

\bibitem[{{Hasselfield} {et~al.}(2013){Hasselfield}, {Hilton}, {Marriage},
  {Addison}, {Barrientos}, {Battaglia}, {Battistelli}, {Bond}, {Crichton},
  {Das}, {Devlin}, {Dicker}, {Dunkley}, {D{\"u}nner}, {Fowler}, {Gralla},
  {Hajian}, {Halpern}, {Hincks}, {Hlozek}, {Hughes}, {Infante}, {Irwin},
  {Kosowsky}, {Marsden}, {Menanteau}, {Moodley}, {Niemack}, {Nolta}, {Page},
  {Partridge}, {Reese}, {Schmitt}, {Sehgal}, {Sherwin}, {Sievers}, {Sif{\'o}n},
  {Spergel}, {Staggs}, {Swetz}, {Switzer}, {Thornton}, {Trac}, \&
  {Wollack}}]{hasselfield13}
{Hasselfield}, M., {Hilton}, M., {Marriage}, T.~A., {et~al.} 2013, \jcap, 7,
  008

\bibitem[{{Hatch} {et~al.}(2017){Hatch}, {Cooke}, {Muldrew}, {Hartley},
  {Almaini}, {Conselice}, \& {Simpson}}]{hatch17}
{Hatch}, N.~A., {Cooke}, E.~A., {Muldrew}, S.~I., {et~al.} 2017, \mnras, 464,
  876

\bibitem[{{Hatch} {et~al.}(2011){Hatch}, {Kurk}, {Pentericci}, {Venemans},
  {Kuiper}, {Miley}, \& {R{\"o}ttgering}}]{hatch11}
{Hatch}, N.~A., {Kurk}, J.~D., {Pentericci}, L., {et~al.} 2011, \mnras, 415,
  2993

\bibitem[{{Hatch} {et~al.}(2014){Hatch}, {Wylezalek}, {Kurk}, {Stern}, {De
  Breuck}, {Jarvis}, {Galametz}, {Gonzalez}, {Hartley}, {Mortlock}, {Seymour},
  \& {Stevens}}]{hatch14}
{Hatch}, N.~A., {Wylezalek}, D., {Kurk}, J.~D., {et~al.} 2014, \mnras, 445, 280

\bibitem[{{Hathi} {et~al.}(2016){Hathi}, {Le F{\`e}vre}, {Ilbert}, {Cassata},
  {Tasca}, {Lemaux}, {Garilli}, {Le Brun}, {Maccagni}, {Pentericci}, {Thomas},
  {Vanzella}, {Zamorani}, {Zucca}, {Amor{\'{\i}}n}, {Bardelli}, {Cassar{\`a}},
  {Castellano}, {Cimatti}, {Cucciati}, {Durkalec}, {Fontana}, {Giavalisco},
  {Grazian}, {Guaita}, {Koekemoer}, {Paltani}, {Pforr}, {Ribeiro}, {Schaerer},
  {Scodeggio}, {Sommariva}, {Talia}, {Tresse}, {Vergani}, {Capak}, {Charlot},
  {Contini}, {Cuby}, {de la Torre}, {Dunlop}, {Fotopoulou},
  {L{\'o}pez-Sanjuan}, {Mellier}, {Salvato}, {Scoville}, {Taniguchi}, \&
  {Wang}}]{nimish16}
{Hathi}, N.~P., {Le F{\`e}vre}, O., {Ilbert}, O., {et~al.} 2016, \aap, 588, A26

\bibitem[{{Hayes} {et~al.}(2011){Hayes}, {Schaerer}, {{\"O}stlin}, {Mas-Hesse},
  {Atek}, \& {Kunth}}]{hayes11}
{Hayes}, M., {Schaerer}, D., {{\"O}stlin}, G., {et~al.} 2011, \apj, 730, 8

\bibitem[{{Heckman}(2001)}]{heckman01}
{Heckman}, T.~M. 2001, in Astronomical Society of the Pacific Conference
  Series, Vol. 240, Gas and Galaxy Evolution, ed. J.~E. {Hibbard}, M.~{Rupen},
  \& J.~H. {van Gorkom}, 345

\bibitem[{{Hilton} {et~al.}(2012){Hilton}, {Romer}, {Kay}, {Mehrtens},
  {Lloyd-Davies}, {Thomas}, {Short}, {Mayers}, {Rooney}, {Stott}, {Collins},
  {Harrison}, {Hoyle}, {Liddle}, {Mann}, {Miller}, {Sahl{\'e}n}, {Viana},
  {Davidson}, {Hosmer}, {Nichol}, {Sabirli}, {Stanford}, \& {West}}]{hilton12}
{Hilton}, M., {Romer}, A.~K., {Kay}, S.~T., {et~al.} 2012, \mnras, 424, 2086

\bibitem[{{Hoaglin} {et~al.}(1983){Hoaglin}, {Mosteller}, \&
  {Tukey}}]{hoaglin83}
{Hoaglin}, D.~C., {Mosteller}, F., \& {Tukey}, J.~W. 1983, {Understanding
  robust and exploratory data anlysis}

\bibitem[{{Hoekstra} {et~al.}(2013){Hoekstra}, {Bartelmann}, {Dahle}, {Israel},
  {Limousin}, \& {Meneghetti}}]{hoekstra13}
{Hoekstra}, H., {Bartelmann}, M., {Dahle}, H., {et~al.} 2013, \ssr, 177, 75

\bibitem[{{Ilbert} {et~al.}(2006){Ilbert}, {Arnouts}, {McCracken},
  {Bolzonella}, {Bertin}, {Le F{\`e}vre}, {Mellier}, {Zamorani}, {Pell{\`o}},
  {Iovino}, {Tresse}, {Le Brun}, {Bottini}, {Garilli}, {Maccagni}, {Picat},
  {Scaramella}, {Scodeggio}, {Vettolani}, {Zanichelli}, {Adami}, {Bardelli},
  {Cappi}, {Charlot}, {Ciliegi}, {Contini}, {Cucciati}, {Foucaud}, {Franzetti},
  {Gavignaud}, {Guzzo}, {Marano}, {Marinoni}, {Mazure}, {Meneux}, {Merighi},
  {Paltani}, {Pollo}, {Pozzetti}, {Radovich}, {Zucca}, {Bondi}, {Bongiorno},
  {Busarello}, {de La Torre}, {Gregorini}, {Lamareille}, {Mathez}, {Merluzzi},
  {Ripepi}, {Rizzo}, \& {Vergani}}]{dreadolivier06}
{Ilbert}, O., {Arnouts}, S., {McCracken}, H.~J., {et~al.} 2006, \aap, 457, 841

\bibitem[{{Ilbert} {et~al.}(2009){Ilbert}, {Capak}, {Salvato}, {Aussel},
  {McCracken}, {Sanders}, {Scoville}, {Kartaltepe}, {Arnouts}, {Le Floc'h},
  {Mobasher}, {Taniguchi}, {Lamareille}, {Leauthaud}, {Sasaki}, {Thompson},
  {Zamojski}, {Zamorani}, {Bardelli}, {Bolzonella}, {Bongiorno}, {Brusa},
  {Caputi}, {Carollo}, {Contini}, {Cook}, {Coppa}, {Cucciati}, {de la Torre},
  {de Ravel}, {Franzetti}, {Garilli}, {Hasinger}, {Iovino}, {Kampczyk},
  {Kneib}, {Knobel}, {Kovac}, {Le Borgne}, {Le Brun}, {F{\`e}vre}, {Lilly},
  {Looper}, {Maier}, {Mainieri}, {Mellier}, {Mignoli}, {Murayama}, {Pell{\`o}},
  {Peng}, {P{\'e}rez-Montero}, {Renzini}, {Ricciardelli}, {Schiminovich},
  {Scodeggio}, {Shioya}, {Silverman}, {Surace}, {Tanaka}, {Tasca}, {Tresse},
  {Vergani}, \& {Zucca}}]{dreadolivier09}
{Ilbert}, O., {Capak}, P., {Salvato}, M., {et~al.} 2009, \apj, 690, 1236

\bibitem[{{Ilbert} {et~al.}(2013){Ilbert}, {McCracken}, {Le F{\`e}vre},
  {Capak}, {Dunlop}, {Karim}, {Renzini}, {Caputi}, {Boissier}, {Arnouts},
  {Aussel}, {Comparat}, {Guo}, {Hudelot}, {Kartaltepe}, {Kneib}, {Krogager},
  {Le Floc'h}, {Lilly}, {Mellier}, {Milvang-Jensen}, {Moutard}, {Onodera},
  {Richard}, {Salvato}, {Sanders}, {Scoville}, {Silverman}, {Taniguchi},
  {Tasca}, {Thomas}, {Toft}, {Tresse}, {Vergani}, {Wolk}, \& {Zirm}}]{ilbert13}
{Ilbert}, O., {McCracken}, H.~J., {Le F{\`e}vre}, O., {et~al.} 2013, \aap, 556,
  A55

\bibitem[{{Ivison} {et~al.}(2000){Ivison}, {Dunlop}, {Smail}, {Dey}, {Liu}, \&
  {Graham}}]{ivison00}
{Ivison}, R.~J., {Dunlop}, J.~S., {Smail}, I., {et~al.} 2000, \apj, 542, 27

\bibitem[{{Jansen} {et~al.}(2001){Jansen}, {Lumb}, {Altieri}, {Clavel}, {Ehle},
  {Erd}, {Gabriel}, {Guainazzi}, {Gondoin}, {Much}, {Munoz}, {Santos},
  {Schartel}, {Texier}, \& {Vacanti}}]{jansen01}
{Jansen}, F., {Lumb}, D., {Altieri}, B., {et~al.} 2001, \aap, 365, L1

\bibitem[{{Jee} {et~al.}(2006){Jee}, {White}, {Ford}, {Illingworth},
  {Blakeslee}, {Holden}, \& {Mei}}]{jee06}
{Jee}, M.~J., {White}, R.~L., {Ford}, H.~C., {et~al.} 2006, \apj, 642, 720

\bibitem[{{Kent} \& {Gunn}(1982)}]{kentngunn82}
{Kent}, S.~M. \& {Gunn}, J.~E. 1982, \aj, 87, 945

\bibitem[{{Kitzbichler} \& {White}(2008)}]{kdubz08}
{Kitzbichler}, M.~G. \& {White}, S.~D.~M. 2008, \mnras, 391, 1489

\bibitem[{{Koekemoer} {et~al.}(2007){Koekemoer}, {Aussel}, {Calzetti}, {Capak},
  {Giavalisco}, {Kneib}, {Leauthaud}, {Le F{\`e}vre}, {McCracken}, {Massey},
  {Mobasher}, {Rhodes}, {Scoville}, \& {Shopbell}}]{anton07}
{Koekemoer}, A.~M., {Aussel}, H., {Calzetti}, D., {et~al.} 2007, \apjs, 172,
  196

\bibitem[{{Koekemoer} {et~al.}(2011){Koekemoer}, {Faber}, {Ferguson}, {Grogin},
  {Kocevski}, {Koo}, {Lai}, {Lotz}, {Lucas}, {McGrath}, {Ogaz}, {Rajan},
  {Riess}, {Rodney}, {Strolger}, {Casertano}, {Castellano}, {Dahlen},
  {Dickinson}, {Dolch}, {Fontana}, {Giavalisco}, {Grazian}, {Guo}, {Hathi},
  {Huang}, {van der Wel}, {Yan}, {Acquaviva}, {Alexander}, {Almaini}, {Ashby},
  {Barden}, {Bell}, {Bournaud}, {Brown}, {Caputi}, {Cassata}, {Challis},
  {Chary}, {Cheung}, {Cirasuolo}, {Conselice}, {Roshan Cooray}, {Croton},
  {Daddi}, {Dav{\'e}}, {de Mello}, {de Ravel}, {Dekel}, {Donley}, {Dunlop},
  {Dutton}, {Elbaz}, {Fazio}, {Filippenko}, {Finkelstein}, {Frazer}, {Gardner},
  {Garnavich}, {Gawiser}, {Gruetzbauch}, {Hartley}, {H{\"a}ussler},
  {Herrington}, {Hopkins}, {Huang}, {Jha}, {Johnson}, {Kartaltepe},
  {Khostovan}, {Kirshner}, {Lani}, {Lee}, {Li}, {Madau}, {McCarthy},
  {McIntosh}, {McLure}, {McPartland}, {Mobasher}, {Moreira}, {Mortlock},
  {Moustakas}, {Mozena}, {Nandra}, {Newman}, {Nielsen}, {Niemi}, {Noeske},
  {Papovich}, {Pentericci}, {Pope}, {Primack}, {Ravindranath}, {Reddy},
  {Renzini}, {Rix}, {Robaina}, {Rosario}, {Rosati}, {Salimbeni}, {Scarlata},
  {Siana}, {Simard}, {Smidt}, {Snyder}, {Somerville}, {Spinrad}, {Straughn},
  {Telford}, {Teplitz}, {Trump}, {Vargas}, {Villforth}, {Wagner}, {Wandro},
  {Wechsler}, {Weiner}, {Wiklind}, {Wild}, {Wilson}, {Wuyts}, \&
  {Yun}}]{anton11}
{Koekemoer}, A.~M., {Faber}, S.~M., {Ferguson}, H.~C., {et~al.} 2011, \apjs,
  197, 36

\bibitem[{{Kravtsov} \& {Borgani}(2012)}]{kravtsovnborgani12}
{Kravtsov}, A.~V. \& {Borgani}, S. 2012, \araa, 50, 353

\bibitem[{{Kuiper} {et~al.}(2011){Kuiper}, {Hatch}, {Venemans}, {Miley},
  {R{\"o}ttgering}, {Kurk}, {Overzier}, {Pentericci}, {Bland-Hawthorn}, \&
  {Cepa}}]{kuiper11}
{Kuiper}, E., {Hatch}, N.~A., {Venemans}, B.~P., {et~al.} 2011, \mnras, 417,
  1088

\bibitem[{{Laigle} {et~al.}(2016){Laigle}, {McCracken}, {Ilbert}, {Hsieh},
  {Davidzon}, {Capak}, {Hasinger}, {Silverman}, {Pichon}, {Coupon}, {Aussel},
  {Le Borgne}, {Caputi}, {Cassata}, {Chang}, {Civano}, {Dunlop}, {Fynbo},
  {Kartaltepe}, {Koekemoer}, {Le F{\`e}vre}, {Le Floc'h}, {Leauthaud}, {Lilly},
  {Lin}, {Marchesi}, {Milvang-Jensen}, {Salvato}, {Sanders}, {Scoville},
  {Smolcic}, {Stockmann}, {Taniguchi}, {Tasca}, {Toft}, {Vaccari}, \&
  {Zabl}}]{laigle16}
{Laigle}, C., {McCracken}, H.~J., {Ilbert}, O., {et~al.} 2016, \apjs, 224, 24

\bibitem[{{Le F{\`e}vre} {et~al.}(2000){Le F{\`e}vre}, {Abraham}, {Lilly},
  {Ellis}, {Brinchmann}, {Schade}, {Tresse}, {Colless}, {Crampton},
  {Glazebrook}, {Hammer}, \& {Broadhurst}}]{dong2000}
{Le F{\`e}vre}, O., {Abraham}, R., {Lilly}, S.~J., {et~al.} 2000, \mnras, 311,
  565

\bibitem[{{Le F{\`e}vre} {et~al.}(2003){Le F{\`e}vre}, {Saisse}, {Mancini},
  {Brau-Nogue}, {Caputi}, {Castinel}, {D'Odorico}, {Garilli}, {Kissler-Patig},
  {Lucuix}, {Mancini}, {Pauget}, {Sciarretta}, {Scodeggio}, {Tresse}, \&
  {Vettolani}}]{dong03}
{Le F{\`e}vre}, O., {Saisse}, M., {Mancini}, D., {et~al.} 2003, in \procspie,
  Vol. 4841, Instrument Design and Performance for Optical/Infrared
  Ground-based Telescopes, ed. M.~{Iye} \& A.~F.~M. {Moorwood}, 1670--1681

\bibitem[{{Le F{\`e}vre} {et~al.}(2015){Le F{\`e}vre}, {Tasca}, {Cassata},
  {Garilli}, {Le Brun}, {Maccagni}, {Pentericci}, {Thomas}, {Vanzella},
  {Zamorani}, {Zucca}, {Amorin}, {Bardelli}, {Capak}, {Cassar{\`a}},
  {Castellano}, {Cimatti}, {Cuby}, {Cucciati}, {de la Torre}, {Durkalec},
  {Fontana}, {Giavalisco}, {Grazian}, {Hathi}, {Ilbert}, {Lemaux}, {Moreau},
  {Paltani}, {Ribeiro}, {Salvato}, {Schaerer}, {Scodeggio}, {Sommariva},
  {Talia}, {Taniguchi}, {Tresse}, {Vergani}, {Wang}, {Charlot}, {Contini},
  {Fotopoulou}, {L{\'o}pez-Sanjuan}, {Mellier}, \& {Scoville}}]{dong15}
{Le F{\`e}vre}, O., {Tasca}, L.~A.~M., {Cassata}, P., {et~al.} 2015, \aap, 576,
  A79

\bibitem[{{Le F{\`e}vre} {et~al.}(2005){Le F{\`e}vre}, {Vettolani}, {Garilli},
  {Tresse}, {Bottini}, {Le Brun}, {Maccagni}, {Picat}, {Scaramella},
  {Scodeggio}, {Zanichelli}, {Adami}, {Arnaboldi}, {Arnouts}, {Bardelli},
  {Bolzonella}, {Cappi}, {Charlot}, {Ciliegi}, {Contini}, {Foucaud},
  {Franzetti}, {Gavignaud}, {Guzzo}, {Ilbert}, {Iovino}, {McCracken}, {Marano},
  {Marinoni}, {Mathez}, {Mazure}, {Meneux}, {Merighi}, {Paltani}, {Pell{\`o}},
  {Pollo}, {Pozzetti}, {Radovich}, {Zamorani}, {Zucca}, {Bondi}, {Bongiorno},
  {Busarello}, {Lamareille}, {Mellier}, {Merluzzi}, {Ripepi}, \&
  {Rizzo}}]{dong05}
{Le F{\`e}vre}, O., {Vettolani}, G., {Garilli}, B., {et~al.} 2005, \aap, 439,
  845

\bibitem[{{Le Floc'h} {et~al.}(2009){Le Floc'h}, {Aussel}, {Ilbert},
  {Riguccini}, {Frayer}, {Salvato}, {Arnouts}, {Surace}, {Feruglio},
  {Rodighiero}, {Capak}, {Kartaltepe}, {Heinis}, {Sheth}, {Yan}, {McCracken},
  {Thompson}, {Sanders}, {Scoville}, \& {Koekemoer}}]{emeric09}
{Le Floc'h}, E., {Aussel}, H., {Ilbert}, O., {et~al.} 2009, \apj, 703, 222

\bibitem[{{Lee} {et~al.}(2016){Lee}, {Hennawi}, {White}, {Prochaska},
  {Font-Ribera}, {Schlegel}, {Rich}, {Suzuki}, {Stark}, {Le F{\`e}vre},
  {Nugent}, {Salvato}, \& {Zamorani}}]{lee16}
{Lee}, K.-G., {Hennawi}, J.~F., {White}, M., {et~al.} 2016, \apj, 817, 160

\bibitem[{{Lemaux} {et~al.}(2014{\natexlab{a}}){Lemaux}, {Cucciati}, {Tasca},
  {Le F{\`e}vre}, {Zamorani}, {Cassata}, {Garilli}, {Le Brun}, {Maccagni},
  {Pentericci}, {Thomas}, {Vanzella}, {Zucca}, {Amor{\'{\i}}n}, {Bardelli},
  {Capak}, {Cassar{\`a}}, {Castellano}, {Cimatti}, {Cuby}, {de la Torre},
  {Durkalec}, {Fontana}, {Giavalisco}, {Grazian}, {Hathi}, {Ilbert}, {Moreau},
  {Paltani}, {Ribeiro}, {Salvato}, {Schaerer}, {Scodeggio}, {Sommariva},
  {Talia}, {Taniguchi}, {Tresse}, {Vergani}, {Wang}, {Charlot}, {Contini},
  {Fotopoulou}, {Gal}, {Kocevski}, {L{\'o}pez-Sanjuan}, {Lubin}, {Mellier},
  {Sadibekova}, \& {Scoville}}]{lem14a}
{Lemaux}, B.~C., {Cucciati}, O., {Tasca}, L.~A.~M., {et~al.}
  2014{\natexlab{a}}, \aap, 572, A41

\bibitem[{{Lemaux} {et~al.}(2012){Lemaux}, {Gal}, {Lubin}, {Kocevski},
  {Fassnacht}, {McGrath}, {Squires}, {Surace}, \& {Lacy}}]{lem12}
{Lemaux}, B.~C., {Gal}, R.~R., {Lubin}, L.~M., {et~al.} 2012, \apj, 745, 106

\bibitem[{{Lemaux} {et~al.}(2014{\natexlab{b}}){Lemaux}, {Le Floc'h}, {Le
  F{\`e}vre}, {Ilbert}, {Tresse}, {Lubin}, {Zamorani}, {Gal}, {Ciliegi},
  {Cassata}, {Kocevski}, {McGrath}, {Bardelli}, {Zucca}, \& {Squires}}]{lem14b}
{Lemaux}, B.~C., {Le Floc'h}, E., {Le F{\`e}vre}, O., {et~al.}
  2014{\natexlab{b}}, \aap, 572, A90

\bibitem[{{Lemaux} {et~al.}(2009){Lemaux}, {Lubin}, {Sawicki}, {Martin},
  {Lagattuta}, {Gal}, {Kocevski}, {Fassnacht}, \& {Squires}}]{lem09}
{Lemaux}, B.~C., {Lubin}, L.~M., {Sawicki}, M., {et~al.} 2009, \apj, 700, 20

\bibitem[{{Lemaux} {et~al.}(2010){Lemaux}, {Lubin}, {Shapley}, {Kocevski},
  {Gal}, \& {Squires}}]{lem10}
{Lemaux}, B.~C., {Lubin}, L.~M., {Shapley}, A., {et~al.} 2010, \apj, 716, 970

\bibitem[{{Lemaux} {et~al.}(2017){Lemaux}, {Tomczak}, {Lubin}, {Wu}, {Gal},
  {Rumbaugh}, {Kocevski}, \& {Squires}}]{lem17}
{Lemaux}, B.~C., {Tomczak}, A.~R., {Lubin}, L.~M., {et~al.} 2017, \mnras, 472,
  419

\bibitem[{{Lilly} {et~al.}(2009){Lilly}, {Le Brun}, {Maier}, {Mainieri},
  {Mignoli}, {Scodeggio}, {Zamorani}, {Carollo}, {Contini}, {Kneib}, {Le
  F{\`e}vre}, {Renzini}, {Bardelli}, {Bolzonella}, {Bongiorno}, {Caputi},
  {Coppa}, {Cucciati}, {de la Torre}, {de Ravel}, {Franzetti}, {Garilli},
  {Iovino}, {Kampczyk}, {Kovac}, {Knobel}, {Lamareille}, {Le Borgne}, {Pello},
  {Peng}, {P{\'e}rez-Montero}, {Ricciardelli}, {Silverman}, {Tanaka}, {Tasca},
  {Tresse}, {Vergani}, {Zucca}, {Ilbert}, {Salvato}, {Oesch}, {Abbas},
  {Bottini}, {Capak}, {Cappi}, {Cassata}, {Cimatti}, {Elvis}, {Fumana},
  {Guzzo}, {Hasinger}, {Koekemoer}, {Leauthaud}, {Maccagni}, {Marinoni},
  {McCracken}, {Memeo}, {Meneux}, {Porciani}, {Pozzetti}, {Sanders},
  {Scaramella}, {Scarlata}, {Scoville}, {Shopbell}, \& {Taniguchi}}]{lilly09}
{Lilly}, S.~J., {Le Brun}, V., {Maier}, C., {et~al.} 2009, \apjs, 184, 218

\bibitem[{{Lilly} {et~al.}(2007){Lilly}, {Le F{\`e}vre}, {Renzini}, {Zamorani},
  {Scodeggio}, {Contini}, {Carollo}, {Hasinger}, {Kneib}, {Iovino}, {Le Brun},
  {Maier}, {Mainieri}, {Mignoli}, {Silverman}, {Tasca}, {Bolzonella},
  {Bongiorno}, {Bottini}, {Capak}, {Caputi}, {Cimatti}, {Cucciati}, {Daddi},
  {Feldmann}, {Franzetti}, {Garilli}, {Guzzo}, {Ilbert}, {Kampczyk}, {Kovac},
  {Lamareille}, {Leauthaud}, {Borgne}, {McCracken}, {Marinoni}, {Pello},
  {Ricciardelli}, {Scarlata}, {Vergani}, {Sanders}, {Schinnerer}, {Scoville},
  {Taniguchi}, {Arnouts}, {Aussel}, {Bardelli}, {Brusa}, {Cappi}, {Ciliegi},
  {Finoguenov}, {Foucaud}, {Franceschini}, {Halliday}, {Impey}, {Knobel},
  {Koekemoer}, {Kurk}, {Maccagni}, {Maddox}, {Marano}, {Marconi}, {Meneux},
  {Mobasher}, {Moreau}, {Peacock}, {Porciani}, {Pozzetti}, {Scaramella},
  {Schiminovich}, {Shopbell}, {Smail}, {Thompson}, {Tresse}, {Vettolani},
  {Zanichelli}, \& {Zucca}}]{lilly07}
{Lilly}, S.~J., {Le F{\`e}vre}, O., {Renzini}, A., {et~al.} 2007, \apjs, 172,
  70

\bibitem[{{L{\'o}pez-Sanjuan} {et~al.}(2015){L{\'o}pez-Sanjuan}, {Cenarro},
  {Varela}, {Viironen}, {Molino}, {Ben{\'{\i}}tez}, {Arnalte-Mur}, {Ascaso},
  {D{\'{\i}}az-Garc{\'{\i}}a}, {Fern{\'a}ndez-Soto}, {Jim{\'e}nez-Teja},
  {M{\'a}rquez}, {Masegosa}, {Moles}, {Povi{\'c}}, {Aguerri}, {Alfaro},
  {Aparicio-Villegas}, {Broadhurst}, {Cabrera-Ca{\~n}o}, {Castander}, {Cepa},
  {Cervi{\~n}o}, {Crist{\'o}bal-Hornillos}, {Del Olmo}, {Gonz{\'a}lez Delgado},
  {Husillos}, {Infante}, {Mart{\'{\i}}nez}, {Perea}, {Prada}, \&
  {Quintana}}]{carlos15}
{L{\'o}pez-Sanjuan}, C., {Cenarro}, A.~J., {Varela}, J., {et~al.} 2015, \aap,
  576, A53

\bibitem[{{L{\'o}pez-Sanjuan} {et~al.}(2013){L{\'o}pez-Sanjuan}, {Le
  F{\`e}vre}, {Tasca}, {Epinat}, {Amram}, {Contini}, {Garilli},
  {Kissler-Patig}, {Moultaka}, {Paioro}, {Perret}, {Queyrel}, {Tresse},
  {Vergani}, \& {Divoy}}]{carlos13}
{L{\'o}pez-Sanjuan}, C., {Le F{\`e}vre}, O., {Tasca}, L.~A.~M., {et~al.} 2013,
  \aap, 553, A78

\bibitem[{{Lotz} {et~al.}(2011){Lotz}, {Jonsson}, {Cox}, {Croton}, {Primack},
  {Somerville}, \& {Stewart}}]{jlotz11}
{Lotz}, J.~M., {Jonsson}, P., {Cox}, T.~J., {et~al.} 2011, \apj, 742, 103

\bibitem[{{Lubin} {et~al.}(2009){Lubin}, {Gal}, {Lemaux}, {Kocevski}, \&
  {Squires}}]{lub09}
{Lubin}, L.~M., {Gal}, R.~R., {Lemaux}, B.~C., {Kocevski}, D.~D., \& {Squires},
  G.~K. 2009, \aj, 137, 4867

\bibitem[{{Lutz} {et~al.}(2011){Lutz}, {Poglitsch}, {Altieri}, {Andreani},
  {Aussel}, {Berta}, {Bongiovanni}, {Brisbin}, {Cava}, {Cepa}, {Cimatti},
  {Daddi}, {Dominguez-Sanchez}, {Elbaz}, {F{\"o}rster Schreiber}, {Genzel},
  {Grazian}, {Gruppioni}, {Harwit}, {Le Floc'h}, {Magdis}, {Magnelli},
  {Maiolino}, {Nordon}, {P{\'e}rez Garc{\'{\i}}a}, {Popesso}, {Pozzi},
  {Riguccini}, {Rodighiero}, {Saintonge}, {Sanchez Portal}, {Santini}, {Shao},
  {Sturm}, {Tacconi}, {Valtchanov}, {Wetzstein}, \& {Wieprecht}}]{lutz11}
{Lutz}, D., {Poglitsch}, A., {Altieri}, B., {et~al.} 2011, \aap, 532, A90

\bibitem[{{Malhotra} \& {Rhoads}(2004)}]{sangeeta04}
{Malhotra}, S. \& {Rhoads}, J.~E. 2004, \apjl, 617, L5

\bibitem[{{Malhotra} {et~al.}(2005){Malhotra}, {Rhoads}, {Pirzkal}, {Haiman},
  {Xu}, {Daddi}, {Yan}, {Bergeron}, {Wang}, {Ferguson}, {Gronwall},
  {Koekemoer}, {Kuemmel}, {Moustakas}, {Panagia}, {Pasquali}, {Stiavelli},
  {Walsh}, {Windhorst}, \& {di Serego Alighieri}}]{sangeeta05}
{Malhotra}, S., {Rhoads}, J.~E., {Pirzkal}, N., {et~al.} 2005, \apj, 626, 666

\bibitem[{{Martin} {et~al.}(2005){Martin}, {Fanson}, {Schiminovich},
  {Morrissey}, {Friedman}, {Barlow}, {Conrow}, {Grange}, {Jelinsky},
  {Milliard}, {Siegmund}, {Bianchi}, {Byun}, {Donas}, {Forster}, {Heckman},
  {Lee}, {Madore}, {Malina}, {Neff}, {Rich}, {Small}, {Surber}, {Szalay},
  {Welsh}, \& {Wyder}}]{martin05}
{Martin}, D.~C., {Fanson}, J., {Schiminovich}, D., {et~al.} 2005, \apjl, 619,
  L1

\bibitem[{{Martini} {et~al.}(2013){Martini}, {Miller}, {Brodwin}, {Stanford},
  {Gonzalez}, {Bautz}, {Hickox}, {Stern}, {Eisenhardt}, {Galametz}, {Norman},
  {Jannuzi}, {Dey}, {Murray}, {Jones}, \& {Brown}}]{martini13}
{Martini}, P., {Miller}, E.~D., {Brodwin}, M., {et~al.} 2013, \apj, 768, 1

\bibitem[{{Matsuda} {et~al.}(2011){Matsuda}, {Smail}, {Geach}, {Best},
  {Sobral}, {Tanaka}, {Nakata}, {Ohta}, {Kurk}, {Iwata}, {Bielby}, {Wardlow},
  {Bower}, {Ivison}, {Kodama}, {Yamada}, {Mawatari}, \& {Casali}}]{matsuda11}
{Matsuda}, Y., {Smail}, I., {Geach}, J.~E., {et~al.} 2011, \mnras, 416, 2041

\bibitem[{{Matsuda} {et~al.}(2012){Matsuda}, {Yamada}, {Hayashino}, {Yamauchi},
  {Nakamura}, {Morimoto}, {Ouchi}, {Ono}, {Umemura}, \& {Mori}}]{matsuda12}
{Matsuda}, Y., {Yamada}, T., {Hayashino}, T., {et~al.} 2012, \mnras, 425, 878

\bibitem[{{Matthee} {et~al.}(2016){Matthee}, {Sobral}, {Oteo}, {Best}, {Smail},
  {R{\"o}ttgering}, \& {Paulino-Afonso}}]{mathee16}
{Matthee}, J., {Sobral}, D., {Oteo}, I., {et~al.} 2016, \mnras, 458, 449

\bibitem[{{McBride} {et~al.}(2009){McBride}, {Fakhouri}, \& {Ma}}]{Mcbride09}
{McBride}, J., {Fakhouri}, O., \& {Ma}, C.-P. 2009, \mnras, 398, 1858

\bibitem[{{McCracken} {et~al.}(2012){McCracken}, {Milvang-Jensen}, {Dunlop},
  {Franx}, {Fynbo}, {Le F{\`e}vre}, {Holt}, {Caputi}, {Goranova}, {Buitrago},
  {Emerson}, {Freudling}, {Hudelot}, {L{\'o}pez-Sanjuan}, {Magnard}, {Mellier},
  {M{\o}ller}, {Nilsson}, {Sutherland}, {Tasca}, \& {Zabl}}]{mccracken12}
{McCracken}, H.~J., {Milvang-Jensen}, B., {Dunlop}, J., {et~al.} 2012, \aap,
  544, A156

\bibitem[{{M{\'e}nard} {et~al.}(2013){M{\'e}nard}, {Scranton}, {Schmidt},
  {Morrison}, {Jeong}, {Budavari}, \& {Rahman}}]{menard13}
{M{\'e}nard}, B., {Scranton}, R., {Schmidt}, S., {et~al.} 2013, ArXiv e-prints
  [\eprint[arXiv]{1303.4722}]

\bibitem[{{Meurer} {et~al.}(1999){Meurer}, {Heckman}, \& {Calzetti}}]{meurer99}
{Meurer}, G.~R., {Heckman}, T.~M., \& {Calzetti}, D. 1999, \apj, 521, 64

\bibitem[{{Miettinen} {et~al.}(2017){Miettinen}, {Novak}, {Smol{\v c}i{\'c}},
  {Delvecchio}, {Aravena}, {Brisbin}, {Karim}, {Murphy}, {Schinnerer},
  {Albrecht}, {Aussel}, {Bertoldi}, {Capak}, {Casey}, {Civano}, {Hayward},
  {Herrera Ruiz}, {Ilbert}, {Jiang}, {Laigle}, {Le F{\`e}vre}, {Magnelli},
  {Marchesi}, {McCracken}, {Middelberg}, {Mu{\~n}oz Arancibia}, {Navarrete},
  {Padilla}, {Riechers}, {Salvato}, {Scott}, {Sheth}, {Tasca}, {Bondi}, \&
  {Zamorani}}]{miettinen17}
{Miettinen}, O., {Novak}, M., {Smol{\v c}i{\'c}}, V., {et~al.} 2017, ArXiv
  e-prints [\eprint[arXiv]{1702.07527}]

\bibitem[{{Miley} \& {De Breuck}(2008)}]{gmileznfriend08}
{Miley}, G. \& {De Breuck}, C. 2008, \aapr, 15, 67

\bibitem[{{Miley} {et~al.}(2004){Miley}, {Overzier}, {Tsvetanov}, {Bouwens},
  {Ben{\'{\i}}tez}, {Blakeslee}, {Ford}, {Illingworth}, {Postman}, {Rosati},
  {Clampin}, {Hartig}, {Zirm}, {R{\"o}ttgering}, {Venemans}, {Ardila},
  {Bartko}, {Broadhurst}, {Brown}, {Burrows}, {Cheng}, {Cross}, {De Breuck},
  {Feldman}, {Franx}, {Golimowski}, {Gronwall}, {Infante}, {Martel},
  {Menanteau}, {Meurer}, {Sirianni}, {Kimble}, {Krist}, {Sparks}, {Tran},
  {White}, \& {Zheng}}]{GMilez04}
{Miley}, G.~K., {Overzier}, R.~A., {Tsvetanov}, Z.~I., {et~al.} 2004, \nat,
  427, 47

\bibitem[{{Momose} {et~al.}(2016){Momose}, {Ouchi}, {Nakajima}, {Ono},
  {Shibuya}, {Shimasaku}, {Yuma}, {Mori}, \& {Umemura}}]{momose16}
{Momose}, R., {Ouchi}, M., {Nakajima}, K., {et~al.} 2016, \mnras, 457, 2318

\bibitem[{{Moran} {et~al.}(2007){Moran}, {Ellis}, {Treu}, {Smith}, {Rich}, \&
  {Smail}}]{moran07}
{Moran}, S.~M., {Ellis}, R.~S., {Treu}, T., {et~al.} 2007, \apj, 671, 1503

\bibitem[{{Mortlock} {et~al.}(2013){Mortlock}, {Conselice}, {Hartley},
  {Ownsworth}, {Lani}, {Bluck}, {Almaini}, {Duncan}, {van der Wel},
  {Koekemoer}, {Dekel}, {Dav{\'e}}, {Ferguson}, {de Mello}, {Newman}, {Faber},
  {Grogin}, {Kocevski}, \& {Lai}}]{mortlock13}
{Mortlock}, A., {Conselice}, C.~J., {Hartley}, W.~G., {et~al.} 2013, \mnras,
  433, 1185

\bibitem[{{Moutard} {et~al.}(2016){Moutard}, {Arnouts}, {Ilbert}, {Coupon},
  {Davidzon}, {Guzzo}, {Hudelot}, {McCracken}, {Van Werbaeke}, {Morrison}, {Le
  F{\`e}vre}, {Comte}, {Bolzonella}, {Fritz}, {Garilli}, \&
  {Scodeggio}}]{thibaud16}
{Moutard}, T., {Arnouts}, S., {Ilbert}, O., {et~al.} 2016, \aap, 590, A103

\bibitem[{{Muchovej} {et~al.}(2007){Muchovej}, {Mroczkowski}, {Carlstrom},
  {Cartwright}, {Greer}, {Hennessy}, {Loh}, {Pryke}, {Reddall}, {Runyan},
  {Sharp}, {Hawkins}, {Lamb}, {Woody}, {Joy}, {Leitch}, \&
  {Miller}}]{muchovej07}
{Muchovej}, S., {Mroczkowski}, T., {Carlstrom}, J.~E., {et~al.} 2007, \apj,
  663, 708

\bibitem[{{Muldrew} {et~al.}(2015){Muldrew}, {Hatch}, \& {Cooke}}]{muldrew15}
{Muldrew}, S.~I., {Hatch}, N.~A., \& {Cooke}, E.~A. 2015, \mnras, 452, 2528

\bibitem[{{M{\"u}ller}(2000)}]{muller00}
{M{\"u}ller}, J.~W. 2000, Journal of Research of the National Institute of
  Standards and Technology, 105, 551

\bibitem[{{Muzzin} {et~al.}(2012){Muzzin}, {Wilson}, {Yee}, {Gilbank},
  {Hoekstra}, {Demarco}, {Balogh}, {van Dokkum}, {Franx}, {Ellingson}, {Hicks},
  {Nantais}, {Noble}, {Lacy}, {Lidman}, {Rettura}, {Surace}, \& {Webb}}]{muz12}
{Muzzin}, A., {Wilson}, G., {Yee}, H.~K.~C., {et~al.} 2012, \apj, 746, 188

\bibitem[{{Nasonova} {et~al.}(2011){Nasonova}, {de Freitas Pacheco}, \&
  {Karachentsev}}]{nasonova11}
{Nasonova}, O.~G., {de Freitas Pacheco}, J.~A., \& {Karachentsev}, I.~D. 2011,
  \aap, 532, A104

\bibitem[{{Novak} {et~al.}(2015){Novak}, {Smol{\v c}i{\'c}}, {Civano}, {Bondi},
  {Ciliegi}, {Wang}, {Loeb}, {Banfield}, {Bourke}, {Elvis}, {Hallinan},
  {Intema}, {Kl{\"o}ckner}, {Mooley}, \& {Navarrete}}]{mladen15}
{Novak}, M., {Smol{\v c}i{\'c}}, V., {Civano}, F., {et~al.} 2015, \mnras, 447,
  1282

\bibitem[{{Oke} \& {Gunn}(1983)}]{okengunn83}
{Oke}, J.~B. \& {Gunn}, J.~E. 1983, \apj, 266, 713

\bibitem[{{Oke} {et~al.}(1998){Oke}, {Postman}, \& {Lubin}}]{oke98}
{Oke}, J.~B., {Postman}, M., \& {Lubin}, L.~M. 1998, \aj, 116, 549

\bibitem[{{Oliver} {et~al.}(2012){Oliver}, {Bock}, {Altieri}, {Amblard},
  {Arumugam}, {Aussel}, {Babbedge}, {Beelen}, {B{\'e}thermin}, {Blain},
  {Boselli}, {Bridge}, {Brisbin}, {Buat}, {Burgarella},
  {Castro-Rodr{\'{\i}}guez}, {Cava}, {Chanial}, {Cirasuolo}, {Clements},
  {Conley}, {Conversi}, {Cooray}, {Dowell}, {Dubois}, {Dwek}, {Dye}, {Eales},
  {Elbaz}, {Farrah}, {Feltre}, {Ferrero}, {Fiolet}, {Fox}, {Franceschini},
  {Gear}, {Giovannoli}, {Glenn}, {Gong}, {Gonz{\'a}lez Solares}, {Griffin},
  {Halpern}, {Harwit}, {Hatziminaoglou}, {Heinis}, {Hurley}, {Hwang}, {Hyde},
  {Ibar}, {Ilbert}, {Isaak}, {Ivison}, {Lagache}, {Le Floc'h}, {Levenson},
  {Faro}, {Lu}, {Madden}, {Maffei}, {Magdis}, {Mainetti}, {Marchetti},
  {Marsden}, {Marshall}, {Mortier}, {Nguyen}, {O'Halloran}, {Omont}, {Page},
  {Panuzzo}, {Papageorgiou}, {Patel}, {Pearson}, {P{\'e}rez-Fournon}, {Pohlen},
  {Rawlings}, {Raymond}, {Rigopoulou}, {Riguccini}, {Rizzo}, {Rodighiero},
  {Roseboom}, {Rowan-Robinson}, {S{\'a}nchez Portal}, {Schulz}, {Scott},
  {Seymour}, {Shupe}, {Smith}, {Stevens}, {Symeonidis}, {Trichas}, {Tugwell},
  {Vaccari}, {Valtchanov}, {Vieira}, {Viero}, {Vigroux}, {Wang}, {Ward},
  {Wardlow}, {Wright}, {Xu}, \& {Zemcov}}]{oliver12}
{Oliver}, S.~J., {Bock}, J., {Altieri}, B., {et~al.} 2012, \mnras, 424, 1614

\bibitem[{{Orsi} {et~al.}(2008){Orsi}, {Lacey}, {Baugh}, \& {Infante}}]{orsi08}
{Orsi}, A., {Lacey}, C.~G., {Baugh}, C.~M., \& {Infante}, L. 2008, \mnras, 391,
  1589

\bibitem[{{Orsi} {et~al.}(2016){Orsi}, {Fanidakis}, {Lacey}, \&
  {Baugh}}]{orsi16}
{Orsi}, {\'A}.~A., {Fanidakis}, N., {Lacey}, C.~G., \& {Baugh}, C.~M. 2016,
  \mnras, 456, 3827

\bibitem[{{Ouchi} {et~al.}(2008){Ouchi}, {Shimasaku}, {Akiyama}, {Simpson},
  {Saito}, {Ueda}, {Furusawa}, {Sekiguchi}, {Yamada}, {Kodama}, {Kashikawa},
  {Okamura}, {Iye}, {Takata}, {Yoshida}, \& {Yoshida}}]{ouchi08}
{Ouchi}, M., {Shimasaku}, K., {Akiyama}, M., {et~al.} 2008, \apjs, 176, 301

\bibitem[{{Ouchi} {et~al.}(2003){Ouchi}, {Shimasaku}, {Furusawa}, {Miyazaki},
  {Doi}, {Hamabe}, {Hayashino}, {Kimura}, {Kodaira}, {Komiyama}, {Matsuda},
  {Miyazaki}, {Nakata}, {Okamura}, {Sekiguchi}, {Shioya}, {Tamura},
  {Taniguchi}, {Yagi}, \& {Yasuda}}]{ouchi03}
{Ouchi}, M., {Shimasaku}, K., {Furusawa}, H., {et~al.} 2003, \apj, 582, 60

\bibitem[{{Ouchi} {et~al.}(2010){Ouchi}, {Shimasaku}, {Furusawa}, {Saito},
  {Yoshida}, {Akiyama}, {Ono}, {Yamada}, {Ota}, {Kashikawa}, {Iye}, {Kodama},
  {Okamura}, {Simpson}, \& {Yoshida}}]{ouchi10}
{Ouchi}, M., {Shimasaku}, K., {Furusawa}, H., {et~al.} 2010, \apj, 723, 869

\bibitem[{{Overzier} {et~al.}(2008){Overzier}, {Bouwens}, {Cross}, {Venemans},
  {Miley}, {Zirm}, {Ben{\'{\i}}tez}, {Blakeslee}, {Coe}, {Demarco}, {Ford},
  {Homeier}, {Illingworth}, {Kurk}, {Martel}, {Mei}, {Oliveira},
  {R{\"o}ttgering}, {Tsvetanov}, \& {Zheng}}]{overzier08}
{Overzier}, R.~A., {Bouwens}, R.~J., {Cross}, N.~J.~G., {et~al.} 2008, \apj,
  673, 143

\bibitem[{{Paczynski}(1987)}]{paczynski87}
{Paczynski}, B. 1987, \nat, 325, 572

\bibitem[{{Parkes} {et~al.}(1994){Parkes}, {Collins}, \& {Joseph}}]{parkes94}
{Parkes}, I.~M., {Collins}, C.~A., \& {Joseph}, R.~D. 1994, \mnras, 266, 983

\bibitem[{{Partridge}(1974)}]{partridge74}
{Partridge}, R.~B. 1974, \apj, 192, 241

\bibitem[{{Patton} {et~al.}(2011){Patton}, {Ellison}, {Simard}, {McConnachie},
  \& {Mendel}}]{patton11}
{Patton}, D.~R., {Ellison}, S.~L., {Simard}, L., {McConnachie}, A.~W., \&
  {Mendel}, J.~T. 2011, \mnras, 412, 591

\bibitem[{{Pentericci} {et~al.}(2000{\natexlab{a}}){Pentericci}, {Kurk},
  {R{\"o}ttgering}, {Miley}, {van Breugel}, {Carilli}, {Ford}, {Heckman},
  {McCarthy}, \& {Moorwood}}]{Lpentz00}
{Pentericci}, L., {Kurk}, J.~D., {R{\"o}ttgering}, H.~J.~A., {et~al.}
  2000{\natexlab{a}}, \aap, 361, L25

\bibitem[{{Pentericci} {et~al.}(2000{\natexlab{b}}){Pentericci}, {Kurk},
  {R{\"o}ttgering}, {Miley}, {van Breugel}, {Carilli}, {Ford}, {Heckman},
  {McCarthy}, \& {Moorwood}}]{lpent00}
{Pentericci}, L., {Kurk}, J.~D., {R{\"o}ttgering}, H.~J.~A., {et~al.}
  2000{\natexlab{b}}, \aap, 361, L25

\bibitem[{{Pierre} {et~al.}(2016){Pierre}, {Pacaud}, {Adami}, {Alis},
  {Altieri}, {Baran}, {Benoist}, {Birkinshaw}, {Bongiorno}, {Bremer}, {Brusa},
  {Butler}, {Ciliegi}, {Chiappetti}, {Clerc}, {Corasaniti}, {Coupon}, {De
  Breuck}, {Democles}, {Desai}, {Delhaize}, {Devriendt}, {Dubois}, {Eckert},
  {Elyiv}, {Ettori}, {Evrard}, {Faccioli}, {Farahi}, {Ferrari}, {Finet},
  {Fotopoulou}, {Fourmanoit}, {Gandhi}, {Gastaldello}, {Gastaud},
  {Georgantopoulos}, {Giles}, {Guennou}, {Guglielmo}, {Horellou}, {Husband},
  {Huynh}, {Iovino}, {Kilbinger}, {Koulouridis}, {Lavoie}, {Le Brun}, {Le
  Fevre}, {Lidman}, {Lieu}, {Lin}, {Mantz}, {Maughan}, {Maurogordato},
  {McCarthy}, {McGee}, {Melin}, {Melnyk}, {Menanteau}, {Novak}, {Paltani},
  {Plionis}, {Poggianti}, {Pomarede}, {Pompei}, {Ponman}, {Ramos-Ceja},
  {Ranalli}, {Rapetti}, {Raychaudury}, {Reiprich}, {Rottgering}, {Rozo},
  {Rykoff}, {Sadibekova}, {Santos}, {Sauvageot}, {Schimd}, {Sereno}, {Smith},
  {Smol{\v c}i{\'c}}, {Snowden}, {Spergel}, {Stanford}, {Surdej}, {Valageas},
  {Valotti}, {Valtchanov}, {Vignali}, {Willis}, \& {Ziparo}}]{pierre16}
{Pierre}, M., {Pacaud}, F., {Adami}, C., {et~al.} 2016, \aap, 592, A1

\bibitem[{{Pierre} {et~al.}(2004){Pierre}, {Valtchanov}, {Altieri}, {Andreon},
  {Bolzonella}, {Bremer}, {Disseau}, {Dos Santos}, {Gandhi}, {Jean}, {Pacaud},
  {Read}, {Refregier}, {Willis}, {Adami}, {Alloin}, {Birkinshaw}, {Chiappetti},
  {Cohen}, {Detal}, {Duc}, {Gosset}, {Hjorth}, {Jones}, {Le F{\`e}vre},
  {Lonsdale}, {Maccagni}, {Mazure}, {McBreen}, {McCracken}, {Mellier},
  {Ponman}, {Quintana}, {Rottgering}, {Smette}, {Surdej}, {Starck}, {Vigroux},
  \& {White}}]{pierre04}
{Pierre}, M., {Valtchanov}, I., {Altieri}, B., {et~al.} 2004, \jcap, 9, 011

\bibitem[{{Piffaretti} {et~al.}(2011){Piffaretti}, {Arnaud}, {Pratt},
  {Pointecouteau}, \& {Melin}}]{piffaretti11}
{Piffaretti}, R., {Arnaud}, M., {Pratt}, G.~W., {Pointecouteau}, E., \&
  {Melin}, J.-B. 2011, \aap, 534, A109

\bibitem[{{Pilbratt} {et~al.}(2010){Pilbratt}, {Riedinger}, {Passvogel},
  {Crone}, {Doyle}, {Gageur}, {Heras}, {Jewell}, {Metcalfe}, {Ott}, \&
  {Schmidt}}]{pilbratt10}
{Pilbratt}, G.~L., {Riedinger}, J.~R., {Passvogel}, T., {et~al.} 2010, \aap,
  518, L1

\bibitem[{{Planck Collaboration} {et~al.}(2016){Planck Collaboration}, {Ade},
  {Aghanim}, {Arnaud}, {Ashdown}, {Aumont}, {Baccigalupi}, {Banday},
  {Barreiro}, {Bartlett}, \& et~al.}]{planck16}
{Planck Collaboration}, {Ade}, P.~A.~R., {Aghanim}, N., {et~al.} 2016, \aap,
  594, A13

\bibitem[{{Poggianti} {et~al.}(2009){Poggianti}, {Arag{\'o}n-Salamanca},
  {Zaritsky}, {De Lucia}, {Milvang-Jensen}, {Desai}, {Jablonka}, {Halliday},
  {Rudnick}, {Varela}, {Bamford}, {Best}, {Clowe}, {Noll}, {Saglia},
  {Pell{\'o}}, {Simard}, {von der Linden}, \& {White}}]{pog09}
{Poggianti}, B.~M., {Arag{\'o}n-Salamanca}, A., {Zaritsky}, D., {et~al.} 2009,
  \apj, 693, 112

\bibitem[{{Poggianti} {et~al.}(2008){Poggianti}, {Desai}, {Finn}, {Bamford},
  {De Lucia}, {Varela}, {Arag{\'o}n-Salamanca}, {Halliday}, {Noll}, {Saglia},
  {Zaritsky}, {Best}, {Clowe}, {Milvang-Jensen}, {Jablonka}, {Pell{\'o}},
  {Rudnick}, {Simard}, {von der Linden}, \& {White}}]{poggianti08}
{Poggianti}, B.~M., {Desai}, V., {Finn}, R., {et~al.} 2008, \apj, 684, 888

\bibitem[{{Poglitsch} {et~al.}(2010){Poglitsch}, {Waelkens}, {Geis},
  {Feuchtgruber}, {Vandenbussche}, {Rodriguez}, {Krause}, {Renotte}, {van
  Hoof}, {Saraceno}, {Cepa}, {Kerschbaum}, {Agn{\`e}se}, {Ali}, {Altieri},
  {Andreani}, {Augueres}, {Balog}, {Barl}, {Bauer}, {Belbachir}, {Benedettini},
  {Billot}, {Boulade}, {Bischof}, {Blommaert}, {Callut}, {Cara}, {Cerulli},
  {Cesarsky}, {Contursi}, {Creten}, {De Meester}, {Doublier}, {Doumayrou},
  {Duband}, {Exter}, {Genzel}, {Gillis}, {Gr{\"o}zinger}, {Henning},
  {Herreros}, {Huygen}, {Inguscio}, {Jakob}, {Jamar}, {Jean}, {de Jong},
  {Katterloher}, {Kiss}, {Klaas}, {Lemke}, {Lutz}, {Madden}, {Marquet},
  {Martignac}, {Mazy}, {Merken}, {Montfort}, {Morbidelli}, {M{\"u}ller},
  {Nielbock}, {Okumura}, {Orfei}, {Ottensamer}, {Pezzuto}, {Popesso},
  {Putzeys}, {Regibo}, {Reveret}, {Royer}, {Sauvage}, {Schreiber}, {Stegmaier},
  {Schmitt}, {Schubert}, {Sturm}, {Thiel}, {Tofani}, {Vavrek}, {Wetzstein},
  {Wieprecht}, \& {Wiezorrek}}]{poglitsch10}
{Poglitsch}, A., {Waelkens}, C., {Geis}, N., {et~al.} 2010, \aap, 518, L2

\bibitem[{{Postman} {et~al.}(2005){Postman}, {Franx}, {Cross}, {Holden},
  {Ford}, {Illingworth}, {Goto}, {Demarco}, {Rosati}, {Blakeslee}, {Tran},
  {Ben{\'{\i}}tez}, {Clampin}, {Hartig}, {Homeier}, {Ardila}, {Bartko},
  {Bouwens}, {Bradley}, {Broadhurst}, {Brown}, {Burrows}, {Cheng}, {Feldman},
  {Golimowski}, {Gronwall}, {Infante}, {Kimble}, {Krist}, {Lesser}, {Martel},
  {Mei}, {Menanteau}, {Meurer}, {Miley}, {Motta}, {Sirianni}, {Sparks}, {Tran},
  {Tsvetanov}, {White}, \& {Zheng}}]{postman05}
{Postman}, M., {Franx}, M., {Cross}, N.~J.~G., {et~al.} 2005, \apj, 623, 721

\bibitem[{{Pritchet} \& {Hartwick}(1987)}]{prichetnhartwick87}
{Pritchet}, C.~J. \& {Hartwick}, F.~D.~A. 1987, \apj, 320, 464

\bibitem[{{Quadri} {et~al.}(2012){Quadri}, {Williams}, {Franx}, \&
  {Hildebrandt}}]{quadri12}
{Quadri}, R.~F., {Williams}, R.~J., {Franx}, M., \& {Hildebrandt}, H. 2012,
  \apj, 744, 88

\bibitem[{{Raichoor} {et~al.}(2012){Raichoor}, {Mei}, {Stanford}, {Holden},
  {Nakata}, {Rosati}, {Shankar}, {Tanaka}, {Ford}, {Huertas-Company},
  {Illingworth}, {Kodama}, {Postman}, {Rettura}, {Blakeslee}, {Demarco}, {Jee},
  \& {White}}]{raichoor12}
{Raichoor}, A., {Mei}, S., {Stanford}, S.~A., {et~al.} 2012, \apj, 745, 130

\bibitem[{{Reddy} {et~al.}(2010){Reddy}, {Erb}, {Pettini}, {Steidel}, \&
  {Shapley}}]{reddy10}
{Reddy}, N.~A., {Erb}, D.~K., {Pettini}, M., {Steidel}, C.~C., \& {Shapley},
  A.~E. 2010, \apj, 712, 1070

\bibitem[{{Reddy} {et~al.}(2016){Reddy}, {Steidel}, {Pettini},
  {Bogosavljevi{\'c}}, \& {Shapley}}]{reddy16}
{Reddy}, N.~A., {Steidel}, C.~C., {Pettini}, M., {Bogosavljevi{\'c}}, M., \&
  {Shapley}, A.~E. 2016, \apj, 828, 108

\bibitem[{{Rettura} {et~al.}(2010){Rettura}, {Rosati}, {Nonino}, {Fosbury},
  {Gobat}, {Menci}, {Strazzullo}, {Mei}, {Demarco}, \& {Ford}}]{rettura10}
{Rettura}, A., {Rosati}, P., {Nonino}, M., {et~al.} 2010, \apj, 709, 512

\bibitem[{{Ribeiro} {et~al.}(2016{\natexlab{a}}){Ribeiro}, {Le F{\`e}vre},
  {Cassata}, {Garilli}, {Lemaux}, {Maccagni}, {Schaerer}, {Tasca}, {Zamorani},
  {Zucca}, {Amor{\'{\i}}n}, {Bardelli}, {Hathi}, {Koekemoer}, \&
  {Pforr}}]{bruno16b}
{Ribeiro}, B., {Le F{\`e}vre}, O., {Cassata}, P., {et~al.} 2016{\natexlab{a}},
  ArXiv e-prints [\eprint[arXiv]{1611.05869}]

\bibitem[{{Ribeiro} {et~al.}(2016{\natexlab{b}}){Ribeiro}, {Le F{\`e}vre},
  {Tasca}, {Lemaux}, {Cassata}, {Garilli}, {Maccagni}, {Zamorani}, {Zucca},
  {Amor{\'{\i}}n}, {Bardelli}, {Fontana}, {Giavalisco}, {Hathi}, {Koekemoer},
  {Pforr}, {Tresse}, \& {Dunlop}}]{bruno16}
{Ribeiro}, B., {Le F{\`e}vre}, O., {Tasca}, L.~A.~M., {et~al.}
  2016{\natexlab{b}}, \aap, 593, A22

\bibitem[{{Rieke} {et~al.}(2004){Rieke}, {Young}, {Engelbracht}, {Kelly},
  {Low}, {Haller}, {Beeman}, {Gordon}, {Stansberry}, {Misselt}, {Cadien},
  {Morrison}, {Rivlis}, {Latter}, {Noriega-Crespo}, {Padgett}, {Stapelfeldt},
  {Hines}, {Egami}, {Muzerolle}, {Alonso-Herrero}, {Blaylock}, {Dole}, {Hinz},
  {Le Floc'h}, {Papovich}, {P{\'e}rez-Gonz{\'a}lez}, {Smith}, {Su}, {Bennett},
  {Frayer}, {Henderson}, {Lu}, {Masci}, {Pesenson}, {Rebull}, {Rho}, {Keene},
  {Stolovy}, {Wachter}, {Wheaton}, {Werner}, \& {Richards}}]{rieke04}
{Rieke}, G.~H., {Young}, E.~T., {Engelbracht}, C.~W., {et~al.} 2004, \apjs,
  154, 25

\bibitem[{{Rumbaugh} {et~al.}(2013){Rumbaugh}, {Kocevski}, {Gal}, {Lemaux},
  {Lubin}, {Fassnacht}, \& {Squires}}]{thecreature13}
{Rumbaugh}, N., {Kocevski}, D.~D., {Gal}, R.~R., {et~al.} 2013, \apj, 763, 124

\bibitem[{{Sanders} {et~al.}(2007){Sanders}, {Salvato}, {Aussel}, {Ilbert},
  {Scoville}, {Surace}, {Frayer}, {Sheth}, {Helou}, {Brooke}, {Bhattacharya},
  {Yan}, {Kartaltepe}, {Barnes}, {Blain}, {Calzetti}, {Capak}, {Carilli},
  {Carollo}, {Comastri}, {Daddi}, {Ellis}, {Elvis}, {Fall}, {Franceschini},
  {Giavalisco}, {Hasinger}, {Impey}, {Koekemoer}, {Le F{\`e}vre}, {Lilly},
  {Liu}, {McCracken}, {Mobasher}, {Renzini}, {Rich}, {Schinnerer}, {Shopbell},
  {Taniguchi}, {Thompson}, {Urry}, \& {Williams}}]{sanders07}
{Sanders}, D.~B., {Salvato}, M., {Aussel}, H., {et~al.} 2007, \apjs, 172, 86

\bibitem[{{Schenker} {et~al.}(2012){Schenker}, {Stark}, {Ellis}, {Robertson},
  {Dunlop}, {McLure}, {Kneib}, \& {Richard}}]{schenker12}
{Schenker}, M.~A., {Stark}, D.~P., {Ellis}, R.~S., {et~al.} 2012, \apj, 744,
  179

\bibitem[{{Schinnerer} {et~al.}(2016){Schinnerer}, {Groves}, {Sargent},
  {Karim}, {Oesch}, {Magnelli}, {LeFevre}, {Tasca}, {Civano}, {Cassata}, \&
  {Smol{\v c}i{\'c}}}]{schinnerer16}
{Schinnerer}, E., {Groves}, B., {Sargent}, M.~T., {et~al.} 2016, \apj, 833, 112

\bibitem[{{Schinnerer} {et~al.}(2010){Schinnerer}, {Sargent}, {Bondi}, {Smol{\v
  c}i{\'c}}, {Datta}, {Carilli}, {Bertoldi}, {Blain}, {Ciliegi}, {Koekemoer},
  \& {Scoville}}]{schinnerer10}
{Schinnerer}, E., {Sargent}, M.~T., {Bondi}, M., {et~al.} 2010, \apjs, 188, 384

\bibitem[{{Schinnerer} {et~al.}(2007){Schinnerer}, {Smol{\v c}i{\'c}},
  {Carilli}, {Bondi}, {Ciliegi}, {Jahnke}, {Scoville}, {Aussel}, {Bertoldi},
  {Blain}, {Impey}, {Koekemoer}, {Le Fevre}, \& {Urry}}]{schinnerer07}
{Schinnerer}, E., {Smol{\v c}i{\'c}}, V., {Carilli}, C.~L., {et~al.} 2007,
  \apjs, 172, 46

\bibitem[{{Schmidt} {et~al.}(2013){Schmidt}, {M{\'e}nard}, {Scranton},
  {Morrison}, \& {McBride}}]{skiddishsam13}
{Schmidt}, S.~J., {M{\'e}nard}, B., {Scranton}, R., {Morrison}, C., \&
  {McBride}, C.~K. 2013, \mnras, 431, 3307

\bibitem[{{Schrabback} {et~al.}(2016){Schrabback}, {Applegate}, {Dietrich},
  {Hoekstra}, {Bocquet}, {Gonzalez}, {von der Linden}, {McDonald}, {Morrison},
  {Raihan}, {Allen}, {Bayliss}, {Benson}, {Bleem}, {Chiu}, {Desai}, {Foley},
  {de Haan}, {High}, {Hilbert}, {Mantz}, {Massey}, {Mohr}, {Reichardt}, {Saro},
  {Simon}, {Stern}, {Stubbs}, \& {Zenteno}}]{schrabback16}
{Schrabback}, T., {Applegate}, D., {Dietrich}, J.~P., {et~al.} 2016, ArXiv
  e-prints [\eprint[arXiv]{1611.03866}]

\bibitem[{{Scott} {et~al.}(2008){Scott}, {Austermann}, {Perera}, {Wilson},
  {Aretxaga}, {Bock}, {Hughes}, {Kang}, {Kim}, {Mauskopf}, {Sanders},
  {Scoville}, \& {Yun}}]{scott08}
{Scott}, K.~S., {Austermann}, J.~E., {Perera}, T.~A., {et~al.} 2008, \mnras,
  385, 2225

\bibitem[{{Scoville} {et~al.}(2007{\natexlab{a}}){Scoville}, {Abraham},
  {Aussel}, {Barnes}, {Benson}, {Blain}, {Calzetti}, {Comastri}, {Capak},
  {Carilli}, {Carlstrom}, {Carollo}, {Colbert}, {Daddi}, {Ellis}, {Elvis},
  {Ewald}, {Fall}, {Franceschini}, {Giavalisco}, {Green}, {Griffiths}, {Guzzo},
  {Hasinger}, {Impey}, {Kneib}, {Koda}, {Koekemoer}, {Lefevre}, {Lilly}, {Liu},
  {McCracken}, {Massey}, {Mellier}, {Miyazaki}, {Mobasher}, {Mould}, {Norman},
  {Refregier}, {Renzini}, {Rhodes}, {Rich}, {Sanders}, {Schiminovich},
  {Schinnerer}, {Scodeggio}, {Sheth}, {Shopbell}, {Taniguchi}, {Tyson}, {Urry},
  {Van Waerbeke}, {Vettolani}, {White}, \& {Yan}}]{scoville07b}
{Scoville}, N., {Abraham}, R.~G., {Aussel}, H., {et~al.} 2007{\natexlab{a}},
  \apjs, 172, 38

\bibitem[{{Scoville} {et~al.}(2013){Scoville}, {Arnouts}, {Aussel}, {Benson},
  {Bongiorno}, {Bundy}, {Calvo}, {Capak}, {Carollo}, {Civano}, {Dunlop},
  {Elvis}, {Faisst}, {Finoguenov}, {Fu}, {Giavalisco}, {Guo}, {Ilbert},
  {Iovino}, {Kajisawa}, {Kartaltepe}, {Leauthaud}, {Le F{\`e}vre}, {LeFloch},
  {Lilly}, {Liu}, {Manohar}, {Massey}, {Masters}, {McCracken}, {Mobasher},
  {Peng}, {Renzini}, {Rhodes}, {Salvato}, {Sanders}, {Sarvestani}, {Scarlata},
  {Schinnerer}, {Sheth}, {Shopbell}, {Smol{\v c}i{\'c}}, {Taniguchi}, {Taylor},
  {White}, \& {Yan}}]{scoville13}
{Scoville}, N., {Arnouts}, S., {Aussel}, H., {et~al.} 2013, \apjs, 206, 3

\bibitem[{{Scoville} {et~al.}(2007{\natexlab{b}}){Scoville}, {Aussel}, {Brusa},
  {Capak}, {Carollo}, {Elvis}, {Giavalisco}, {Guzzo}, {Hasinger}, {Impey},
  {Kneib}, {LeFevre}, {Lilly}, {Mobasher}, {Renzini}, {Rich}, {Sanders},
  {Schinnerer}, {Schminovich}, {Shopbell}, {Taniguchi}, \&
  {Tyson}}]{scoville07a}
{Scoville}, N., {Aussel}, H., {Brusa}, M., {et~al.} 2007{\natexlab{b}}, \apjs,
  172, 1

\bibitem[{{Scoville} {et~al.}(2016){Scoville}, {Sheth}, {Aussel}, {Vanden
  Bout}, {Capak}, {Bongiorno}, {Casey}, {Murchikova}, {Koda},
  {{\'A}lvarez-M{\'a}rquez}, {Lee}, {Laigle}, {McCracken}, {Ilbert}, {Pope},
  {Sanders}, {Chu}, {Toft}, {Ivison}, \& {Manohar}}]{scoville16}
{Scoville}, N., {Sheth}, K., {Aussel}, H., {et~al.} 2016, \apj, 820, 83

\bibitem[{{Shapley} {et~al.}(2003){Shapley}, {Steidel}, {Pettini}, \&
  {Adelberger}}]{alice03}
{Shapley}, A.~E., {Steidel}, C.~C., {Pettini}, M., \& {Adelberger}, K.~L. 2003,
  \apj, 588, 65

\bibitem[{{Shapley}(1930)}]{shapley30}
{Shapley}, H. 1930, Harvard College Observatory Bulletin, 874, 9

\bibitem[{{Shapley} \& {Ames}(1926)}]{shapleynames26}
{Shapley}, H. \& {Ames}, A. 1926, Harvard College Observatory Circular, 294, 1

\bibitem[{{Shattow} {et~al.}(2013){Shattow}, {Croton}, {Skibba}, {Muldrew},
  {Pearce}, \& {Abbas}}]{shattow13}
{Shattow}, G.~M., {Croton}, D.~J., {Skibba}, R.~A., {et~al.} 2013, \mnras, 433,
  3314

\bibitem[{{Smol{\v c}i{\'c}} {et~al.}(2014){Smol{\v c}i{\'c}}, {Ciliegi},
  {Jeli{\'c}}, {Bondi}, {Schinnerer}, {Carilli}, {Riechers}, {Salvato},
  {Brkovi{\'c}}, {Capak}, {Ilbert}, {Karim}, {McCracken}, \&
  {Scoville}}]{vernesa14}
{Smol{\v c}i{\'c}}, V., {Ciliegi}, P., {Jeli{\'c}}, V., {et~al.} 2014, \mnras,
  443, 2590

\bibitem[{{Smol{\v c}i{\'c}} {et~al.}(2015){Smol{\v c}i{\'c}}, {Karim},
  {Miettinen}, {Novak}, {Magnelli}, {Riechers}, {Schinnerer}, {Capak}, {Bondi},
  {Ciliegi}, {Aravena}, {Bertoldi}, {Bourke}, {Banfield}, {Carilli}, {Civano},
  {Ilbert}, {Intema}, {Le F{\`e}vre}, {Finoguenov}, {Hallinan}, {Kl{\"o}ckner},
  {Koekemoer}, {Laigle}, {Masters}, {McCracken}, {Mooley}, {Murphy},
  {Navarette}, {Salvato}, {Sargent}, {Sheth}, {Toft}, \&
  {Zamorani}}]{vernesa15}
{Smol{\v c}i{\'c}}, V., {Karim}, A., {Miettinen}, O., {et~al.} 2015, \aap, 576,
  A127

\bibitem[{{Smol{\v c}i{\'c}} {et~al.}(2017{\natexlab{a}}){Smol{\v c}i{\'c}},
  {Miettinen}, {Tomi{\v c}i{\'c}}, {Zamorani}, {Finoguenov}, {Lemaux},
  {Aravena}, {Capak}, {Chiang}, {Civano}, {Delvecchio}, {Ilbert}, {Jurlin},
  {Karim}, {Laigle}, {Le F{\`e}vre}, {Marchesi}, {McCracken}, {Riechers},
  {Salvato}, {Schinnerer}, {Tasca}, \& {Toft}}]{vernesa17}
{Smol{\v c}i{\'c}}, V., {Miettinen}, O., {Tomi{\v c}i{\'c}}, N., {et~al.}
  2017{\natexlab{a}}, \aap, 597, A4

\bibitem[{{Smol{\v c}i{\'c}} {et~al.}(2017{\natexlab{b}}){Smol{\v c}i{\'c}},
  {Novak}, {Bondi}, {Ciliegi}, {Mooley}, {Schinnerer}, {Zamorani}, {Navarrete},
  {Bourke}, {Karim}, {Vardoulaki}, {Leslie}, {Delhaize}, {Carilli}, {Myers},
  {Baran}, {Delvecchio}, {Miettinen}, {Banfield}, {Balokovi{\'c}}, {Bertoldi},
  {Capak}, {Frail}, {Hallinan}, {Hao}, {Herrera Ruiz}, {Horesh}, {Ilbert},
  {Intema}, {Jeli{\'c}}, {Kl{\"o}ckner}, {Krpan}, {Kulkarni}, {McCracken},
  {Laigle}, {Middleberg}, {Murphy}, {Sargent}, {Scoville}, \&
  {Sheth}}]{vernesa17b}
{Smol{\v c}i{\'c}}, V., {Novak}, M., {Bondi}, M., {et~al.} 2017{\natexlab{b}},
  \aap, 602, A1

\bibitem[{{Soucail} {et~al.}(1987){Soucail}, {Fort}, {Mellier}, \&
  {Picat}}]{soucail87}
{Soucail}, G., {Fort}, B., {Mellier}, Y., \& {Picat}, J.~P. 1987, \aap, 172,
  L14

\bibitem[{{Stanway} {et~al.}(2004){Stanway}, {Bunker}, {McMahon}, {Ellis},
  {Treu}, \& {McCarthy}}]{stanway04}
{Stanway}, E.~R., {Bunker}, A.~J., {McMahon}, R.~G., {et~al.} 2004, \apj, 607,
  704

\bibitem[{{Steidel} {et~al.}(1998){Steidel}, {Adelberger}, {Dickinson},
  {Giavalisco}, {Pettini}, \& {Kellogg}}]{steidel98}
{Steidel}, C.~C., {Adelberger}, K.~L., {Dickinson}, M., {et~al.} 1998, \apj,
  492, 428

\bibitem[{{Steidel} {et~al.}(1999){Steidel}, {Adelberger}, {Giavalisco},
  {Dickinson}, \& {Pettini}}]{steidel99}
{Steidel}, C.~C., {Adelberger}, K.~L., {Giavalisco}, M., {Dickinson}, M., \&
  {Pettini}, M. 1999, \apj, 519, 1

\bibitem[{{Szalay} {et~al.}(1999){Szalay}, {Connolly}, \& {Szokoly}}]{szalay99}
{Szalay}, A.~S., {Connolly}, A.~J., \& {Szokoly}, G.~P. 1999, \aj, 117, 68

\bibitem[{{Tacconi} {et~al.}(2010){Tacconi}, {Genzel}, {Neri}, {Cox}, {Cooper},
  {Shapiro}, {Bolatto}, {Bouch{\'e}}, {Bournaud}, {Burkert}, {Combes},
  {Comerford}, {Davis}, {Schreiber}, {Garcia-Burillo}, {Gracia-Carpio}, {Lutz},
  {Naab}, {Omont}, {Shapley}, {Sternberg}, \& {Weiner}}]{tacconi10}
{Tacconi}, L.~J., {Genzel}, R., {Neri}, R., {et~al.} 2010, \nat, 463, 781

\bibitem[{{Talia} {et~al.}(2016){Talia}, {Cimatti}, {Brusa}, {Lemaux},
  {Amorin}, {Bardelli}, {Cassar{\`a}}, {Cucciati}, {Garilli}, {Grazian},
  {Guaita}, {Hathi}, {Koekemoer}, {Le F{\`e}vre}, {Maccagni}, {Nakajima},
  {Pentericci}, {Pforr}, {Schaerer}, {Vanzella}, {Vergani}, {Zamorani}, \&
  {Zucca}}]{marghe16}
{Talia}, M., {Cimatti}, A., {Brusa}, M., {et~al.} 2016, ArXiv e-prints
  [\eprint[arXiv]{1611.05884}]

\bibitem[{{Talia} {et~al.}(2012){Talia}, {Mignoli}, {Cimatti}, {Kurk}, {Berta},
  {Bolzonella}, {Cassata}, {Daddi}, {Dickinson}, {Franceschini}, {Halliday},
  {Pozzetti}, {Renzini}, {Rodighiero}, {Rosati}, \& {Zamorani}}]{marghe12}
{Talia}, M., {Mignoli}, M., {Cimatti}, A., {et~al.} 2012, \aap, 539, A61

\bibitem[{{Taniguchi} {et~al.}(2007){Taniguchi}, {Scoville}, {Murayama},
  {Sanders}, {Mobasher}, {Aussel}, {Capak}, {Ajiki}, {Miyazaki}, {Komiyama},
  {Shioya}, {Nagao}, {Sasaki}, {Koda}, {Carilli}, {Giavalisco}, {Guzzo},
  {Hasinger}, {Impey}, {LeFevre}, {Lilly}, {Renzini}, {Rich}, {Schinnerer},
  {Shopbell}, {Kaifu}, {Karoji}, {Arimoto}, {Okamura}, \& {Ohta}}]{taniguchi07}
{Taniguchi}, Y., {Scoville}, N., {Murayama}, T., {et~al.} 2007, \apjs, 172, 9

\bibitem[{{Tasca} {et~al.}(2016){Tasca}, {Le Fevre}, {Ribeiro}, {Thomas},
  {Moreau}, {Cassata}, {Garilli}, {Le Brun}, {Lemaux}, {Maccagni},
  {Pentericci}, {Schaerer}, {Vanzella}, {Zamorani}, {Zucca}, {Amorin},
  {Bardelli}, {Cassara}, {Castellano}, {Cimatti}, {Cucciati}, {Durkalec},
  {Fontana}, {Giavalisco}, {Grazian}, {Hathi}, {Ilbert}, {Paltani}, {Pforr},
  {Scodeggio}, {Sommariva}, {Talia}, {Tresse}, {Vergani}, {Capak}, {Charlot},
  {Contini}, {de la Torre}, {Dunlop}, {Fotopoulou}, {Guaita}, {Koekemoer},
  {Lopez-Sanjuan}, {Mellier}, {Salvato}, {Scoville}, {Taniguchi}, \&
  {Wang}}]{tizzytasca16}
{Tasca}, L.~A.~M., {Le Fevre}, O., {Ribeiro}, B., {et~al.} 2016, ArXiv e-prints
  [\eprint[arXiv]{1602.01842}]

\bibitem[{{Thomas} {et~al.}(2017{\natexlab{a}}){Thomas}, {Le F{\`e}vre}, {Le
  Brun}, {Cassata}, {Garilli}, {Lemaux}, {Maccagni}, {Pentericci}, {Tasca},
  {Zamorani}, {Zucca}, {Amorin}, {Bardelli}, {Cassar{\`a}}, {Castellano},
  {Cimatti}, {Cucciati}, {Durkalec}, {Fontana}, {Giavalisco}, {Grazian},
  {Hathi}, {Ilbert}, {Paltani}, {Pforr}, {Ribeiro}, {Schaerer}, {Scodeggio},
  {Sommariva}, {Talia}, {Tresse}, {Vanzella}, {Vergani}, {Capak}, {Charlot},
  {Contini}, {Cuby}, {de la Torre}, {Dunlop}, {Fotopoulou}, {Koekemoer},
  {L{\'o}pez-Sanjuan}, {Mellier}, {Salvato}, {Scoville}, {Taniguchi}, \&
  {Wang}}]{romain17a}
{Thomas}, R., {Le F{\`e}vre}, O., {Le Brun}, V., {et~al.} 2017{\natexlab{a}},
  \aap, 597, A88

\bibitem[{{Thomas} {et~al.}(2017{\natexlab{b}}){Thomas}, {Le F{\`e}vre},
  {Scodeggio}, {Cassata}, {Garilli}, {Le Brun}, {Lemaux}, {Maccagni}, {Pforr},
  {Tasca}, {Zamorani}, {Bardelli}, {Hathi}, {Tresse}, {Zucca}, \&
  {Koekemoer}}]{romain17b}
{Thomas}, R., {Le F{\`e}vre}, O., {Scodeggio}, M., {et~al.} 2017{\natexlab{b}},
  \aap, 602, A35

\bibitem[{{Tomczak} {et~al.}(2014){Tomczak}, {Quadri}, {Tran}, {Labb{\'e}},
  {Straatman}, {Papovich}, {Glazebrook}, {Allen}, {Brammer}, {Kacprzak},
  {Kawinwanichakij}, {Kelson}, {McCarthy}, {Mehrtens}, {Monson}, {Persson},
  {Spitler}, {Tilvi}, \& {van Dokkum}}]{AA14}
{Tomczak}, A.~R., {Quadri}, R.~F., {Tran}, K.-V.~H., {et~al.} 2014, \apj, 783,
  85

\bibitem[{{Toshikawa} {et~al.}(2016){Toshikawa}, {Kashikawa}, {Overzier},
  {Malkan}, {Furusawa}, {Ishikawa}, {Onoue}, {Ota}, {Tanaka}, {Niino}, \&
  {Uchiyama}}]{tersetoshikawa16}
{Toshikawa}, J., {Kashikawa}, N., {Overzier}, R., {et~al.} 2016, \apj, 826, 114

\bibitem[{{Toshikawa} {et~al.}(2014){Toshikawa}, {Kashikawa}, {Overzier},
  {Shibuya}, {Ishikawa}, {Ota}, {Shimasaku}, {Tanaka}, {Hayashi}, {Niino}, \&
  {Onoue}}]{tersetoshikawa14}
{Toshikawa}, J., {Kashikawa}, N., {Overzier}, R., {et~al.} 2014, \apj, 792, 15

\bibitem[{{Trainor} {et~al.}(2015){Trainor}, {Steidel}, {Strom}, \&
  {Rudie}}]{trainor15}
{Trainor}, R.~F., {Steidel}, C.~C., {Strom}, A.~L., \& {Rudie}, G.~C. 2015,
  \apj, 809, 89

\bibitem[{{Ueda} {et~al.}(2014){Ueda}, {Akiyama}, {Hasinger}, {Miyaji}, \&
  {Watson}}]{ueda14}
{Ueda}, Y., {Akiyama}, M., {Hasinger}, G., {Miyaji}, T., \& {Watson}, M.~G.
  2014, \apj, 786, 104

\bibitem[{{van der Burg} {et~al.}(2014){van der Burg}, {Muzzin}, {Hoekstra},
  {Wilson}, {Lidman}, \& {Yee}}]{vanderburg14}
{van der Burg}, R.~F.~J., {Muzzin}, A., {Hoekstra}, H., {et~al.} 2014, \aap,
  561, A79

\bibitem[{{Vanzella} {et~al.}(2005){Vanzella}, {Cristiani}, {Dickinson},
  {Kuntschner}, {Moustakas}, {Nonino}, {Rosati}, {Stern}, {Cesarsky}, {Ettori},
  {Ferguson}, {Fosbury}, {Giavalisco}, {Haase}, {Renzini}, {Rettura}, {Serra},
  \& {GOODS Team}}]{vanzella05}
{Vanzella}, E., {Cristiani}, S., {Dickinson}, M., {et~al.} 2005, \aap, 434, 53

\bibitem[{{Venemans} {et~al.}(2002{\natexlab{a}}){Venemans}, {Kurk}, {Miley},
  {R{\"o}ttgering}, {van Breugel}, {Carilli}, {De Breuck}, {Ford}, {Heckman},
  {McCarthy}, \& {Pentericci}}]{venemens02}
{Venemans}, B.~P., {Kurk}, J.~D., {Miley}, G.~K., {et~al.} 2002{\natexlab{a}},
  \apjl, 569, L11

\bibitem[{{Venemans} {et~al.}(2002{\natexlab{b}}){Venemans}, {Kurk}, {Miley},
  {R{\"o}ttgering}, {van Breugel}, {Carilli}, {De Breuck}, {Ford}, {Heckman},
  {McCarthy}, \& {Pentericci}}]{venemans02}
{Venemans}, B.~P., {Kurk}, J.~D., {Miley}, G.~K., {et~al.} 2002{\natexlab{b}},
  \apjl, 569, L11

\bibitem[{{Wang} {et~al.}(2016){Wang}, {Elbaz}, {Daddi}, {Finoguenov}, {Liu},
  {Schreiber}, {Mart{\'{\i}}n}, {Strazzullo}, {Valentino}, {van der Burg},
  {Zanella}, {Ciesla}, {Gobat}, {Le Brun}, {Pannella}, {Sargent}, {Shu}, {Tan},
  {Cappelluti}, \& {Li}}]{wang16}
{Wang}, T., {Elbaz}, D., {Daddi}, E., {et~al.} 2016, \apj, 828, 56

\bibitem[{{Wardlow} {et~al.}(2014){Wardlow}, {Malhotra}, {Zheng},
  {Finkelstein}, {Bock}, {Bridge}, {Calanog}, {Ciardullo}, {Conley}, {Cooray},
  {Farrah}, {Gawiser}, {Gronwall}, {Heinis}, {Ibar}, {Ivison}, {Marsden},
  {Oliver}, {Rhoads}, {Riechers}, {Schulz}, {Smith}, {Viero}, {Wang}, \&
  {Zemcov}}]{wardlow14}
{Wardlow}, J.~L., {Malhotra}, S., {Zheng}, Z., {et~al.} 2014, \apj, 787, 9

\bibitem[{{Weiner} {et~al.}(2009){Weiner}, {Coil}, {Prochaska}, {Newman},
  {Cooper}, {Bundy}, {Conselice}, {Dutton}, {Faber}, {Koo}, {Lotz}, {Rieke}, \&
  {Rubin}}]{weiner09}
{Weiner}, B.~J., {Coil}, A.~L., {Prochaska}, J.~X., {et~al.} 2009, \apj, 692,
  187

\bibitem[{{Wilkinson} {et~al.}(2017){Wilkinson}, {Almaini}, {Chen}, {Smail},
  {Arumugam}, {Blain}, {Chapin}, {Chapman}, {Conselice}, {Cowley}, {Dunlop},
  {Farrah}, {Geach}, {Hartley}, {Ivison}, {Maltby}, {Micha{\l}owski},
  {Mortlock}, {Scott}, {Simpson}, {Simpson}, {van der Werf}, \&
  {Wild}}]{wilkinson17}
{Wilkinson}, A., {Almaini}, O., {Chen}, C.-C., {et~al.} 2017, \mnras, 464, 1380

\bibitem[{{Wylezalek} {et~al.}(2013){Wylezalek}, {Galametz}, {Stern}, {Vernet},
  {De Breuck}, {Seymour}, {Brodwin}, {Eisenhardt}, {Gonzalez}, {Hatch},
  {Jarvis}, {Rettura}, {Stanford}, \& {Stevens}}]{wylezalek13}
{Wylezalek}, D., {Galametz}, A., {Stern}, D., {et~al.} 2013, \apj, 769, 79

\bibitem[{{Zamojski} {et~al.}(2007){Zamojski}, {Schiminovich}, {Rich},
  {Mobasher}, {Koekemoer}, {Capak}, {Taniguchi}, {Sasaki}, {McCracken},
  {Mellier}, {Bertin}, {Aussel}, {Sanders}, {Le F{\`e}vre}, {Ilbert},
  {Salvato}, {Thompson}, {Kartaltepe}, {Scoville}, {Barlow}, {Forster},
  {Friedman}, {Martin}, {Morrissey}, {Neff}, {Seibert}, {Small}, {Wyder},
  {Bianchi}, {Donas}, {Heckman}, {Lee}, {Madore}, {Milliard}, {Szalay},
  {Welsh}, \& {Yi}}]{zamojski07}
{Zamojski}, M.~A., {Schiminovich}, D., {Rich}, R.~M., {et~al.} 2007, \apjs,
  172, 468

\bibitem[{{Zeimann} {et~al.}(2012){Zeimann}, {Stanford}, {Brodwin}, {Gonzalez},
  {Snyder}, {Stern}, {Eisenhardt}, {Mancone}, \& {Dey}}]{greg12}
{Zeimann}, G.~R., {Stanford}, S.~A., {Brodwin}, M., {et~al.} 2012, \apj, 756,
  115

\bibitem[{{Zheng} {et~al.}(2006){Zheng}, {Overzier}, {Bouwens}, {White},
  {Ford}, {Ben{\'{\i}}tez}, {Blakeslee}, {Bradley}, {Jee}, {Martel}, {Mei},
  {Zirm}, {Illingworth}, {Clampin}, {Hartig}, {Ardila}, {Bartko}, {Broadhurst},
  {Brown}, {Burrows}, {Cheng}, {Cross}, {Demarco}, {Feldman}, {Franx},
  {Golimowski}, {Goto}, {Gronwall}, {Holden}, {Homeier}, {Infante}, {Kimble},
  {Krist}, {Lesser}, {Menanteau}, {Meurer}, {Miley}, {Motta}, {Postman},
  {Rosati}, {Sirianni}, {Sparks}, {Tran}, \& {Tsvetanov}}]{zheng06}
{Zheng}, W., {Overzier}, R.~A., {Bouwens}, R.~J., {et~al.} 2006, \apj, 640, 574

\bibitem[{{Zheng} {et~al.}(2013){Zheng}, {Finkelstein}, {Finkelstein}, {Tilvi},
  {Rhoads}, {Malhotra}, {Wang}, {Miller}, {Hibon}, \& {Xia}}]{zheng13}
{Zheng}, Z.-Y., {Finkelstein}, S.~L., {Finkelstein}, K., {et~al.} 2013, \mnras,
  431, 3589

\bibitem[{{Zheng} {et~al.}(2016){Zheng}, {Malhotra}, {Rhoads}, {Finkelstein},
  {Wang}, {Jiang}, \& {Cai}}]{zheng16}
{Zheng}, Z.-Y., {Malhotra}, S., {Rhoads}, J.~E., {et~al.} 2016, \apjs, 226, 23

\bibitem[{{Zwicky}(1937)}]{zwicky37}
{Zwicky}, F. 1937, \apj, 86, 217

\bibitem[{{Zwicky} {et~al.}(1961){Zwicky}, {Herzog}, {Wild}, {Karpowicz}, \&
  {Kowal}}]{zwicky61}
{Zwicky}, F., {Herzog}, E., {Wild}, P., {Karpowicz}, M., \& {Kowal}, C.~T.
  1961, {Catalogue of galaxies and of clusters of galaxies, Vol. I}

\end{thebibliography}

\appendix

\section{COSMOS multiwavelength imaging data}

The COSMOS field, selected initially for its equatorial location, paucity of 
bright foreground objects, and extremely low Galactic extinction, first began to distinguish itself more than a decade ago following an unprecedented campaign 
with the \emph{Hubble} Space Telescope (\emph{HST}) which resulted in a 1.64 $\Box^{\circ}$
mosaic of the field compiled from 583 single-orbit pointings (5$\sigma$ depth $m_{F814W}\sim27.2$) with
the Advanced Camera for Surveys (ACS; \citealt{ford98}) in the $F814W$ band \citep{scoville07b,anton07}. In parallel, a 1 $\Box^{\circ}$
overlapping sub-section of the field was selected to be the second deep field (D2) of the Canada-France-Hawai'i Telescope Legacy Survey
(CFHTLS\footnote{http://www.cfht.hawaii.edu/Science/CFHTLS/}) which was, in conjunction with a proprietary
campaign \citep{boulade03}, to eventually provide imaging in $u^{\ast}/g^{\prime}/r^{\prime}/i^{\prime}/z^{\prime}$ to a 80\% point source completeness depth of
26.26/26.31/25.91/25.51/25.14\footnote{http://www.cfht.hawaii.edu/Science/CFHTLS/T0007/}, respectively, under exquisite conditions
(median seeing 0.58$\arcsec$). These two datasets would form the basis to motivate the subsequent training of almost every major telescope on the field, including:

\begin{itemize}
\renewcommand{\labelitemi}{--}

\item At radio wavelengths (10$-$90 cm) with the Very Large Array (\citealt{schinnerer07, schinnerer10, vernesa14}), the Giant Metrewave
Radio Telescope (\citealt{vernesa15}; Karim et al., \emph{in prep}), and the Jansky Very Large Array\footnote{http://jvla-cosmos.phy.hr/Home.html}
(\citealt{mladen15, vernesa17b}) to 5$\sigma$ depths ranging from $\sim$11.5-400 $\mu$Jy/beam (from 10$-$90 cm, respectively).

\item In the far-infrared (FIR; 100$-$500$\mu m$) with the \emph{Herschel} space observatory \citep{pilbratt10} Photodetector Array Camera and Spectrometer
(PACS; \citealt{poglitsch10}) as part of the PACS Evolutionary Probe (PEP; \citealt{lutz11}) and Spectral and Photometric and Imaging REceiver (SPIRE; \citealt{griffin10})
as part of the \emph{Herschel} Multi-tiered Extragalactic Survey (HerMES; \citealt{oliver12}) to $3\sigma$ depths of 4.5, 10.0, 18.4, 19.3, \& 21.2 mJy (including
confusion noise) in the 100, 160, 250, 350, \& 500 $\mu m$ channels, respectively.
%14.8, 13.9, 13.3, 13.2, \& 13.1 mags (including
% Mostly Javier's paper, but checked with Rodighiero et al. 2011 and scaling VVDS

\item In the near-, mid-, and far-infrared (3.6-160$\mu m$) with the all seven channels of the \emph{Spitzer} InfraRed Array Camera (IRAC; \citealt{fazio04}) and Multiband
Imaging Photometer for \emph{Spitzer} (MIPS; \citealt{rieke04}) for the COSMOS \emph{Spitzer} survey (S-COSMOS; \citealt{sanders07}) to 5$\sigma$ depths of 24.0, 23.5, 22.0, 21.2,
\& 19.0 mags in the [3.6], [4.5], [5.8], [8.0], \& 24$\mu m$ bands, respectively, and, recently, extremely deeply in the non-cryogenic IRAC bands as part of the Spitzer
Large Area Survey with Hyper-Suprime-Cam (SPLASH; \citealt{laigle16}) to $3\sigma$ depths of $\sim$25.4 in the both the [3.6] and [4.5] bands.
% Done with a combination of the VUDS-COSMOS photometric catalog and the S-COSMOS paper, Clotilde's catalog, and Emeric's paper in 2007 for MIPS. 

\item In the near-infrared (NIR) $Y/J/H/K_s$ bands (1.0-2.1$\mu m$) as the subject of the UltraVISTA survey \citep{mccracken12} to $3\sigma$ depths of
$m_{AB}\sim24.5$ in the ``Deep" 0.88 $\Box^{\circ}$ region of the survey and $m_{AB}\sim25$ in the smaller 0.62 $\Box^{\circ}$ "UltraDeep" portion of the survey,
manifest as a series of four equally spaced and almost equally sized N-S vertical strips.

\item A small sub-section of the field (0.056 $\Box^{\circ}$) in the optical/NIR (0.6-1.6$\mu m$) as part of the Cosmic Assembly Near-infrared Deep Extragalactic
Survey (CANDELS; \citealt{grogin11,anton11}) to 5$\sigma$ depths of 28.2, 28.4, 27.0, and 26.9 in the $F606W$, $F814W$, $F125W$, and $F160W$ bands, respectively.

\item At optical wavelengths (0.4-0.9$\mu m$) with Subaru/Suprime-Cam employing a large variety of narrow, medium, and broadband ($B/V/r^{+}/i^{+}/z^{+}$) filters
\citep{taniguchi07} as well as new imaging in $z^{++}$ as part of the SPLASH survey reaching 5$\sigma$ depths of $\sim26-27$ mags in all broadbands and
$\sim25-26$ mags in all medium/narrow bands.

\item In the ultraviolet (UV) with both channels (1500-2300\AA) of the GALaxy Evolution eXplorer (\emph{GALEX}; \citealt{martin05}) to 3$\sigma$ depths of 25.8 and
25.5 in the far-UV ($FUV$) and near-UV ($NUV$) bands, respectively \citep{zamojski07}.
% This was done from the VUDS-COSMOS master photometric catalog, 3sigma depth = mean by eye of the locus of points at sigma_m = 0.36, which is the sigma_m of a S/N=3 f_v det.

\item With the X-Ray Multi-mirror Mission space telescope (\emph{XMM-Newton}; \citealt{jansen01}) as the subject of the \emph{XMM-COSMOS} survey
\citep{cappelluti09, brusa10} and the \emph{Chandra} X-ray observatory Advanced CCD Imaging Spectrometer (ACIS; \citealt{garmire03}) as
the subject of the \emph{Chandra} COSMOS survey (C-COSMOS; \citealt{elvis09}) and the more recent \emph{Chandra} COSMOS-Legacy Survey \citep{civano16} to $\sim$3$\sigma$
point source depths ranging from $4\times10^{-16}-$1$\times10^{-14}$ ergs s$^{-1}$ cm$^{2}$ depending on the band and the telescope.

\end{itemize}

\noindent as well as a variety of other observations including those at sub-millimeter wavelengths (e.g., \citealt{scott08, Aretxaga11, connivingcasey13, miettinen17, geach17}).
For a complete listing of all observations taken in the field see the COSMOS website\footnote{http://cosmos.astro.caltech.edu/page/astronomers}.
Details of each observation, data-reduction processes, the depth of the imaging, and the matching of various images are related in the referenced studies.
For an overview of the current state of the data covering the COSMOS field see \citet{laigle16}.
%for that of far-IR see \citet{javier16}, for that of the X-ray data see \citet{salvato11}, and for that of radio see \citet{vernesa15}. 

\section{Proto-cluster member and coeval field sample definitions}

Analysis of the VUDS spectroscopic sample throughout \S\ref{properties} is limited to those galaxies with secure spectroscopic 
redshifts (see \S\ref{dongdong}) in the redshift range $ 4.23 \le z_{spec} \le 4.88$. This redshift window is imposed to create a roughly $\pm$100 Myr  % 1.26148e+10 -> 1.25092e+10 -> 1.23752e+10 yr
window centered on the epoch at which the proto-cluster is observed. We additionally impose the criteria $M_r<-22$, 
$\log(\mathcal{M}_{\ast}/M_{\odot})\ge9.7$, and $M_{NUV}-M_{r} < -14.4- 0.72M_r$ (see Fig. \ref{fig:restframeCMDnCSMD}) for 
any galaxy with a secure spectroscopic redshift to enter our analysis. From one- and two-dimensional Kolmogorov-Smirnov (KS) tests it was 
found that in this region of phase space the distributions of the secure VUDS spectral sample in $M_{NUV}-M_r$/$M_r-M_J$, $M_r$, and 
$\mathcal{M}_{\ast}$ are indistinguishable from all $z_{phot}$ objects at $ 4.23 \le z_{phot} \le 4.88$ subject to the same criteria and those 
additional criteria given in \S\ref{voronoi}. In other words, in this region of four-dimensional phase space, the secure 
$z_{spec}$ sample appears \emph{representative} of the entire field population at these redshifts. Secure $z_{spec}$ members are those galaxies
at $4.53\le z_{spec}\le 4.60$ which fall within $R_{proj}\le 2$ Mpc from the number-weighted center. All nine secure
$z_{spec}$ members (hereafter simply members) persist in the sample after applying these criteria. The remaining 54 galaxies comprise the
$z_{spec}$ coeval field sample (hereafter coeval field).

For all comparisons involving photometric redshifts we adopt the redshift range $4.53-1.5\sigma_{NMAD}(1+4.53) \le z_{phot} \le 4.60+1.5\sigma_{NMAD}*(1+4.60)$ and
impose a stellar mass cut making no rest-frame color cut. An additional rest-frame color cut as was, for example, done for the spectral sample is not needed
as once we impose a stellar mass cut above which the data are highly complete, the resulting sample will be, by definition, representative
of the overall galaxy population at these redshifts and above the stellar mass limit imposed.
In order to determine the appropriate stellar mass cut for the $z_{phot}$ sample we choose not to adopt the limits of \citet{laigle16},
$\log(\mathcal{M}_{\ast}/M_{\odot})\ge10.7$ and 10.3 in the UltraVISTA Deep and UltraDeep regions, respectively, as these limits are
based solely on the $K_{s}$ band magnitude at which 90\% of all sources could be recovered. In order to estimate the stellar mass limit of our
(broadly) [3.6]-selected %\footnote{Despite the fact that the COSMOS2015 uses a $z^{++}YJHK_{s}$ $\chi^2$ detection image, due in part to the 
%extremely deep $z^{++}$ imaging, are large number of objects while being undetected in the individual NIR bands in UltraVISTA are cataloged and
%detected significantly at [3.6] (see \citealt{laigle16, idolatrousiary17}).}
$z_{phot}$ sample we adopted approach identical to that employed in \citet{quadri12} and \citet{AA14}. This approach is as follows.
%Adopting slightly more advanced methodologies \citep{vanderBurg13} results in a limit of $\log(\mathcal{M}_{\ast}/M_{\odot})\ga9.3$ for the sample selected in this study} % I did something similar in notes_on_VUDS_representativeness_for_COSMOSproto-structure26.notes and got 10^10.2 and checked a maximally old CB07 template for IRAC1 < 25.3 and got ~10^9 and verified that a 10^10.3 galaxy corresponds to a K_band ~ 24.8 and a 10^10.7 corresponds to K_s ~ 24.0, use K_correct_wz.pro with K-band and IRAC1. In the histogram of SM in this redshift range the fall off occurs somewhere between 10^9.3 and 10^9.7 

We began by selecting all objects in the same $z_{phot}$ range which were $\sim$2.5$\times$ brighter in flux density (i.e., 1 mag) than the
3$\sigma$ magnitude limit at [3.6] within a narrow window of $\pm$0.1 mags. Because of their brightness relative to the 3$\sigma$ detection limit,
essentially all objects at these magnitudes in this $z_{phot}$ range should be detected in the COSMOS2015 catalog. The stellar mass distribution
of the resultant sample was then scaled down by exactly the same factor as the ratio of the average flux density of the sample and the flux density
corresponding to the 3$\sigma$ completeness limit. By scaling in such a way, we artificially reproduce the stellar mass distribution (though not the number)
of objects appearing right at the 3$\sigma$ magnitude limit modulo changes in the average $\mathcal{M}_{\ast}/L$ ratio. Since we are working with
a sample almost exclusively comprised of star-forming galaxies by virtue of the redshift considered, a constant $\mathcal{M}_{\ast}/L$ as a function
of stellar mass within the range considered here is at least a plausible assumption. The stellar mass completeness limit (hereafter $\mathcal{M}_{lim}$)
was then defined as the stellar mass for which 80\% of the objects in the scaled stellar mass distribution were more massive. In other words, since the
scaled mass distribution is intended to represent the true stellar mass distribution at the $3\sigma$ completeness limit of the image, and since objects
with lower stellar masses are more likely to be missed given the selection band and the redshift, removing the bottom 20\% of scaled stellar masses
returns the 80\% $\mathcal{M}_{lim}$. This exercise resulted in a $\mathcal{M}_{lim}$
of $\log(\mathcal{M}_{\ast}/M_{\odot})\ge9.5$, a limit which was not sensitive to the exact magnitude range chosen in this exercise. This
limit corresponds to the stellar mass at which the number counts of all $z_{phot}$ objects in the COSMOS2015 catalog in this redshift range
subject to the constraints given in \S\ref{voronoi} begin to turn over. Additionally, this number is similar to the recent $\mathcal{M}_{lim}$
estimated by \citet{idolatrousiary17} at the same redshift for a [3.6]-selected COSMOS2015 sample using entirely different methodologies.
%all comparisons made are internal, it is likely that any bias induced by using this lower stellar mass threshold is minimal. Regardless, all conclusions
%presented in this paper are robust against adopting a more conservative stellar mass threshold. 

Photometric redshift members (hereafter $z_{phot}$ members) were defined as objects which satisfy all above criteria and fall within $R_{proj}\le 3$ Mpc
from the spectral-number-weighted center (for the rationale behind this choice see \S\ref{zphotoverdens}). There are 82 such objects. The remaining
1975 objects comprise the $z_{phot}$ coeval field sample. We note that             % 34 and 661 at logM>10.1
the ($z_{spec}$) members and coeval field galaxies are not removed from these samples as to do so would impose unnecessary spatial and color bias on the
$z_{phot}$ sample. For all comparisons in this section and throughout the remainder of the paper, except where otherwise noted, we use the term average
synonymously with median, preferred over the mean to minimize the effects of outliers in what are admittedly relatively sparsely sampled distributions,
and denote the median of any parameter by the notation $\tilde{x}$. Uncertainties on the median are in all cases given by $\sigma_{NMAD}/\sqrt{n-1}$
(see \citealt{muller00} for a discussion on adopting this type of estimate on the uncertainty on the median). We proceed, for most comparisons throughout
the remainder of the paper, by diligently testing differences in various subsets of galaxies by both one-dimensional KS tests and by comparing medians
and their associated uncertainties. These are broadly conservative choices, as if we instead use the mean as an estimator of the average value or if
we substitute one-dimensional KS tests with Student's $t$ or a Mann-Whitney tests the significance of differences between subsets are generally seen to
increase.

% Results of KS/median tests (the KS results do not appear to change in the cases where I used a Mann-Whitney test or a Student's t-test instead):
% 
% Photo-z members vs. non (I did this outside of a program myself in notes_on_VUDS_representativeness_for_COSMOSproto-structure26.notes) Mr, MNUV-Mr, SM
% 0.749870, 0.540502, 0.451020, with median log(SM) values of 10.151100, 10.091300, respectively
%
% Photo-z members vs. non MNUV-Mr >= 1.5, Mr, MNUV-Mr, SM
% 0.632260, 0.685035, 0.317670, with median log(SM) values of 10.671100, 10.314400, respectively and mean log(SM) values of 11.00+/-0.10 and 10.84+/-0.20
%
% Spec-z members vs. non MNUV-Mr >= 1.5
% mean log(SM) values of 10.77+/-0.14 and 10.41+/-0.08, respectively
%
% Spectral members vs. non Mr, MNUV-Mr, Mr-MJ, SM, beta slope, L(Lya), EW(Lya), r_T^50/SM^alpha
% 0.787849, 0.518581, 0.435101, 0.0138056, 0.785759, 0.00422153, 0.0116006, I dunno but something high, with median log(SM) of 10.3493, 10.0977, respectively
% median log(SM) is 10.35 +/- 0.15 and 9.96 +/- 0.05, respectively
% mean log(SM) is 10.48 +/- 0.14 and 9.96 +/- 0.07
%
% Romain Spectral Mass, Spectral Age
% 0.000905295, 0.0156202, median log(SM) is 10.508000 and 10.047000, respectively
% mean SM and age for the member and coeval field is 10.622491, 10.131985 and 8.9156342, 8.7240196, respectively
 
\end{document}